\documentclass[aps,prc,reprint,superscriptaddress,nofootinbib]{revtex4-2}
\usepackage{graphicx}
\usepackage{ifpdf}
\usepackage{ulem}
\usepackage{xspace}
\usepackage{xcolor}
\usepackage{dsfont}
\usepackage{bm}
\usepackage{amsfonts}
\usepackage{amsmath}
\usepackage{amssymb}
\usepackage{bbold}
\usepackage{xurl} 
\usepackage{enumitem}
\usepackage{hyperref}
\usepackage{axodraw2}
\usepackage{pstricks}
\usepackage{relsize,exscale,scalefnt,anyfontsize,cases}
\usepackage[utf8]{inputenc}
\usepackage{slashed}

\graphicspath{{./figs/}}

\begin{document} 

\title{Formulation of fully covariant Quantum-Molecular Dynamics for an N-body system with scalar and vector potentials}

\author{Jiaxing Zhao}
\email{jzhao@itp.uni-frankfurt.de}
\affiliation{Institute for Theoretical Physics, Johann Wolfgang Goethe Universit\"{a}t, Frankfurt am Main, Germany}
\affiliation{Helmholtz Research Academy Hessen for FAIR (HFHF),GSI Helmholtz Center for Heavy Ion Research. Campus Frankfurt, 60438 Frankfurt, Germany}

\author{Joerg Aichelin}
\email{aichelin@subatech.in2p3.fr}
\affiliation{SUBATECH UMR 6457 (IMT Atlantique,  Universit\'{e} de Nantes, IN2P3/CNRS), 4 Rue Alfred Kastler, F-44307 Nantes, France}
\affiliation{Frankfurt Institute for Advanced Studies, Ruth-Moufang-Strasse 1, 60438 Frankfurt am Main, Germany}

\author{Elena Bratkovskaya}
\email{E.Bratkovskaya@gsi.de}
\affiliation{GSI Helmholtzzentrum f\"{u}r Schwerionenforschung GmbH, Planckstrasse 1, 64291 Darmstadt, Germany}
\affiliation{Institute for Theoretical Physics, Johann Wolfgang Goethe Universit\"{a}t, Frankfurt am Main, Germany}
\affiliation{Helmholtz Research Academy Hessen for FAIR (HFHF),GSI Helmholtz Center for Heavy Ion Research. Campus Frankfurt, 60438 Frankfurt, Germany}

\begin{abstract}
We present a fully covariant transport framework for Molecular Dynamics that enables a consistent description of the evolution of relativistic N-body systems. We derive relativistic equations of motion of a system of particles interacting with both scalar and vector two body interactions within a manifestly covariant formulation. This approach, without approximations, addresses several fundamental issues in relativistic many-body dynamics: the implications of different choices of time-constraints, the emergence of the non-relativistic limit, the frame independence of the system’s evolution, and the distinct dynamical roles of scalar and vector potentials. These aspects are investigated in detail for the scattering of two- and four-body systems, offering new insights into the consistency and physical interpretation of relativistic interactions in a covariant setting.
\end{abstract}
\date{\today}

\maketitle

\section{Introduction}

The description of the time evolution of interacting many-body systems of particles with relativistic velocities is a central problem across nuclear, hadronic, and particle physics. Whenever strongly interacting systems are created under extreme conditions---such as in heavy-ion collisions, astrophysical environments, or the early universe---their microscopic constituents evolve far from equilibrium while exchanging energy, momentum, and quantum numbers. The challenge lies in bridging relativistic quantum field theory (QFT) with transport and kinetic approaches that remain tractable for many-body systems. At the most fundamental level, the evolution is governed by QFT, where fields interact according to the Lagrangian of the underlying theory. However, a direct solution of the full many-body QFT dynamics is not feasible except for very small systems.  

Even in classical systems, the description of many-body dynamics poses formidable challenges, although such systems frequently occur in nuclear and astrophysical contexts. In nuclear physics, essentially two families of relativistic transport approaches have been developed: (a) the Boltzmann-Vlasov type, usually referred to as the Boltzmann-Uehling-Uhlenbeck (BUU) method, and (b) the molecular dynamics type, usually called the Quantum Molecular Dynamics (QMD) method. For detailed comparisons, we refer the reader to the comprehensive reviews in Refs.~\cite{Danielewicz:1982kk,Bertsch:1988ik,Botermans:1990qi,Cassing:1990dr,Bonasera:1994zz,Cassing:1999es,Xu:2019hqg,TMEP:2016tup,TMEP:2022xjg,Bleicher:2022kcu}. 

The BUU approach can be explicitly derived from the Bogoliubov-Born-Green-Kirkwood-Yvon (BBGKY) hierarchy by truncation at the two-body level. We mention that the formulation of many-body dynamics up to infinite order has been proposed in Refs. \cite{Cassing:1990dr, Cassing:2025pdm} based on correlation dynamics within the cluster expansion method, however, the numerical solution of this bound equations are still out of power of modern computers, thus a truncation scheme has to be applied.
The truncation procedure on the 2PI (2-particle-irreducible) level  leads to a transport equation for the one-body phase-space distribution function $f({\bf r},{\bf p},t)$, which evolves under the influence of a mean-field potential and a collision integral describing two-body scattering, including Pauli blocking for fermions and Bose enhancement for bosons. Covariant formulations of BUU models exist~\cite{Cassing:1990dr,Cassing:1999es,Cassing:2021fkc}, as well as its extension to $n\leftrightarrow m$ collision integrals  \cite{Cassing:2001ds}.
Further development of the transport approach for the one-body phase-space distribution function has been presented in \cite{Cassing:1999wx,Cassing:1999mh} by the realization of a microscopic covariant off-shell dynamical approach PHSD  (Parton-Hadron-String Dynamics) for strongly interacting systems, formulated as the
first order gradient expansion~\cite{Cassing:2008sv,Moreau:2019vhw}
of Kadanoff-Baym equations for Green's functions in phase-space representation. The BUU models are computationally efficient and well-suited for describing the time evolution of single-particle observables. However, the higher order correlations are lost due to the truncation. 
There are several flavors of BUU type transport approaches that incorporate relativistic features either through relativistic kinematics and/or by introducing a relativistic mean-field (RMF) potential. Representative implementations include pBUU~\cite{Danielewicz:1991dh}, RVUU~\cite{Ko:1987gp},
RBUU~\cite{Cassing:1990dr}, 
HSD~\cite{Cassing:1999es},
 IBUU~\cite{Li:2008gp}, GiBUU~\cite{Buss:2011mx}, 
 SMASH~\cite{SMASH:2016zqf}, 
 DJBUU~\cite{Kim:2020sjy}, AMPT~\cite{Lin:2004en}, and BAMPS \cite{Xu:2004mz}.  

The QMD approach, by contrast, follows a different philosophy. Instead of working with a phase-space distribution function, QMD models represent nucleons (or partons) as localized Gaussian wave packets whose centroids in coordinate and momentum space evolve according to a time-dependent variational principle \cite{raab:2000,broeck:1988},  supplemented by stochastic two-body scattering processes. This n-body framework allows to retain correlations beyond the mean-field level, including event-by-event fluctuations, cluster formation, and multifragmentation phenomena, which are conceptually not to capture within BUU-type models. 

QMD models have been successfully applied to the study of heavy-ion reactions at beam energies up to approximately $1A~\mathrm{GeV}$. At higher beam energies, where the Lorentz factor $\gamma$ becomes significantly larger than unity, such as those explored in the Beam Energy Scan program at RHIC and in future experiments at FAIR, the non-relativistic formulation of QMD is no longer adequate. A relativistic treatment requires a consistent synchronization of the time coordinates of the nucleons, as well as a proper distinction between Lorentz-scalar densities and the baryon four-current. These considerations motivate an extension of the QMD framework to relativistic energies.

Using constrained Hamiltonian dynamics, the first fully relativistic version of QMD was developed as RQMD~\cite{Sorge:1989dy,Maruyama:1991bp,Lehmann:1992us}, 
which constructed relativistic equations of motion with Lorentz-scalar interactions. Later JAM~\cite{Nara:1999dz} has been advanced, which is based on the same theoretical approach and uses different approximations. Although not fully covariant, this approach has been applied to relativistic heavy-ion collisions~\cite{Nara:2019qfd,Nara:2020ztb}.
Also UrQMD~\cite{Bass:1998ca,Bleicher:1999xi} is a relativistic QMD approach, however, in numerical calculations it is mostly used in a cascade mode or with non-relativistic potentials.  In addition, a relativistic QMD formulation based on the Nambu–Jona-Lasinio (NJL) model has been developed to describe the expansion and hadronization of a quark-antiquark plasma. In this framework, the NJL model generates a density- and time-dependent constituent mass, which effectively acts as a relativistic scalar potential~\cite{Marty:2012vs}.

Consequently, BUU and QMD approaches represent complementary strategies. BUU excels in computational efficiency and the description of bulk single-particle observables under a mean-field approximation, making it particularly suited for studying average properties of nuclear reactions. In contrast, QMD naturally incorporates correlations and fluctuations that play a crucial role in fragment production, event-by-event observables, and complex many-body dynamics. Together, these frameworks provide complementary insights and form the theoretical backbone of modern transport theory in nuclear physics.  

In relativistic theories, there are two types of potentials: a scalar potential $U_{S}$,
which is invariant under Lorentz transformations, and a vector potential $V^{\mu}$, 
which transforms as a four-vector.  The analysis of elastic proton--heavy-ion scattering data indicates that the absolute magnitudes of $U_S$ and $V^\mu$ are of comparable order \cite{Cooper:1993nx}. Moreover, the elementary nucleon–nucleon interaction, on which the nuclear relativistic mean-field (RMF) approach is based, naturally distinguishes between scalar and vector fields, which are associated with the different mesons exchanged between nucleons~\cite{Walecka:1974qa,Boguta:1977xi,Reinhard:1989zi,Serot:1997xg}.

One of the primary goals of the upcoming relativistic heavy-ion experiments at RHIC-BESII, CBM, HIAF, and NICA is to provide insight in the hadronic
equation of state (EoS) or more generally, in the phase structure of strongly interacting matter. The EoS depends on the baryon density and hence on the zero component of the current of the interacting baryons in the local rest frame. For the study of relativistic heavy-ion collisions and the theoretical analysis of the experimental data it is therefore necessary to have a fully covariant transport approach, which includes both scalar and vector potentials and provides covariant equations of motion. These depend on the appropriate formulation of time constraints that preserve cluster separability. 

The purpose of this paper is to present the first step of such a framework which can serve as an analysis tool for relativistic heavy-ion experiments. We derive relativistic transport equations for an interacting  many-body system that are manifestly covariant and explicitly include both scalar and vector interactions.  We follow the constraint Dirac dynamics approach in reducing the 8N dimensional phase space to 6N+1 dimensions using energy and time constraints. The latter connect the particle times, the zero components of the particle position 4-vector, with a Lorentz invariant parameter $\tau$, which characterizes the position on the 6N+1 dimensional hypersurface. Depending on its definition, this parameter can be identified with the time in the non-relativistic limit.  Formulating the evolution equation of the particle 4-vectors  as a function of $\tau$ allows to move the particles in a covariant way, respecting the time (and energy) constraints. The individual particle times, the zero component of the particle 4-vectors $x^\mu$ , can be recovered as a function of $\tau$ by exploiting the time constraints.  This allows finally to calculate the particle trajectories {\bf x}(t) and {\bf p}(t). These are independent of the choice of the Lorentz frame in which the evolution equations are solved.

We  demonstrate the frame independence of our approach by solving the equations of motion in two distinct Lorentz frames and separately investigate the influence of scalar and vector potentials on particle trajectories. We begin with a systematic overview and clarification of several fundamental concepts in relativistic dynamics. Subsequently, we derive the equations of motion for two- and many-body systems and compare with different formulations proposed in earlier relativistic QMD approaches, with particular attention to their underlying assumptions. The extension to nuclear cluster formation in heavy-ion collisions, together with a detailed discussion of the corresponding observables, will be presented in a forthcoming paper.

The structure of the paper is as follows. In Section~\ref{sec.preparation}, we introduce the basic concepts of the relativistic dynamical framework, including dynamical constraints, Poincaré transformations, world-line invariance, Dirac brackets, and related topics. The non-relativistic limit of an interacting N-body system is presented in Section \ref{sec.nonrela}. Section~\ref{sec.2body} is devoted to the derivation of the relativistic two-body evolution equations, where we discuss key issues such as the role of time constraints and how it approaches the non-relativistic limit. The frameworks for four-body and general N-body systems are presented in Sections~\ref{sec.nbody}, respectively. Finally, we summarize our findings in Section~\ref{sec.summary}.

\section{Basic concepts}
\label{sec.preparation}
There is a long history of research on relativistic particle dynamics~\cite{Dirac:1949cp}. A major milestone was Dirac’s pioneering formulation of constrained Hamiltonian dynamics~\cite{Dirac:1949cp}, which provided a systematic framework for incorporating relativistic invariance into dynamical systems. Building on this foundation, important extensions were developed by Komar~\cite{Komar:1978hc,Komar:1978hd}, Samuel~\cite{Samuel:1982jn,Samuel:1982jk}, Sudarshan, Mukunda, and Goldberg~\cite{Sudarshan:1981pp}, and Todorov~\cite{Fiziev:2000mh}, who generalized the formalism to relativistic two- and many-body systems and analyzed different choices of constraints.  

A further development towards a numerically tractable transport approach has been advanced by Sorge, Stöcker, and Greiner~\cite{Sorge:1989dy,Sorge:1989vt}, who laid the groundwork for relativistic quantum molecular dynamics. In this approach they used a scalar potential, which is a function of the scalar density, to describe the interaction among nucleons.  This interaction cannot be related to an equation of state, which is a function of the baryon density, the zero component of the 
4-vector of the baryon current in the local rest frame. Later Faessler and collaborators~\cite{Maruyama:1991bp,Lehmann:1992us,Fuchs:1996uv} published an independent but very similar method. More recently Aichelin and Marty~\cite{Marty:2012vs} developed further this method to describe the time evolution of a quark - antiquark plasma based on the  Nambu  Jona-Lasinio Lagrangian. Later, Nara and his collaborators continued to derive time evolution equations for particles in both local mean-field scalar and vector potentials in non-covariant and covariant ways ~\cite{Nara:2020ztb,Nara:2021fuu,Nara:2023vrq}. Recently they developed a relativistic approach for wave packets in a local potential ~\cite{Nara:2025pkg}.

In this section, we provide a systematic overview of these developments, tracing the evolution of relativistic dynamical frameworks from their fundamental theoretical origins to modern implementations in transport and molecular dynamics models.

\subsection{Poincar\'e Group and algebra}

We start out with a review of the common features and basic requirements of a relativistic theory. Starting point is the relativistic extension of the Poisson brackets 
\begin{eqnarray}
\{A,B\} \equiv \sum_{k=1}^N \left( \frac{\partial A}{\partial q_k^\mu} \frac{\partial B}{\partial p_{k,\mu}}-\frac{\partial A}{\partial p_k^\mu} \frac{\partial B}{\partial q_{k,\mu}} \right).
\end{eqnarray}
for  position and momentum four-vectors , $q^\mu, p^\mu$, with $\mu=0,...,3$.
It's easy to check that the position and momentum four-vectors of each particle satisfy the Poisson brackets,
\begin{eqnarray}
&&\{q_i^\mu, q_j^\mu\}=\{p_i^\mu, p_j^\mu\}=0,\nonumber\\
&&\{q_i^\mu, p_j^\nu\}=\delta_{ij}g^{\mu \nu},
\end{eqnarray}
with $g^{\mu \nu}$ being the Minkowski metric with the diagonal $\{1,-1,-1,-1\}$ and zero otherwise.

A relativistic theory has to be invariant under Lorentz transformations $\Lambda$ and space-time translations $a$. Both transformations form the Poincar\'e group with the group element, $R(\Lambda,a)$. Its action on the position vector is given by
\begin{eqnarray}
R(\Lambda,a):\quad q'^\mu=R(\Lambda,a)q^\mu =\Lambda_\nu^\mu q^\nu +a^\mu.
\end{eqnarray} 
Finite transformations can be built with help of the infinitesimal ones.
The algebra associated with the continuous symmetry group is given by the algebra of the generators of infinitesimal transformations.
For a N-particle system, generators for the translation group and for the Lorentz group can be expressed as,
 \begin{eqnarray}
P_{\mu}&=&\sum_n p_{n,\mu},\nonumber\\
M_{\mu \nu}&=&\sum_n q_{n,\mu}p_{n,\nu}-q_{n,\nu}p_{n,\mu}.
\label{eq.generator10}
\end{eqnarray} 
These ten generators respect the algebra of the group, which is called Poincar\'e algebra, which satisfies the Poisson brackets:
\begin{eqnarray}
\{P_\mu, P_\nu \}&=&0,  \nonumber\\
\{M_{\mu,\nu},P_\rho\}&=&g_{\mu \rho}P_\nu-g_{\nu \rho}P_\mu,  \nonumber\\
 \{M_{\mu,\nu},M_{\rho,\sigma}\}&=&g_{\mu \rho}M_{\nu \sigma}-g_{\mu \sigma}M_{\nu \rho}-g_{\nu \rho}M_{\mu \sigma} \nonumber\\
&+&g_{\nu \sigma}M_{\mu \rho}. 
\label{eq.generator10PB}
\end{eqnarray} 
The generator of the Poincar\'e group is given by the combination of that of the Lorentz transformation and of a translation, 
\begin{eqnarray}
G\equiv a^\mu P_\mu+\frac{1}{2}\omega^{\mu \nu}M_{\mu \nu}
\end{eqnarray} 
with the infinitesimal translation parameter $a^\mu$ and infinitesimal Lorentz transformation parameter $\omega^{\mu \nu}$. 
For a infinitesimal Poincar\'e transformation, the space-time coordinates of the same event in frame $O$ and $O'$ can be expressed using the Poisson bracket,
 \begin{eqnarray}
q'^\mu=q^\mu+\{q^\mu, G\}=q^\mu+\omega^\mu_\nu q^\nu +a^\mu,
\label{eq.potrans}
\end{eqnarray} 
where $q^\mu$ and $q'^\mu$ are four vectors in the frames $O$ and $O'$, respectively. 

\subsection{Constraints and world line condition}
Relativistic theories are based on four-vectors whose transformation between two inertial systems is given by elements of the Poincar\'e group. As a consequence, the phase space of an N-particle system has no longer 6N dimensions as in non-relativistic dynamics but 8N.
However, the physical phase-space trajectories have 6N+1 dimensions, which are the positions and momenta of the particles as a function of a evolution parameter $\tau$, which is called the computational time. Thus we need constraints to reduce the number of degrees of freedom of the relativistic phase space. 

The constraints concern those variables, which do not appear in the equation of particle trajectories, Usually N-constraints express  the on-shell conditions (also called first class constraints)~\cite{Komar:1978hc,Komar:1978hd}, which read for free particles as $H_i=p_{i,\mu}p_i^{\mu}-m_i^2$. We can see that these constraints satisfy the Poisson bracket, $\{H_i,H_j\}=0$ as well as the  Poincar\'e invariance, $\{H_i,G\}=0$. 
With the constraints $H_i$, the 8N-dimensional space reduces to a 7N dimensional hypersurface $\Sigma$. Because the $H_i$ are Poincar\'e invariant, the 7N dimensional hypersurface $\Sigma$ is also Poincar\'e invariant.  
One reduces the number of degrees of freedom further by introducing additional N-constraints $\chi_i$ (also called second class constraints), where N-1 constraints serve to pick a one-dimensional line on each sheet by fixing the relation between the time components of the N four-vectors in coordinate space. The last constraint connects these fixed time components with $\tau$ (in this paper, $\tau$ is not the proper time), the parameter, which characterizes the trajectory in the 6N dimensional phase space.   

The selection of the second-class time constraints is not unique. The consequences of different choices has been discussed in many previous works~\cite{Samuel:1982jn,Samuel:1982jk,Sudarshan:1981pp}. The time constraints are chosen to fulfill the world line invariance.
Given an initial value in the physical phase space one can solve the equation of motion (EoM) to get the corresponding phase space trajectory in a Minkowski space. This determines the world lines of all particles. Each point of a particle world line is a space-time event that can be Poincar\'e transformed according to the geometric standard rules of the Poincar\'e group.
World line invariance means that the Poincar\'e-transformed particle world lines are the same as those that can be constructed by projecting the phase space trajectory on the Minkowski space after applying the canonical Poincar\'e transformation on the phase space variables. To clarify this concept, the concept of the Dirac bracket is introduced~\cite{Dirac:1949cp}. It is a generalization of the Poisson bracket developed by Dirac and used to describe the time evolution of any function of the phase-space variables along the trajectory determined by the constraints $\varphi_i$ (including the first class $H_i$ and second class $\chi_i$). 
The definition of the the Dirac bracket for any two phase-space functions $A$ and $B$ is 
\begin{eqnarray}
\{A,B\}^*\equiv \{A,B\}-\sum_{i,j}^{2N}\{A,\varphi_i\}C_{ij}\{\varphi_j, B\},
\end{eqnarray} 
with $C_{ij}^{-1}=\{\varphi_i,\varphi_j\}$.

One can easily arrive at the conclusion that the Dirac brackets yield the same result as the Poisson brackets~\eqref{eq.generator10PB} for the ten generators of the Poincar\'e group~\eqref{eq.generator10}. Therefore we can also use the Dirac brackets to construct a transformation between the two inertial systems $O$ and $O'$. This transformation is noted as $R^*(\Lambda,a)$. Both transformations, $R(\Lambda,a)$ as well as $R^*(\Lambda,a)$, map the hypersurface $\Sigma$ to $\Sigma$, but $R(\Lambda,a)$ and $R^*(\Lambda,a)$ map the same point in the frame $O$ to different points in frame $O'$, which both conserved all constraints. Because $\{\chi_i,G\}^*=0$, $\chi_i$ is unchanged under a transformation $R^*(\Lambda,a)$ and the Dirac brackets transform a phase-space point on $O$ to a phase-space point on $O'$ , which has the same value of $\tau$. For the transformation using Poisson brackets this is generally not the case~\cite{Kihlberg:1980fz}. Therefore, Dirac brackets are the proper tool for dealing with transformations between two inertial frames.

Now, let us take the one body case as an example to see the condition of the World line invariance.  For one free particle, the on-shell constraint is,
\begin{eqnarray}
H&=&p_{\mu} p^\mu -m^2.
\label{eq.1bodyh}
\end{eqnarray} 
This constraint reduces the phase space from eight to seven dimensions by relating the energy of the particle with its three-momentum. The EoM in phase space on which this constraint is satisfied is given by the solution of,
\begin{eqnarray}
{dq^\mu \over d\tau}&=&\lambda\{q^\mu(\tau), H\},\nonumber\\
{dp^\mu \over d\tau}&=&\lambda\{p^\mu(\tau), H\},
\label{eq.1bodyeom}
\end{eqnarray} 
where $\lambda$ is a free parameter, which can be fixed by the second order constraint $\chi$. This time constraint $\chi$ has been chosen quite differently what gives different equations of motion. The preservation of the constraint $\chi$ in $\tau$ gives, 
\begin{eqnarray}
{d\chi \over d\tau}={\partial \chi \over \partial \tau}+\lambda \{\chi, H\}=0,
\label{eq.chi1body}
\end{eqnarray} 
and consequently,
\begin{eqnarray}
\lambda=-{\partial \chi\over \partial \tau}\{\chi,H\}^{-1}.
\end{eqnarray} 
Applying this to the above EoM, we get
\begin{eqnarray}
{dq^\mu \over d\tau}&=&-{\partial \chi\over \partial \tau}{ \{q^\mu,H\}\over \{\chi,H\} },\nonumber\\
{dp^\mu \over d\tau}&=&-{\partial \chi\over \partial \tau}{ \{p^\mu,H\}\over \{\chi,H\} }.
\label{eq.eom1body}
\end{eqnarray} 

Assuming $O$ and $O'$ are two inertial frames related by an infinitesimal element of the Poincar\'e group. The transformation can be realized with the Poisson bracket (where $\tau$ is not conserved), as given by equation   Eq.~\eqref{eq.potrans} for a small time step $\delta \tau$,
\begin{eqnarray}
q'^\mu(\tau)&=&q^\mu(\tau+\delta \tau)+\{q^\mu(\tau+\delta \tau),G\},\nonumber\\
&\approx& q^\mu(\tau)+{dq^\mu \over d\tau}\delta \tau+\{q^\mu(\tau),G\}.
\label{eq.xprim}
\end{eqnarray} 
This is called geometrical transformation.
With the EoM~\eqref{eq.eom1body}, the $q'^\mu$ can be further expressed as,
\begin{eqnarray}
q'^\mu(\tau)=q^\mu(\tau)-{\partial \chi\over \partial \tau}{ \{q^\mu,H\}\over \{\chi,H\} }\delta \tau+\{q^\mu(\tau),G\}.
\label{eq.xprimp}
\end{eqnarray} 
On the other hand, the transformation between the inertial frames $O$ and $O'$ can be realized by $R^*(\Lambda, a)$, which is expressed using the Dirac bracket,
\begin{eqnarray}
q'^\mu(\tau)&=&q^\mu(\tau)+\{q^\mu(\tau),G\}^* \nonumber\\
&=&q^\mu(\tau)+\{q^\mu(\tau),G\}-{ \{q^\mu, H \}\{\chi, G\} \over \{H, \chi\}}.
\label{eq.xprimd}
\end{eqnarray} 
This equation is called a canonical transformation to distinguish it from the geometrical transformation. The last step of the above equation is fulfilled due to $\{H,G\}=0$, as obtained before.

Comparing Eq.~\eqref{eq.xprimp} with Eq.~\eqref{eq.xprimd}, we find that if 
\begin{eqnarray}
\delta \tau=\{\chi, G\} \left( {\partial \chi \over \partial \tau} \right)^{-1}.
\label{eq.dt1}
\end{eqnarray}
is fulfilled, the transformation between two inertial systems using Dirac brackets (the canonical transformation) becomes identical to that using Poisson brackets (the geometrical transformation).  This means that the world lines of particles remain the same under the two transformations. They are therefore frame independent. This requirement of frame independence of the trajectories is called the world line condition.

Time constraints, which can be chosen differently, have to fulfill Eq.~\eqref{eq.dt1}
to be eligible for a relativistic dynamics.  If they respect this equation they guaranty the World line invariance. 
\begin{figure}[!htb]
\centering
\includegraphics[width=0.5\textwidth]{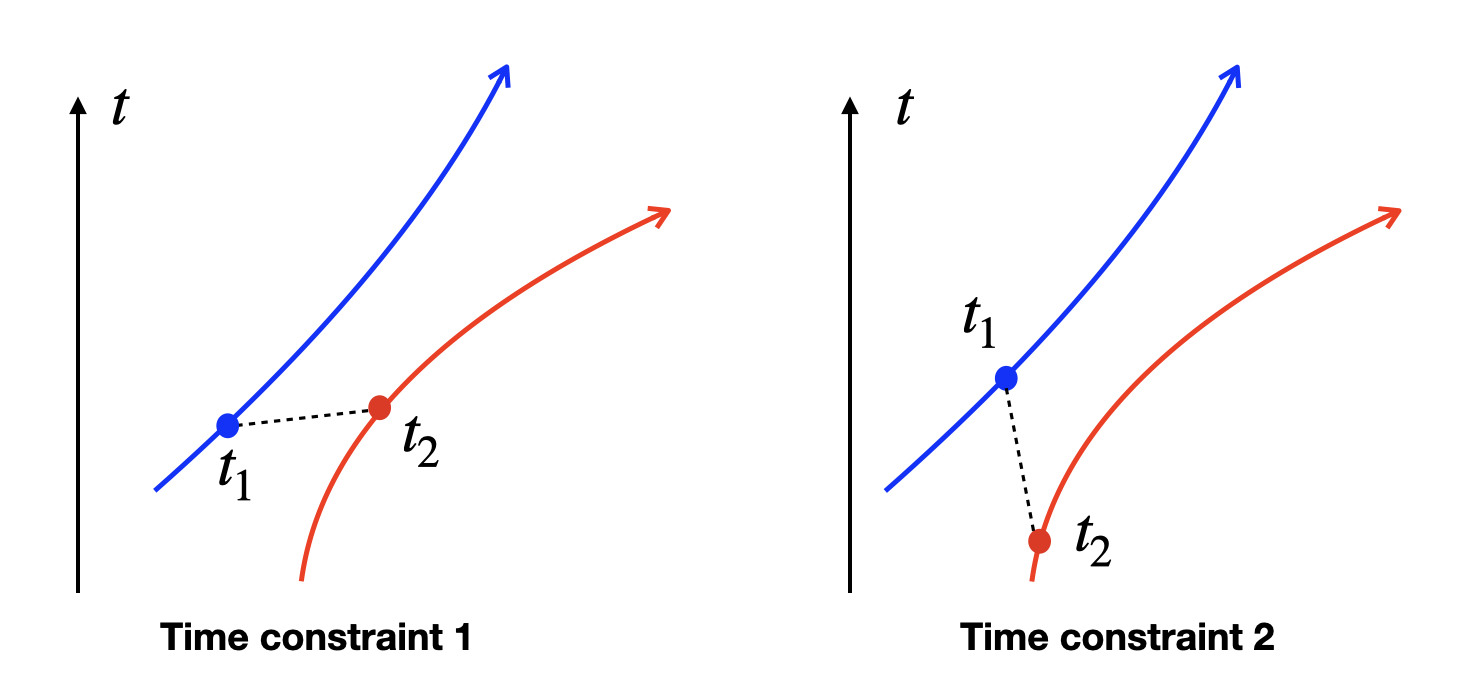}
\caption{The sketch of the two particle system with different kinds of time constraints. The blue and red curves are trajectories of particle 1 and 2, respectively.} 
\label{fig.tcs}
\end{figure}
Extending this to the N-body case we find
\begin{eqnarray}
\delta \tau_i=\{\chi_i, G\} \left( {\partial \chi_i \over \partial \tau} \right)^{-1}, \quad i=1,...,N.
\label{eq.dtn}
\end{eqnarray}
A simple way to choose the time constraints is to let N-1 of them be independent of  $\tau$ (that means ${\partial \chi_i \over \partial \tau}=0$)   and to demand  Poincar\'e invariance, $\{\chi_i,G\}=0$ for $i=1,..,N-1$. The last time constraint is $\tau$-dependent and satisfies the world line invariance condition~\eqref{eq.dt1}. However, there are many choices of time constraints that satisfy the above requirements. Different time constraints result in  different equations of motion (EoM) and evolution times $\tau$~\cite{Sudarshan:1981pp,Marty:2012vs}.
These different time constraints connect different points in Minkowski space, as shown in Fig.~\ref{fig.tcs}.
For a given $\tau$ the time constraints allow to determine the time component of the position four-vector of each of the particles. 
Whether the physical trajectory depends on the choice of time constraints has not yet been established rigorously. This issue will be investigated in the next section through numerical calculations for the two-body case, and in Appendix~\ref{app.b} for the one-body free-particle case.

\section{Non-relativistic time evolution of N-body systems} 
\label{sec.nonrela}
To compare the relativistic with the non-relativistic trajectories, we first examine the non-relativistic framework. The canonical momentum ${\bf p}_i$ is the conjugated variable to the position, the mechanical momentum ${\bf p}^*_i$ is the momentum, which is changed by the forces. In the non-relativistic case, the total Hamiltonian of an $N$-body system that interacts with a four-potential $A_{ij}^\mu=(A_{0,ij},-{\bf A}_{ij})$ (analogous to a electromagnetic potential) and an additional scalar potential $V_{ij}$ can be expressed as,
\begin{eqnarray}
H=\sum_{i}^N \left[{1 \over 2m_i}({\bf p}_i-{\bf A}_i)^2+A_{0,i} \right]+\sum_{i<j} V_{ij},
\end{eqnarray}
where ${\bf A}_i=\sum_{i<j} {\bf A}_{ij}$ and $A_{0,i}=\sum_{i<j} A_{0,ij}$ is the common action of all other particles on particle $i$.
The mechanical momentum ${\bf p}^*_i$ is related to the canonical momentum ${\bf p}_i$, via e.g. ${\bf p}^*_i\equiv{\bf p}_i-{\bf A}_i$. 
Following the Hamilton-Jacobi approach, the equations of motion can be expressed by the non-relativistic Poisson equations as
\begin{eqnarray}
\label{eq.nonrela2as}
{d{\bf q}_i\over dt}&=&{\partial H\over \partial {\bf p}_i}=\sum_{j=1}^N \left[{{\bf p}^*_j\over m_j}{\partial {\bf p}^*_j\over \partial {\bf p}_i}+{\partial A_{0,j}\over \partial {\bf p}_i}\right],\\
{d{\bf p}_i\over dt}&=&-{\partial H\over \partial  {\bf q}_i}=-\sum_{j=1}^N\left [{{\bf p}^*_j\over m_j}{\partial {\bf p}^*_j\over \partial {\bf q}_i}+{\partial A_{0,j}\over \partial {\bf q}_i}\right]-\sum_{j\neq i}{\partial V_{ij} \over \partial {\bf q}_i}.\nonumber
\end{eqnarray} 
Now, let's check the time derivative of the mechanical momentum and the ``Lorentz-like'' force,
\begin{eqnarray}
{d {\bf p}_i^*\over dt}&=&{d {\bf p}_i \over dt}-{\partial {\bf A}_i \over \partial t}-\sum_{j=1}^N (\dot{\bf q}_j\cdot \nabla_j){\bf A}_i\nonumber\\
&=&\sum_{j=1}^N ({\bf v}_j \nabla_i{\bf A}_j- \nabla_i A_{0,j})-\sum_{j\neq i} \nabla_iV_{ij}-\partial_t {\bf A}_i\nonumber\\
&-&\sum_{j=1}^N ({\bf v}_j\cdot \nabla_j){\bf A}_i,
\end{eqnarray} 
where ${\bf v}_j=d{\bf q}_j/dt$ as shown in Eq.~\eqref{eq.nonrela2as}. If the vector potential $A^\mu$ is momentum-independent, ${\bf v}_j={\bf p}^*_j/m_j$.
The right side of the equation corresponds to the force that can be separated into three parts. The term $j=i$ gives the self-``Lorentz force''. $F_i^{\rm Lorentz}=-\nabla_i A_{0,i}-\partial_t {\bf A}_i+{\bf v}_i\times (\nabla_i\times {\bf A}_i)$, where $-\nabla_i A_{0,i}-\partial_t {\bf A}_i={\bf E}_i$ and $\nabla_i\times {\bf A}_i={\bf B}_i$ can be interpreted as the ``electric'' and ``magnetic'' field, respectively. Then we obtain the familiar form $F_i^{\rm Lorentz}={\bf E}_i+{\bf v}_i\times{\bf B}_i$.
This is the ``self-force'', which means particle $i$ interacting with its own potentials $A_i^\mu$.
The term $j\neq i$ gives the mutual interaction force, $F_i^{\rm Mutual}=\sum_{j\neq i}[{\bf v}_j \nabla_i{\bf A}_j- \nabla_i A_{0,j}-({\bf v}_j\cdot \nabla_j ){\bf A}_i]$, which represents the force on particle $i$ due to all other particles $j$ through their interactions. 
${\bf v}_j \nabla_i{\bf A}_j$ is the direct coupling of the current of particle $j$ to the motion of particle $i$. It describes how the vector potential of $i$ changes due to the presence of particle $j$. $-({\bf v}_j\cdot \nabla_j ){\bf A}_i$ looks like a back-reaction term: it accounts for how the potential of $i$ is influenced by the spatial derivative with respect to the source coordinate $j$. Physically, it enforces action–reaction consistency between the pair $i$ and $j$.
$- \nabla_i A_{0,j}$ represents the electric field force at $i$ generated by particle $j$.
Together, these terms represent the electromagnetic interaction between different particles mediated by the vector potential.
The last one is the force induced by the potential $V$, $F_i^V=-\sum_{j\neq i}\nabla_iV_{ij}$.

\section{Relativistic time evolution of two-body systems}
\label{sec.2body}
\subsection{Formal development of the time evolution equations with scalar and vector potentials}
Now, let's examine the relativisitc two-body case. First-class constraints are usually linked to mass shell conditions.
We begin with the mass shell constraints of a two-particle system,
 \begin{eqnarray}
H_1&=&p_{1,\mu} p_1^\mu -m_1^2 ,\nonumber\\
H_2&=&p_{2,\mu} p_2^\mu -m_2^2.
\end{eqnarray} 
The vector interaction, $A_\mu$, and the scalar interaction, $\Phi_i$, can be introduced in a system of two interacting particles,
 \begin{eqnarray}
H_1&=&p^*_{1,\mu}p^{*\mu}_{1} -m_1^2+\Phi_1 ,\nonumber\\
H_2&=&p^*_{2,\mu}p^{*\mu}_{2} -m_2^2+\Phi_2 .
\label{enconst}
\end{eqnarray} 
where $p^*_{i,\mu}$ is the mechanical momentum, which is related to the canonical momentum $p_{i,\mu}$ via, e.g. $p^*_{i,\mu}\equiv p_{i,\mu}-A_{i,\mu}$. 
The requirement that $H_i$ is relativistic invariant implied that the potential
$\Phi_i$ is a function of relativistic invariants only. Although the dimension of $H_i$ is energy square, we call $H_i$ Hamiltonian. The total Dirac Hamiltonian is $H=\lambda_1 H_1+\lambda_2 H_2$.
In order to make sure that $H_1$
and $H_2$ are conserved in the evolution, which means $\dot H_i =\{H_i,H\}= 0$, the $H_i$ should satisfy the constraint,
\begin{eqnarray}
\{H_1,H_2\}&=&4p_{1\mu}^*p_{2\nu}^*\{p_1^{*\mu},p_2^{*\nu}\} \nonumber\\
&-&2p_1^{*\mu} {\partial \Phi_2\over \partial q_1^\mu}+ 2p_2^{*\mu} {\partial \Phi_1\over \partial q_2^\mu}+\{\Phi_1,\Phi_2\}.
\label{eq:H1H2}
\end{eqnarray} 

With vanishing the vector potential, $\{H_1, H_2\}=0$ can be fulfilled, as discussed in the previous study~\cite{Marty:2012vs}. 
Due to the relativistic counterpart of Newton's third law, which leads to,
\begin{eqnarray}
\Phi_1=\Phi_2=\Phi(\sqrt{-q_T^2}),
\label{eq.phi12}
\end{eqnarray} 
where $\Phi_i$ depends on $q_T^2$,
\begin{eqnarray}
q_T^2\equiv q^\mu_Tq_{T,\mu}=q^2-{(q_\nu P^{\nu})^2\over P^{2}}
\end{eqnarray} 
and
\begin{eqnarray}
q^\mu_T=\left(g^{\mu \nu} -{P^{\mu} P^{\nu} \over P^{2}}\right)q_\nu=q^\mu-{q_\nu P^{\nu
}\over P^{2}}P^{\mu}
\end{eqnarray} 
with $q^\mu=q_1^\mu-q_2^\mu$ and $P^{\mu}=p_1^{\mu}+p_2^{\mu}$. 
In the center-of-mass frame the transverse distance $q_T^i$ represents the ordinary distance $q^i_1-q^i_2$ between the two particles and $q_T^0=0$. With this choice we obtain for Eq.~\eqref{eq:H1H2} 
\begin{eqnarray}
\{H_1,H_2\}=-4P^{\mu} q_{T,\mu} {\partial \Phi(q_T) \over \partial q_T^2}=0.
\end{eqnarray} 
so the constraint is fulfilled.

For arbitrary vector potential $A_{i,\mu}$ and scalar potential $\Phi_i$, the compatibility condition is generally not satisfied, $\{H_1, H_2\}\neq 0$. 
Therefore, the scalar and vector interactions cannot be chosen independently, and a special form of the potentials is required to ensure the mutual compatibility of the mass-shell constraints. As demonstrated in the previous section, the compatibility condition is fulfilled for a scalar interaction of the form $\Phi(q_T)$. Motivated by this result, we require the vector interaction to satisfy $4p_{1\mu}^*p_{2\nu}^*\{p_1^{*\mu},p_2^{*\nu}\}=0$, such that the compatibility condition reduces to the scalar case. With the introduction of the vector potential, the transverse relative coordinate is redefined as
\begin{eqnarray}
q_T^2\equiv q^\mu_Tq_{T,\mu}=q^2-{(q_\nu P^{*\nu})^2\over P^{*2}}
\end{eqnarray} 
and
\begin{eqnarray}
q^\mu_T=\left(g^{\mu \nu} -{P^{*\mu} P^{*\nu} \over P^{*2}}\right)q_\nu=q^\mu-{q_\nu P^{*\nu
}\over P^{*2}}P^{*\mu}
\end{eqnarray} 
with total mechanical momentum $P^{*\mu}=p_1^{*\mu}+p_2^{*\mu}$. 
Consequently, 
\begin{eqnarray}
\{H_1,H_2\}=4p_{1\mu}^*p_{2\nu}^*\{p_1^{*\mu},p_2^{*\nu}\}-4P^{*\mu} q_{T,\mu} {\partial \Phi(q_T) \over \partial q_T^2}=0.\nonumber\\
\end{eqnarray}
In this section, we derive the most general form of the vector potential $A_{i,\mu}$ that satisfies the compatibility condition. The explicit form of the vector potential adopted in the present work will be given in Sec.~\ref{subsec.nucleon}.

The equations of motion can be expressed as,
\begin{eqnarray}
{dq_i^\mu \over d\tau}&=&\{ q_i^\mu, H \}=\lambda_1 \{q_i^\mu, H_1\}+\lambda_2 \{q_i^\mu, H_2\}, \nonumber\\
{dp_i^\mu \over d\tau}&=&\{ p_i^\mu, H \}=\lambda_1 \{p_i^\mu, H_1\}+\lambda_2 \{p_i^\mu, H_2\}.
\label{eq.eom21}
\end{eqnarray} 
$\lambda_1$ and $\lambda_2$ are parameters, which can be fixed by introducing the second-class constraints. This will be discussed below. To obtain a world line of the two-body system, one needs still two constraints, $\chi_1(q_1, q_2, p_1, p_2)=0$ and $\chi_2(q_1, q_2, p_1, p_2,\tau)=0$. These two additional constraints reduce the 14-dimensional phase space to a 12-dimensional phase space with a parameter $\tau$ so effectively to a 13-dimensional phase space. The conservation of these constraints in $\tau$ gives, 
 \begin{eqnarray}
 {d\chi_1 \over d\tau}={\partial \chi_1 \over \partial \tau}+\lambda_1\{\chi_1,H_1\}+\lambda_2\{\chi_1,H_2\}=0,\nonumber\\
{d\chi_2 \over d\tau}={\partial \chi_2 \over \partial \tau}+\lambda_1\{\chi_2,H_1\}+\lambda_2\{\chi_2,H_2\}=0.
\end{eqnarray} 
This equation can be written in a matrix form,
\begin{equation}
\left(\begin{array}{cc} 
\{\chi_1, H_1\} & \{\chi_1, H_2\}  \\
\{\chi_2, H_1\}  & \{\chi_2, H_2\}  \\
 \end{array}\right)\left(\begin{array}{c} 
\lambda_1\\
\lambda_2 \\
 \end{array}\right)=\left(\begin{array}{c} 
-{\partial \chi_1 \over \partial \tau}\\
-{\partial \chi_2 \over \partial \tau} \\
 \end{array}\right).
\label{eq.2bodychiH} 
\end{equation}
We can obtain the parameters $\lambda_i$ by solving this eigen equation. 
If one defines,
\begin{equation}
S_{ij}\equiv \left(\begin{array}{cc} 
\{\chi_1, H_1\} & \{\chi_1, H_2\}  \\
\{\chi_2, H_1\}  & \{\chi_2, H_2\}  \\
 \end{array}\right)
\label{eq.defS}
\end{equation}
the inverse $2\times 2$ matrix can be obtained analytically, 
\begin{eqnarray}
S_{ij}^{-1} &=& {1\over \{\chi_1, H_1\}\{\chi_2, H_2\} -\{\chi_1, H_2\}\{\chi_2, H_1\} }\nonumber\\
&\times&
\left(\begin{array}{cc} 
\{\chi_2, H_2\} & -\{\chi_1, H_2\}  \\
-\{\chi_2, H_1\}  & \{\chi_1, H_1\}  \\
 \end{array}\right).
\end{eqnarray}
and the $\lambda_i$ can be expressed as,
\begin{eqnarray}
\lambda_i=-\sum_{j=1}^2S_{ij}^{-1}{\partial \chi_j\over \partial\tau}.
\label{eq.lambda12e}
\end{eqnarray} 
So for given time constraints, $\chi_i$, the EoMs, Eq.~\eqref{eq.eom21}, are determined once $\lambda_1,\lambda_2$
are calculated. It is important to realize that different time constraints may lead to different EoMs.

For the second-class constraints of the two particles, we first assume: 
\begin{eqnarray}
\chi_1&=&{1\over 2}(q_1^\mu-q_2^\mu) U_\mu =0 ,  \nonumber\\ 
\chi_2&=&{1\over 2}(q_1^\mu+q_2^\mu) U_\mu-\tau =0,
\label{eq.const2}
\end{eqnarray} 
where $U_\mu=P^*_\mu/\sqrt{P^{*2}}$ is the four velocity of the center of mass system. We can see immediately that the constraint $\chi_1$ is time independent because $\{\chi_1,G\}=0$. The above constraint implies that $q_T^\mu = q^\mu$ on the constraint surface, but the use of $q_T^\mu$ is not redundant. It provides a covariant definition of the interaction prior to imposing the constraints and ensures that the potential is independent of the relative time component. The second constraint defines a uniform evolution time $\tau$, which is, in the center-of-mass frame, equal to $\tau=(t_1^{com}+t_2^{com})/2$. The evolution is calculated as a function of $\tau$ but we can reconstruct the times $t_1$ and $t_2$ via the time constraints,
\begin{eqnarray}
t_1&=&{1\over U^0}(\tau+x_1U_x+y_1U_y+z_1U_z),\nonumber\\ 
t_2&=&{1\over U^0}(\tau+x_2U_x+y_2U_y+z_2U_z).
\label{eq.t1t2}
\end{eqnarray} 
Thus, the times of the two particles, $t_1$ and $t_2$, are not free parameters, but connected by the constraints and are functions to the coordinates and momenta of the two particles. If the total momentum is zero, then $t_1 = t_2 = \tau$. In this case, the zero (time) component of the coordinate space 4-vector coincides with the the system time $\tau$. 

The Poisson brackets of the time-constraints and the Hamiltonian can be expressed as,
\begin{eqnarray}
&&\{\chi_1,H_1\}\nonumber\\
&&=U_\mu p_1^{*\nu}\left({\partial p^*_{1,\nu}\over \partial p_{1,\mu}}-{\partial p^*_{1,\nu}\over \partial p_{2,\mu}}\right)+ {U_\mu\over 2} \left({\partial \Phi_1\over \partial p_{1,\mu}}-{\partial \Phi_1\over \partial p_{2,\mu}}\right), \nonumber\\
&&\{\chi_1,H_2\}=\nonumber\\
&&U_\mu p_2^{*\nu}\left({\partial p^*_{2,\nu}\over \partial p_{1,\mu}}-{\partial p^*_{2,\nu}\over \partial p_{2,\mu}}\right)+ {U_\mu\over 2} \left({\partial \Phi_2\over \partial p_{1,\mu}}-{\partial \Phi_2\over \partial p_{2,\mu}}\right), \nonumber\\
&&\{\chi_2,H_1\}=\nonumber\\
&&U_\mu p_1^{*\nu} \left({\partial p^*_{1,\nu} \over \partial p_{1,\mu}}+{\partial p^*_{1,\nu} \over \partial p_{2,\mu}} \right)+ {U_\mu\over 2} \left({\partial \Phi_1\over \partial p_{1,\mu}}+{\partial \Phi_1\over \partial p_{2,\mu}}\right),\nonumber\\
&&\{\chi_2,H_2\}=\nonumber\\
&&U_\mu p_2^{*\nu} \left({\partial p^*_{2,\nu} \over \partial p_{1,\mu}}+{\partial p^*_{2,\nu} \over \partial p_{2,\mu}} \right)+ {U_\mu\over 2} \left({\partial \Phi_2\over \partial p_{1,\mu}}+{\partial \Phi_2\over \partial p_{2,\mu}}\right).
\end{eqnarray} 
If the scalar potential, $\Phi_i$, does not depend explicitly on the particle's momentum, then the derivative of $\Phi_i$ with respect to momentum gives
\begin{eqnarray}
{\partial \Phi_i \over \partial p_{i,\mu}}={\partial \Phi_i \over \partial q_T^2}{\partial q_T^2 \over \partial p_{i,\mu}},
\end{eqnarray}
where 
\begin{eqnarray}
{\partial q_T^2 \over \partial p_{i,\mu}}=-2{(q_1^\nu-q_2^\nu) P^*_\nu \over P^{*2}}q_{T}^{\mu}=0,
\label{eq.dq2dp}
\end{eqnarray}
due to the constraint $\chi_1$. Therefore, the momentum derivative of the $\Phi_i$ disappears when second-class constraints like Eq.~\eqref{eq.const2} are chosen.

With these preparations, we can calculate the $\lambda_i$~\eqref{eq.lambda12e} explicitly
\begin{eqnarray}
\lambda_1&=&{U_\mu p_2^{*\nu}\left({\partial p^*_{2,\nu} \over \partial p_{1,\mu}}-{\partial p^*_{2,\nu} \over \partial p_{2,\mu}} \right) \over 2(\mathcal{P}_1-\mathcal{P}_2)}, \nonumber\\
\lambda_2&=&{U_\mu p_1^{*\nu}\left({\partial p^*_{1,\nu} \over \partial p_{2,\mu}}-{\partial p^*_{1,\nu} \over \partial p_{1,\mu}} \right)  \over 2(\mathcal{P}_1-\mathcal{P}_2)},
\label{eq.rela2lambda}
\end{eqnarray} 
where
\begin{eqnarray}
\mathcal{P}_1&=&\left(U_\mu p_1^{*\nu}{\partial p^*_{1,\nu} \over \partial p_{2,\mu}}\right) \left(U_\mu p_2^{*\nu}{\partial p^*_{2,\nu} \over \partial p_{1,\mu}}\right),\nonumber\\
\mathcal{P}_2&=&\left(U_\mu p_1^{*\nu}{\partial p^*_{1,\nu} \over \partial p_{1,\mu}}\right) \left(U_\mu p_2^{*\nu}{\partial p^*_{2,\nu} \over \partial p_{2,\mu}}\right).
\end{eqnarray} 
Moreover, we find for the Poisson brackets of the coordinates $q_i$ and momenta $p_i$ with the Hamiltonian
\begin{eqnarray}
&&\{q_{i}^\mu,H_i\}=2p_i^{*\nu}{\partial p^*_{i,\nu} \over \partial p_{i,\mu}}+{\partial \Phi_i \over \partial p_{i,\mu}}, \nonumber\\
&&\{q_{i}^\mu,H_j\}=2p_j^{*\nu}{\partial p^*_{j,\nu} \over \partial p_{i,\mu}}+{\partial \Phi_j \over \partial p_{i,\mu}}, \nonumber\\
&&\{p_{i}^\mu,H_i\}=-2p_i^{*\nu}{\partial p^*_{i,\nu} \over \partial q_{i,\mu}}-{\partial \Phi_i \over \partial q_{i,\mu}},\nonumber\\
&&\{p_{i}^\mu,H_j\}=-2p_j^{*\nu}{\partial p^*_{j,\nu} \over \partial q_{i,\mu}}-{\partial \Phi_j \over \partial q_{i,\mu}}.
\end{eqnarray} 
Then, the equations of motion (EoMs) for the interacting two-body system~\eqref{eq.eom21} can be written as,
\begin{eqnarray}
{dq_i^\mu \over d\tau}&=&\sum_{k=1}^2 \lambda_k \left(2p^{*\nu}_{k}{\partial p^*_{k,\nu} \over \partial p_{i,\mu}}\right),\nonumber\\
{dp_i^\mu \over d\tau}&=&-\sum_{k=1}^2\lambda_k \left(2p_k^{*\nu}{\partial p^*_{k,\nu} \over \partial q_{i,\mu}}+{\partial \Phi_k \over \partial q_{i,\mu}} \right).
\label{eq.rela2as}
\end{eqnarray} 

An additional issue arises when both scalar and vector interactions are
included, namely the causal propagation of the particles. The trajectory
of each particle must remain time-like, requiring
\begin{equation}
u_i^\mu u_{i\mu}
=
{dq_i^\mu\over d\tau}{dq_{i,\mu}\over d\tau}>0.
\end{equation}
For a purely scalar interaction, the four-velocity remains proportional
to the kinetic four-momentum, and the time-like character of the
trajectory follows directly from the mass-shell constraint. A similar
situation is obtained for a momentum-independent vector potential,
for which
$\partial p_{k,\nu}^*/\partial p_{i,\mu}$ does not contain additional
contributions from the interaction.

For a momentum-dependent vector potential, however, the situation is
more involved. Since
$p_k^{*\mu}=p_k^\mu-A_k^\mu(q,p)$, one has
\begin{equation}
{\partial p^*_{k,\nu}\over\partial p_{i,\mu}}
=
\delta_{ki}\delta^\mu_{\nu}
-
{\partial A_{k,\nu}\over\partial p_{i,\mu}},
\end{equation}
and consequently the four-velocity is, in general, no longer parallel
to the kinetic four-momentum $p_i^{*\mu}$. Therefore, the time-like
mass-shell condition $p_i^{*2}>0$ alone does not automatically guarantee
that $u_i^2>0$. The momentum dependence of the vector interaction must
thus be constrained such that the resulting equations of motion preserve
the time-like character of the particle trajectories throughout the
evolution. In practical implementations, we explicitly verify this
condition during the dynamical evolution, thereby ensuring causal
propagation of all particles.

\subsection{Relation to the RQMD, UrQMD, and JAM}
\label{sec.IVb}
Now, let us examine the connection between the equations of motion derived above and those employed in transport models such as RQMD, UrQMD, and JAM.

For a system with a scalar potential only, the $\lambda_i$ is simplified, 
\begin{eqnarray}
\lambda_1&=&(2p_1^\mu U_\mu)^{-1}, \nonumber\\
\lambda_2&=&(2p_2^\mu U_\mu)^{-1}, 
\end{eqnarray} 
and the above EoMs read as, 
\begin{eqnarray}
{dq_i^\mu \over d\tau}&=&{p_i^\mu \over p_i^\mu U_\mu},\nonumber\\
{dp_i^\mu \over d\tau}&=&-\sum_{k=1}^2{1\over 2p_k^\mu U_\mu} {\partial \Phi_k \over \partial q_{i,\mu}},
\label{eq.rela2onlys}
\end{eqnarray} 
These are exactly the EoMs used in RQMD~\cite{Sorge:1989dy,Maruyama:1991bp,Lehmann:1992us}.
RQMD~\cite{Sorge:1989dy} applied a scalar potential of the same functional form as used in non-relativistic QMD calculations, in which the non-relativistic density is replaced by the scalar density. In practice UrQMD calculations use non-relativistic potentials \cite{Hillmann:2018nmd}.

In the JAM approach~\cite{Nara:2019qfd,Nara:2023vrq,Nara:2020ztb} the time constraints
\begin{eqnarray}
\chi_1&=&(q_1^\mu-q_2^\mu) U_\mu =0 ,  \nonumber\\
\chi_2&=&q_2^\mu U_\mu-\tau =0,
\label{eq:jamtime}
\end{eqnarray} 
have been employed, which is a linear combination of aforementioned constraints~\eqref{eq.const2}.
The equations of motion can be obtained using similar procedures as before
\begin{eqnarray}
{dq_i^\mu \over d\tau}&=&\lambda_i 2p_{i}^{*\mu}-g^{\mu \sigma}\sum_{k=1}^2 \lambda_k \left[2p_{k}^{*v}{\partial V_{k,v} \over \partial p_{i}^{\sigma}} + {\partial m_k^{*2} \over \partial p_{i}^{\sigma}}\right],\nonumber\\
{dp_i^\mu \over d\tau}&=&g^{\mu \sigma}\sum_{k=1}^2\lambda_k \left[2 p_k^{*v}{\partial V_{k,v} \over \partial q_{i}^{\sigma}}+{\partial m_k^{*2} \over \partial q_{i}^{\sigma}} \right].
\end{eqnarray}
The notation of $p_i^*=p_i-V_i$ and $m_i^*=m_i-S_i$ are used. $V_i$ and $S_i$ denote the vector and scalar potential in their paper, respectively. The scalar potential modifies the particle’s mass. There are two common ways for its implementation. The first is to redefine the mass directly as $m^*\equiv m-S$, leading to the constraint 
$H_i=p^*_{1,\mu}p^{*\mu}_{1}-(m_i-S_i)^2$.
The second approach is to add the scalar part explicitly in the Hamiltonian,  $H_i=p^*_{1,\mu}p^{*\mu}_{1}-m_i^2+2m_iS$.
Both formulations yield the same non-relativistic limit if 
$|S_i|\ll m_i$, $|A_i^\mu|\ll m_i$, and $|{\bf p}|\ll m_i$. The second formulation
give the non-relativistic limit also for arbitrary $|S_i|$ and therefore we apply it here. 

If one neglects the momentum-dependent derivatives of the vector potential in the expression of $\lambda_i$ (Eq.~\eqref{eq.rela2lambda}) (${\partial p^*_{2,\nu} \over \partial p_{1,\mu}}=0,{\partial p^*_{2,\nu} \over \partial p_{2,\mu}}=1$), one arrives at
\begin{eqnarray}
\lambda_1&=&{1 \over 2p_1^{*\mu} U_\mu }, \nonumber\\
\lambda_2&=&{1 \over 2p_2^{*\mu} U_\mu }. 
\end{eqnarray}
If one further assumes $2p_i^{*\mu} U_\mu=2p_i^{*0}$, what is  fulfilled in the CoM frame but not in other reference frames, the expressions of $\lambda_i$ simplify further and one obtains the approximate EoM
\begin{eqnarray}
{dq_i^\mu \over d\tau}&=& {p_{i}^{*\mu}\over p_i^{*0}}-g^{\mu \sigma}\sum_{k=1}^2 \left[{p_{k}^{*v}\over p_k^{*0}}{\partial V_{k,v} \over \partial p_{i}^{\sigma}} + {m_k^{*}\over p_k^{*0}}{\partial m_k^{*} \over \partial p_{i}^{\sigma}}\right],\nonumber\\
{dp_i^\mu \over d\tau}&=&g^{\mu \sigma}\sum_{k=1}^2 \left[{p_{k}^{*v}\over p_k^{*0}}{\partial V_{k,v} \over \partial q_{i}^{\sigma}}+{m_k^{*}\over p_k^{*0}}{\partial m_k^{*} \over \partial q_{i}^{\sigma}} \right].
\end{eqnarray} 
These relativistic EoMs have been already obtained by Nara and his collaborators~\cite{Nara:2021fuu, Nara:2023vrq}.
The 3-components of the above equation can be expressed with $g^{\mu v}=(+,-,-,-)$:
\begin{eqnarray}
{d{\bf q}_i \over d\tau}&=& {{\bf p}_{i}^{*}\over p_i^{*0}}+\sum_{k=1}^2 \left[{ {p}_{k}^{*v}\over p_k^{*0}}{\partial V_{k,v} \over \partial {\bf p}_{i}} + {m_k^{*}\over p_k^{*0}}{\partial m_k^{*} \over \partial {\bf p}_{i}}\right],\nonumber\\
{d{\bf p}_i \over d\tau}&=&-\sum_{k=1}^2 \left[{ { p}_{k}^{*v}\over p_k^{*0}}{\partial V_{k,v} \over \partial {\bf q}_{i}}+{m_k^{*}\over p_k^{*0}}{\partial m_k^{*} \over \partial {\bf q}_{i}} \right].
\label{eq:eomfull}
\end{eqnarray} 
The equations above are the basis for the numerical calculations in JAM~\cite{Nara:2019qfd,Nara:2020ztb}. For actual calculations the scalar and vector densities are converted into $\sigma$ and $\omega$ mean fields, which are subsequently used in these EoMs.

\subsection{The non-relativistic limit}

After having defined the EoMs one can numerically calculate the trajectories of the two particles. 
We will consider here only the scalar potential~\eqref{eq.rela2onlys} (the vector potential we will discuss in subsection~\ref{subsec.nucleon}) to check the non-relativistic limit. In this test, we use $V(q_T)=-\frac{c}{2} m\omega^2 q_T^2$, which reduces in the center of mass to the harmonic oscillator potential $V(r)=\frac{c}{2}  m\omega^2 r^2$. For the following numerical calculation, we set $m_1=m_2=m=1~\rm GeV$ and set the frequency $\omega$ to be equal to $0.05~\rm GeV$. In the relativistic case, $\Phi_1=\Phi_2=2m_\mu V(q_T)$ with the reduced mass $m_\mu=m/2$. The parameter $c=\pm 1$ corresponds to an repulsive or attractive potential, respectively.

\begin{figure}[!htb]
\centering
\includegraphics[width=0.23\textwidth]{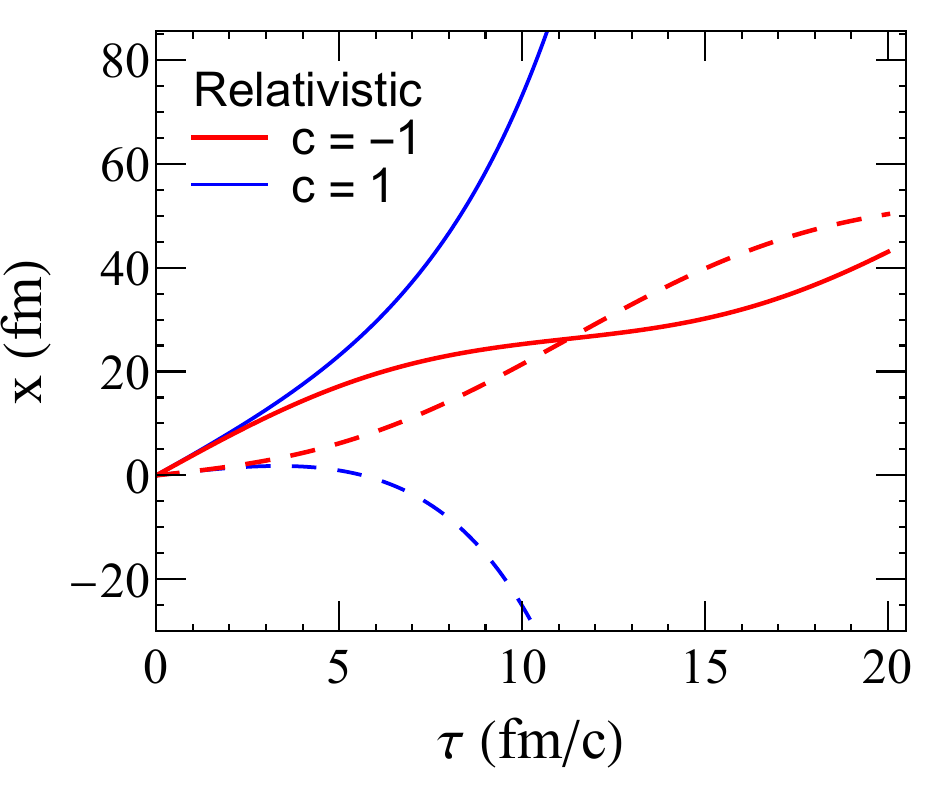}
\includegraphics[width=0.23\textwidth]{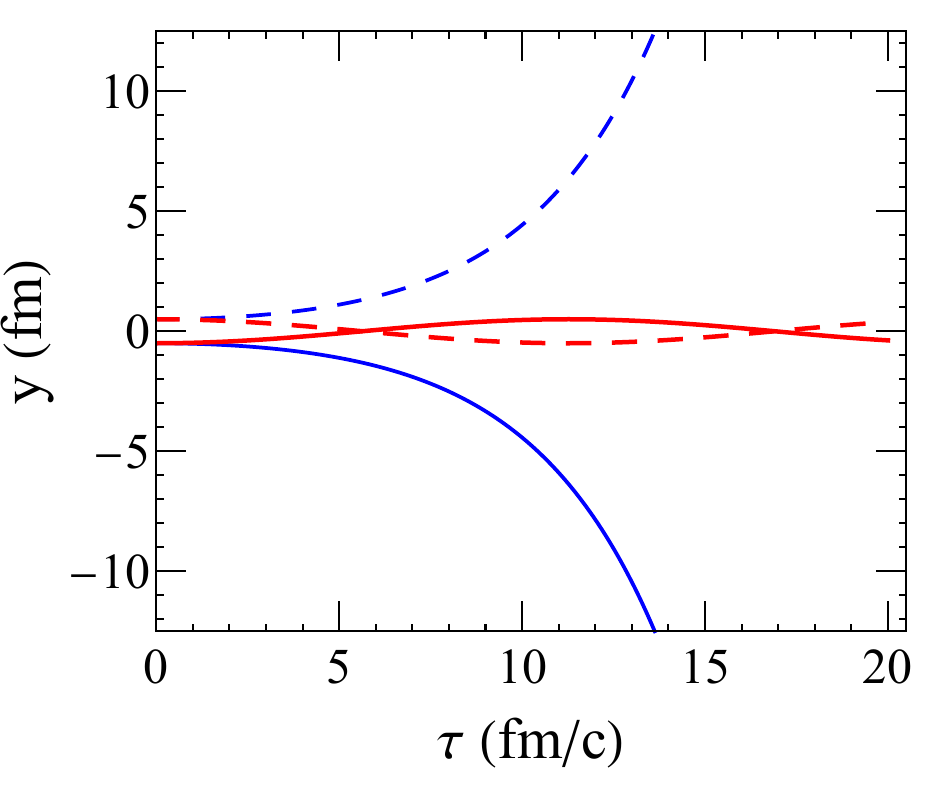}\\
\includegraphics[width=0.23\textwidth]{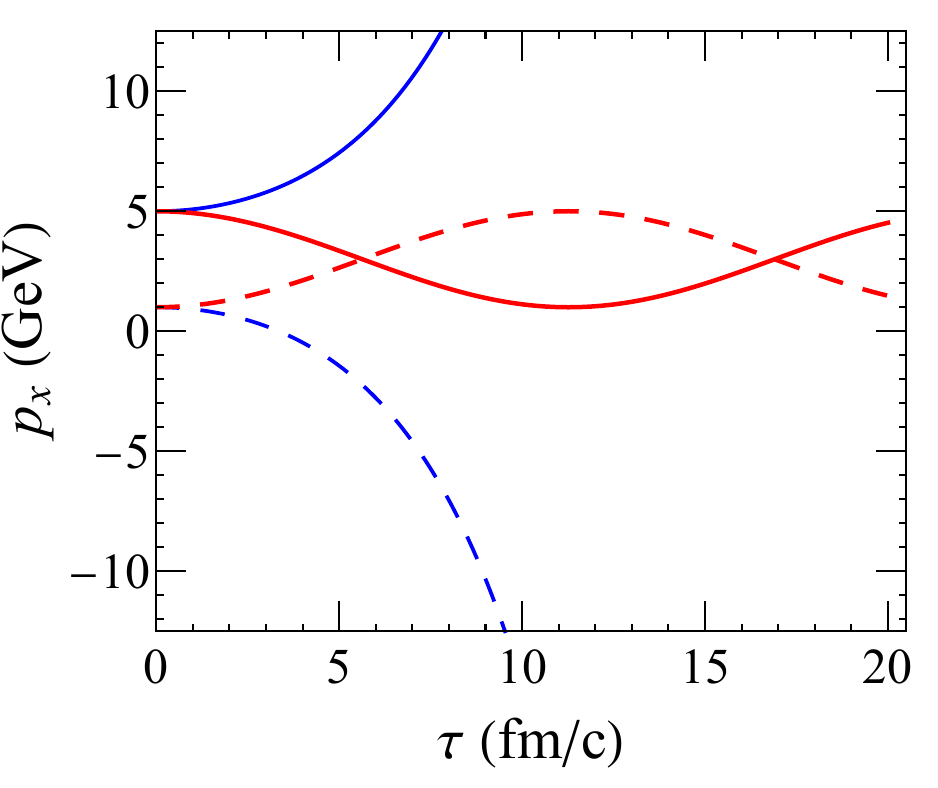}
\includegraphics[width=0.23\textwidth]{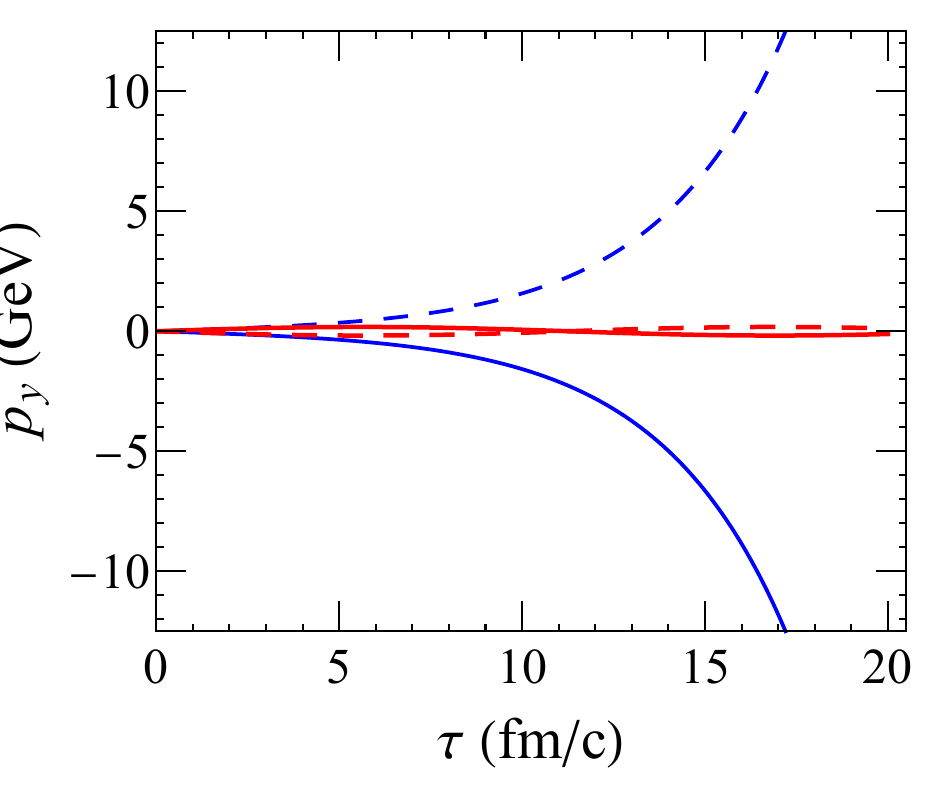}\\
\includegraphics[width=0.23\textwidth]{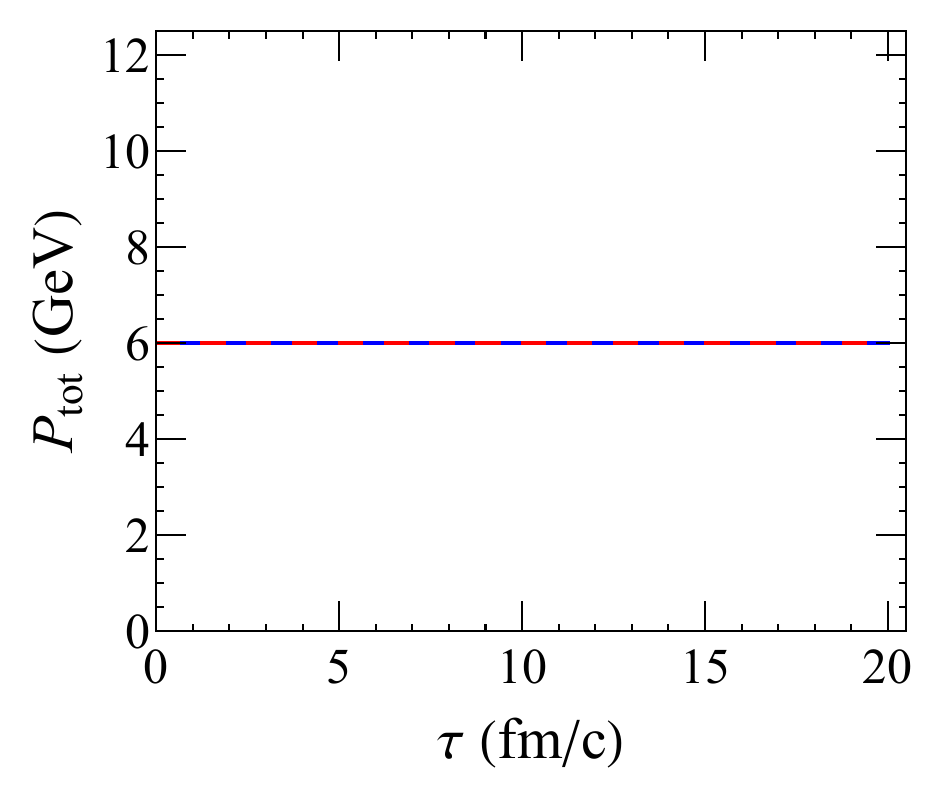}
\includegraphics[width=0.23\textwidth]{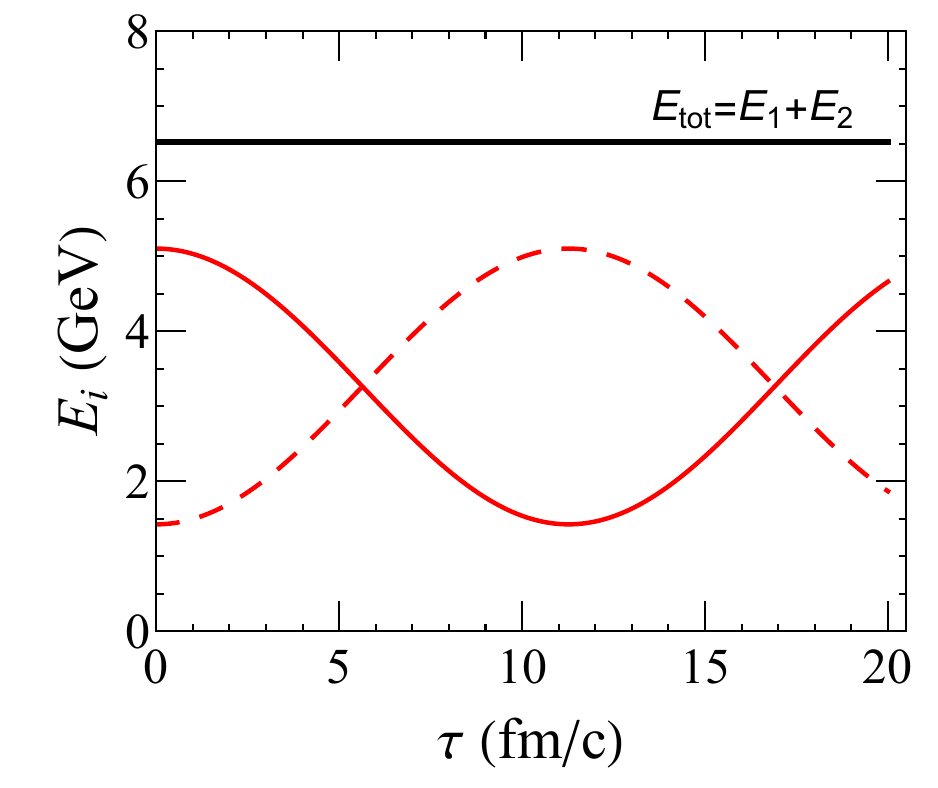}
\caption{The coordinates and momenta of particles 1 (solid lines) and 2 (dashed lines) as a function of computational time $\tau$ in relativistic evolutions. The red and blue lines represent an attractive ($c=-1$) and a repulsive interaction ($c=1$), respectively. The initial condition I is taken.} 
\label{fig.rela}
\end{figure}
We use two simple initial conditions for the two particles:
\begin{enumerate}[label=\Roman*.]
\item ${\bf x}_1=(0,-0.5,0)~\rm fm$, ${\bf x}_2=(0,0.5,0)~\rm fm$, ${\bf p}_1=(5,0,0)~\rm GeV$, ${\bf p}_2=(1,0,0)~\rm GeV$;
\item ${\bf x}_1=(-1.0,0,0)~\rm fm$, ${\bf x}_2=(1.0,0,0)~\rm fm$, ${\bf p}_1=(5,0,0)~\rm GeV$, ${\bf p}_2=(-5,0,0)~\rm GeV$.
\end{enumerate}
The first initial condition gives a total momentum is $P_{\rm tot}=6~\rm GeV$, while for the second one is $P_{\rm tot}=0$. The evolution starts from $\tau=0$. Given the spatial coordinates and momenta, the individual particle times $t_i$ are then uniquely determined from the time constraint, as shown in Eq.~\eqref{eq.t1t2}.

We start with the initial condition I and evolve the two-particle system with either a repulsive or an attractive potential. The results are shown in Fig.~\ref{fig.rela}. For the attractive potential (red curves), the two particles exhibit a spiral motion toward each other, while for the repulsive potential (blue curves), they move apart, as expected.
Since the initial momenta and positions in the $z$-directions are set to zero $p_z=0$ and $z=0$, these components remain zero during the entire evolution. Throughout the evolution, the total momentum ${\bf P}_{tot}={\bf p}_1+{\bf p}_2$ and energy $E_{tot}=E_1+E_2$ of the system is conserved, as seen on the bottom panel. Here, the initial energy of each particle in the presence of the interaction potential is defined as $E_i\equiv \sqrt{p_i^2+m_i^2-\Phi}$. When the total momentum is nonzero, the energy of each particle changes during the evolution, while the total energy remains conserved.

\begin{figure}[!htb]
\centering
\includegraphics[width=0.23\textwidth]{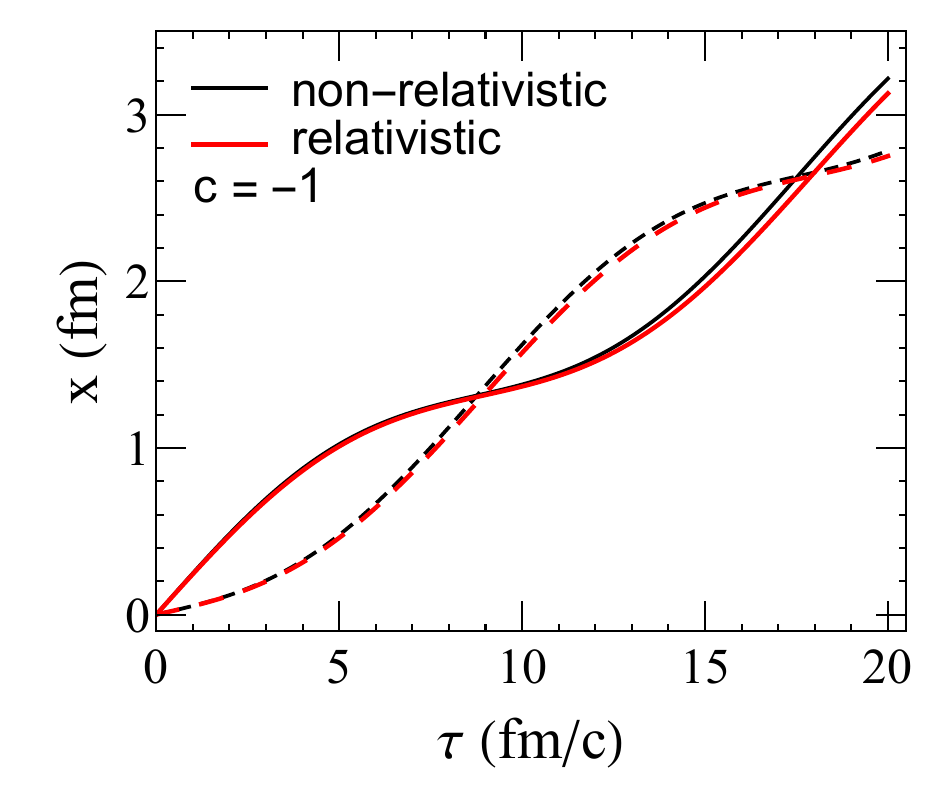}\includegraphics[width=0.23\textwidth]{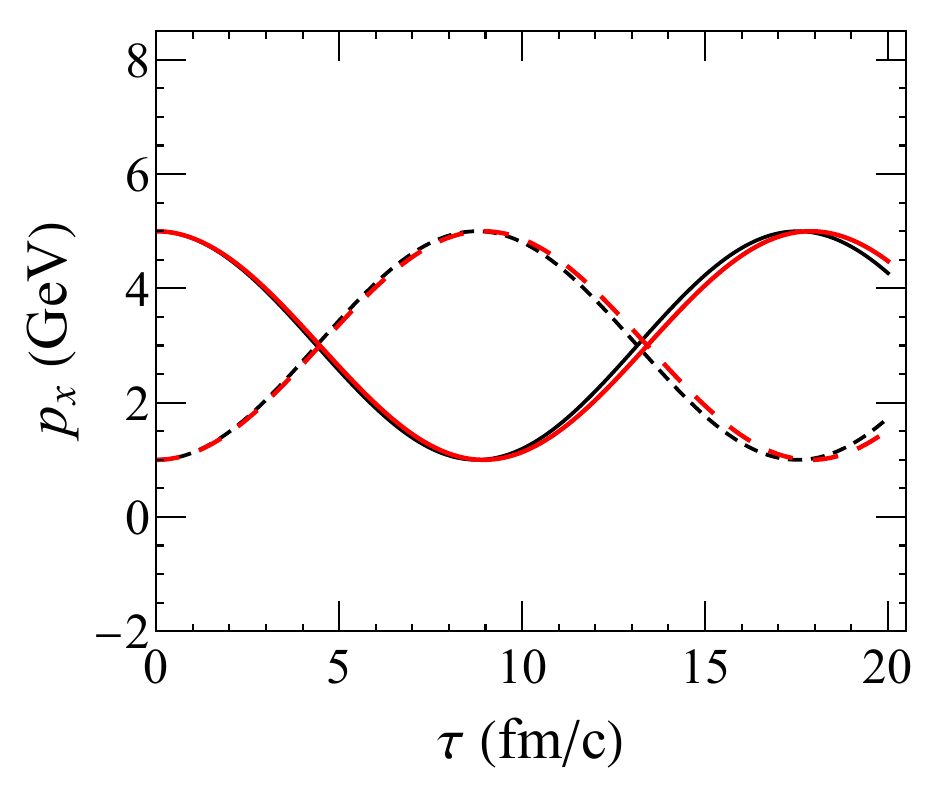}\\
\includegraphics[width=0.23\textwidth]{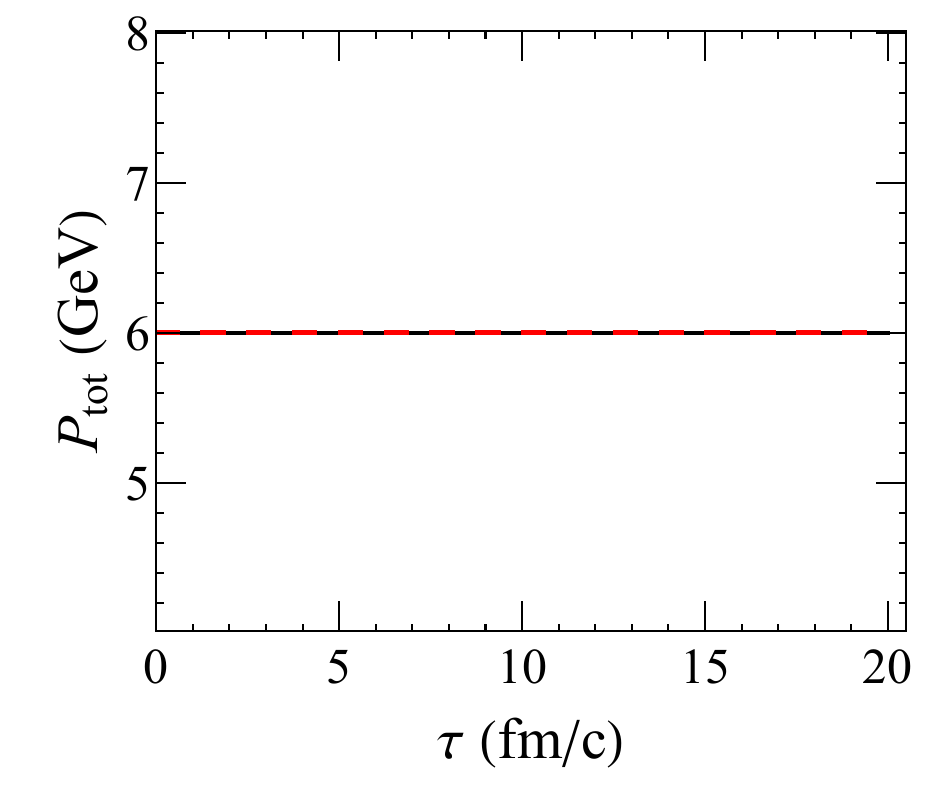}\includegraphics[width=0.23\textwidth]{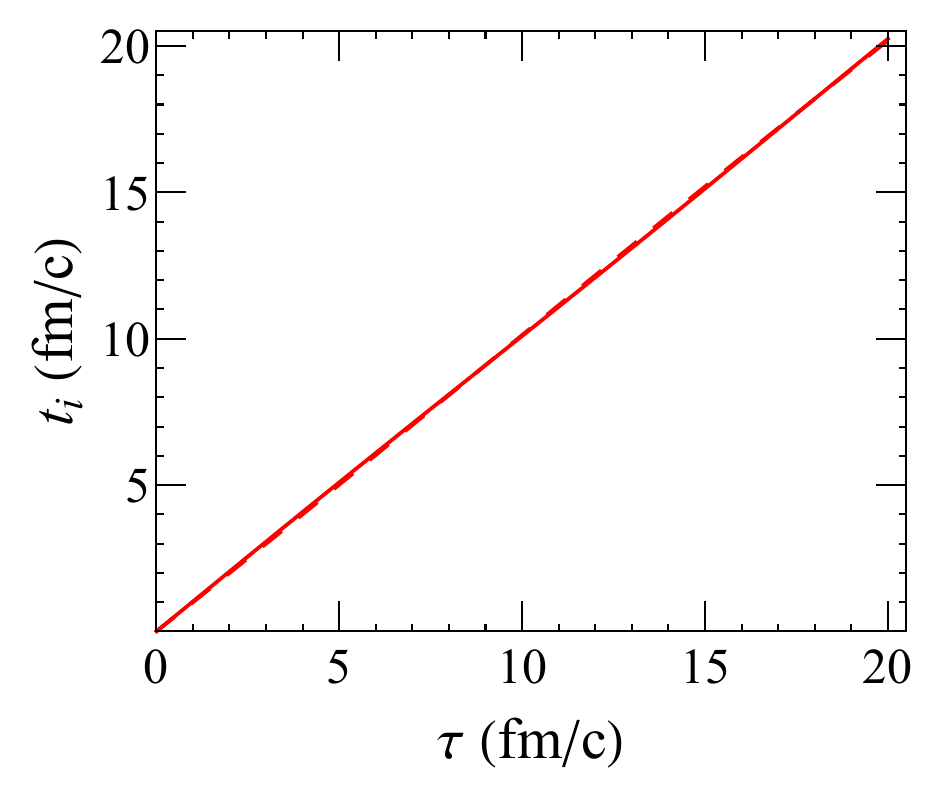}
\caption{The coordinates and momenta of particle 1 (solid lines) and 2 (dashed lines). The black and red line represent the non-relativistic and relativistic evolution with a mass $m=20~\rm GeV$, respectively. Here we take $c=-1$, which means an attractive interaction between two particles. The initial condition I is taken.} 
\label{fig.nonrela}
\end{figure}

We can see that the solution of the relativistic equations of motion~\eqref{eq.rela2as} approaches that of the non-relativistic ones~\eqref{eq.nonrela2as} in the non-relativistic limit, obtained for $m\to \infty$.  We now verify this limit numerically using the harmonic oscillator potential and set the particle mass to $m=20~\rm GeV$, which is large enough to expect non relativistic kinematics. The results are shown in Fig.~\ref{fig.nonrela}. In this case, the computational time $\tau\approx t_1\approx t_2$ from Eq.~\eqref{eq.t1t2}, due to $U^\mu \approx (1,0,0,0)$.
Consequently, the trajectories can be plotted together for direct comparison. As expected, the trajectories from both the relativistic and non-relativistic calculations are very close to each other in this limit.
The tiny difference comes from the potential correction on the evolution in the relative case. 

\begin{figure}[!htb]
\centering
\includegraphics[width=0.23\textwidth]{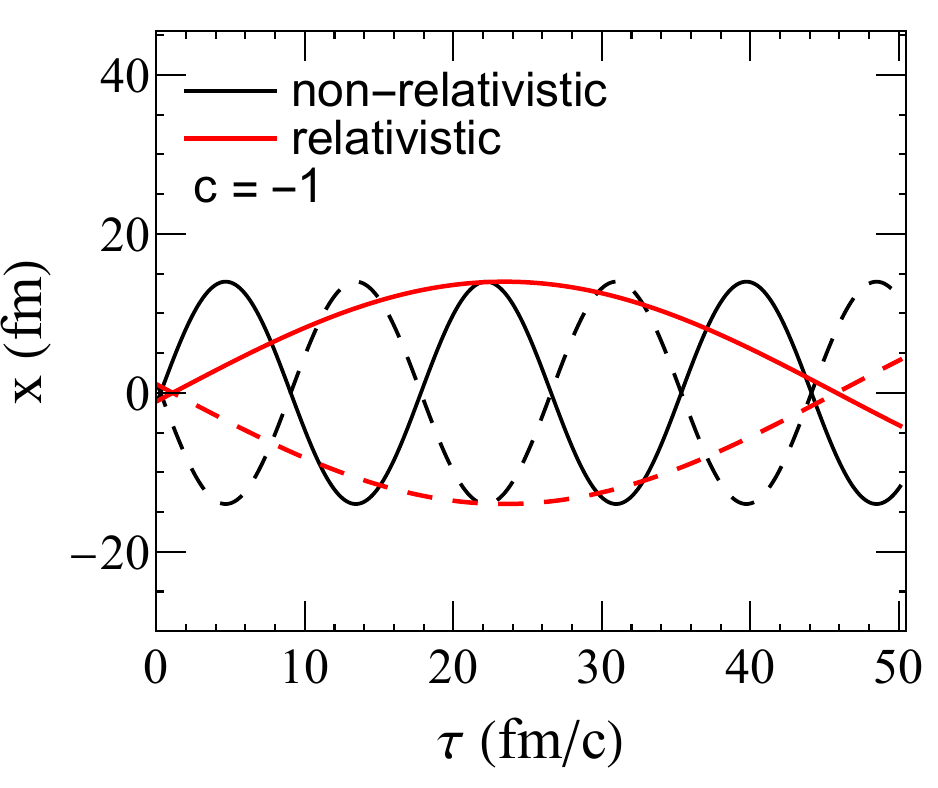}\includegraphics[width=0.23\textwidth]{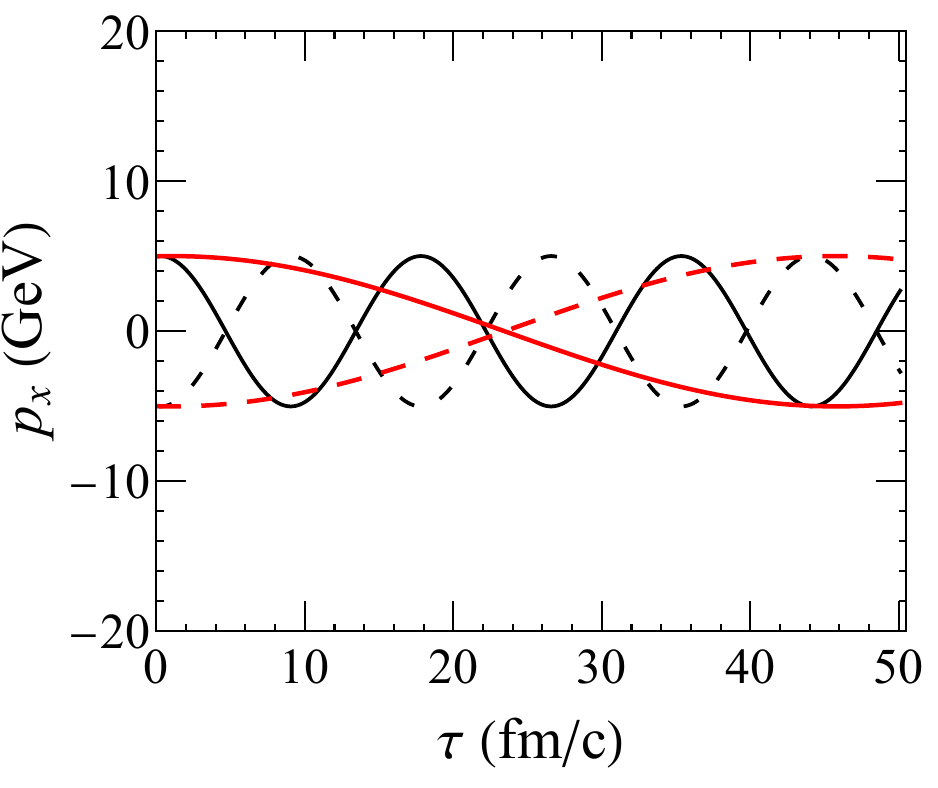}\\
\includegraphics[width=0.23\textwidth]{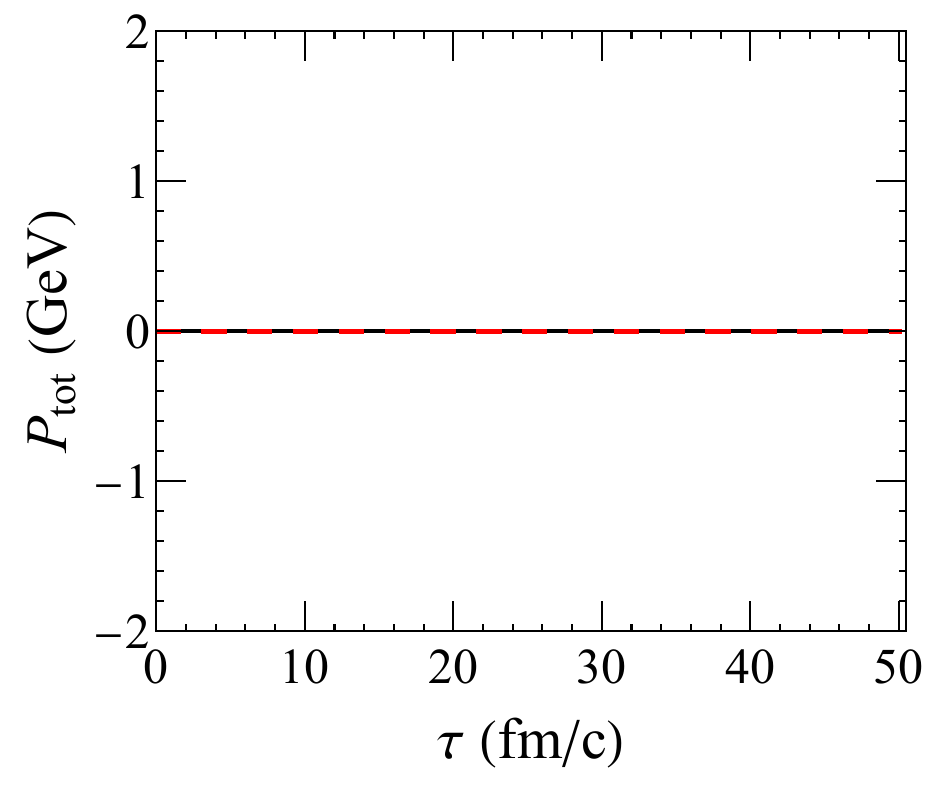}\includegraphics[width=0.23\textwidth]{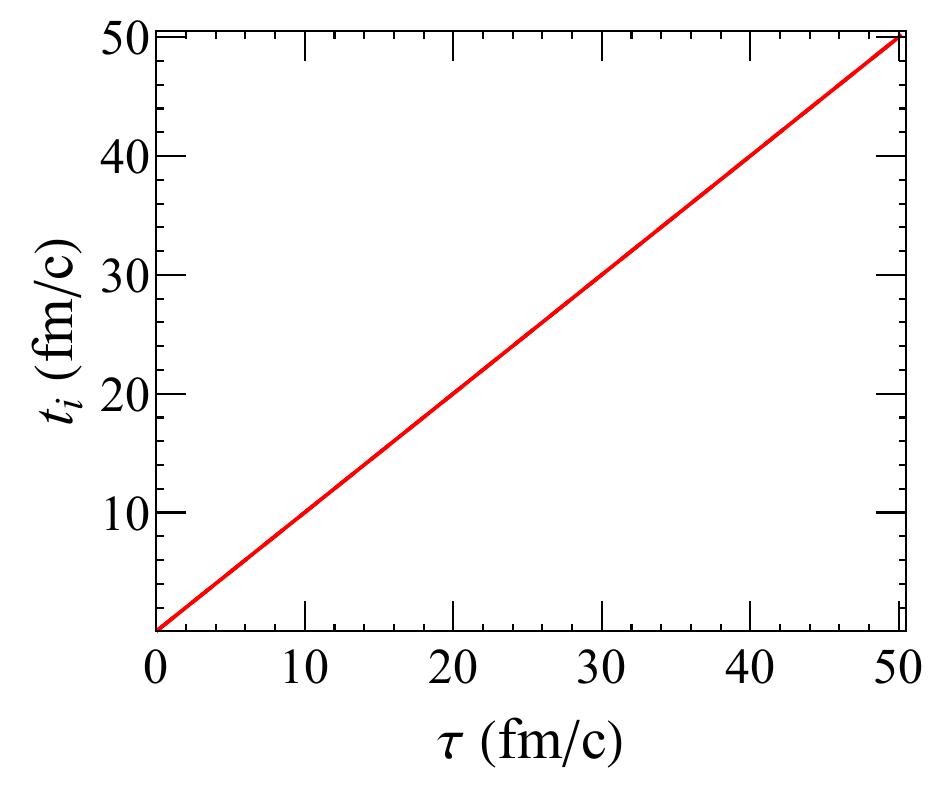}
\caption{The coordinates and momentum of particle 1 (solid lines) and 2 (dashed lines). The black and red line represent the non-relativistic and relativistic evolution, respectively. Here we take $c=-1$, which means an attractive interaction between the two particles. As usual, $m=1$ GeV. The initial condition II is chosen with $P_{\rm tot}=0$.} 
\label{fig.relaP0}
\end{figure}

Now, let's consider a special case in which the total momentum of the system is zero, yet each particle has significant initial momentum. To illustrate this, we performed a simulation using initial condition II, as shown in Fig.~\ref{fig.relaP0}. In this scenario, $\tau= t_1=t_2$ holds exactly because $U^\mu = (1,0,0,0)$. This allows for a direct comparison between the relativistic and non-relativistic trajectories within the same figure. 
The key difference in the equations of motion is that, in the non-relativistic case, the velocity term takes the form $p/m$, whereas in the relativistic case it becomes $p/E$.
The results clearly show that the ``frequency'' of the spiral motion is larger in the non-relativistic case compared to the relativistic one. This highlights that relativistic corrections can significantly modify the dynamical behavior of the system, even when the total momentum vanishes.

\subsection{Different choices of time constraints}

We examine now the influence of the time constraints. In principle, we can choose any time constraint as long as it satisfies the world line invariance and Poincar\'e transformation, discussed before. Two of them~\eqref{eq.const2} 
and~\eqref{eq:jamtime} we discussed already in previous section. For a scalar potential only, both yield the equations of motion
\begin{eqnarray}
{dq_i^\mu \over d\tau}&=&{p_i^\mu \over p_i^\mu U_\mu},\nonumber\\
{dp_i^\mu \over d\tau}&=&-\sum_{k=1}^2{1\over 2p_k^\mu U_\mu} {\partial \Phi_k \over \partial q_{i,\mu}}.
\end{eqnarray} 
as already discussed. 
One may also obtain different EoMs. This is shown by using other time-constraints, which also satisfies the world line invariance and Poincar\'e transformation, 
\begin{eqnarray}
\chi_1&=&{1\over 2}(q_1^\mu-q_2^\mu) U_\mu =0 , \nonumber\\
\chi_2&=&{1\over 2}q_1^\mu U_\mu-\tau =0.
\label{eq.timec2}
\end{eqnarray} 
The related $\lambda_i$ can be calculated as
\begin{eqnarray}
\lambda_1&=&(p_1^\mu U_\mu)^{-1}, \nonumber\\
\lambda_2&=&(p_2^\mu U_\mu)^{-1}. 
\end{eqnarray} 
Then, the equations of motion become,
\begin{eqnarray}
{dq_i^\mu \over d\tau}&=&{2p_i^\mu \over p_i^\mu U_\mu},\nonumber\\
{dp_i^\mu \over d\tau}&=&-\sum_{k=1}^2{1\over p_k^\mu U_\mu} {\partial \Phi_k \over \partial q_{i,\mu}}.
\end{eqnarray} 
Also in this case, the times $t_1$ and $t_2$ can be calculated via the time constraints,
\begin{eqnarray}
t_1&=&{1\over U^0}(2\tau+x_1U_x+y_1U_y+z_1U_z),\nonumber\\ 
t_2&=&{1\over U^0}(2\tau+x_2U_x+y_2U_y+z_2U_z).
\end{eqnarray} 
\begin{figure}[!htb]
\centering
\includegraphics[width=0.23\textwidth]{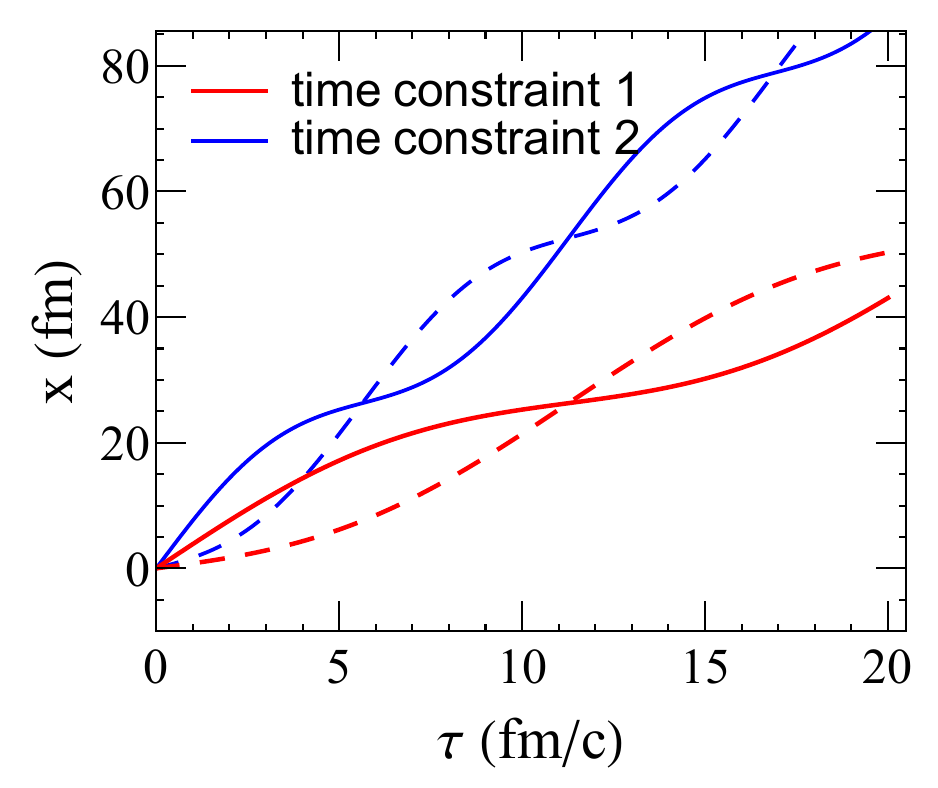}\includegraphics[width=0.23\textwidth]{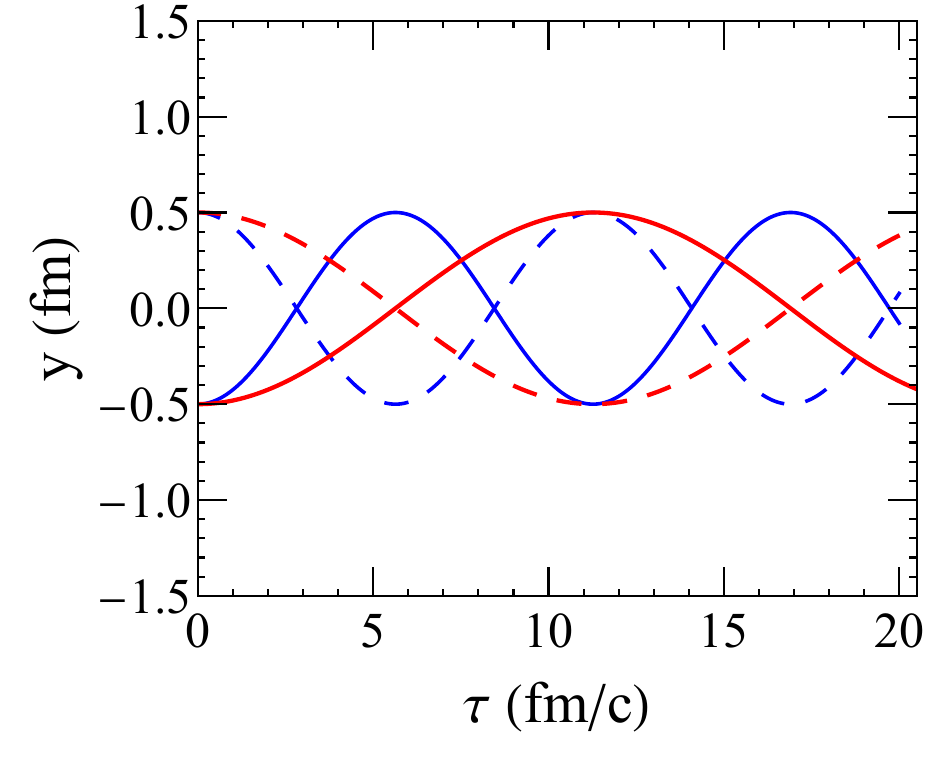}\\
\includegraphics[width=0.23\textwidth]{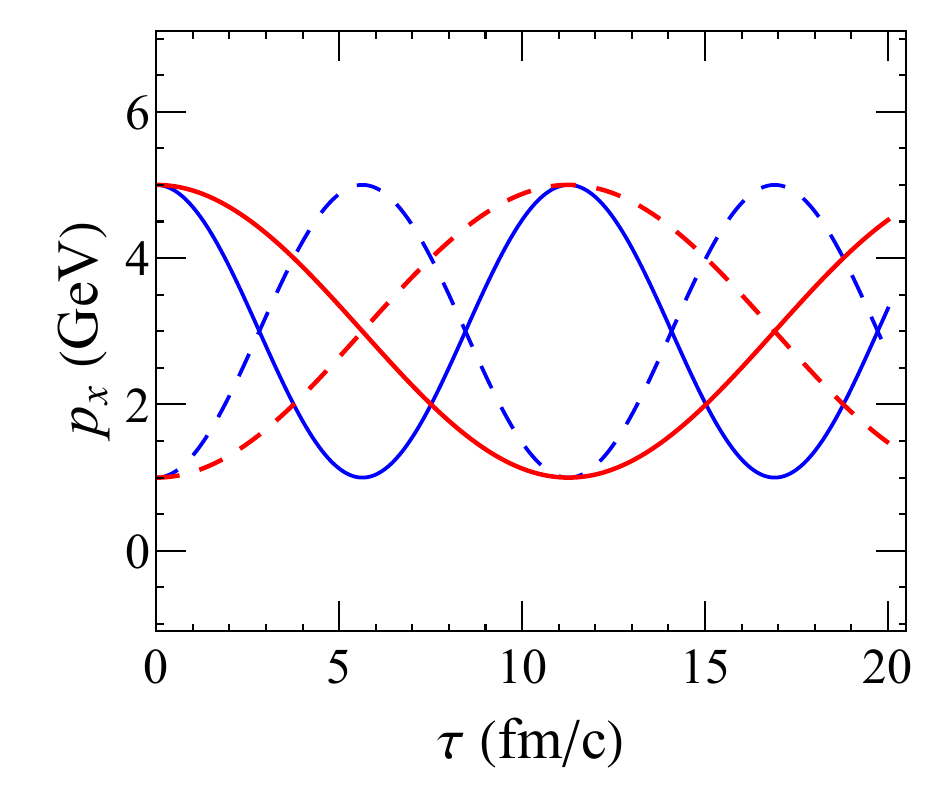}\includegraphics[width=0.23\textwidth]{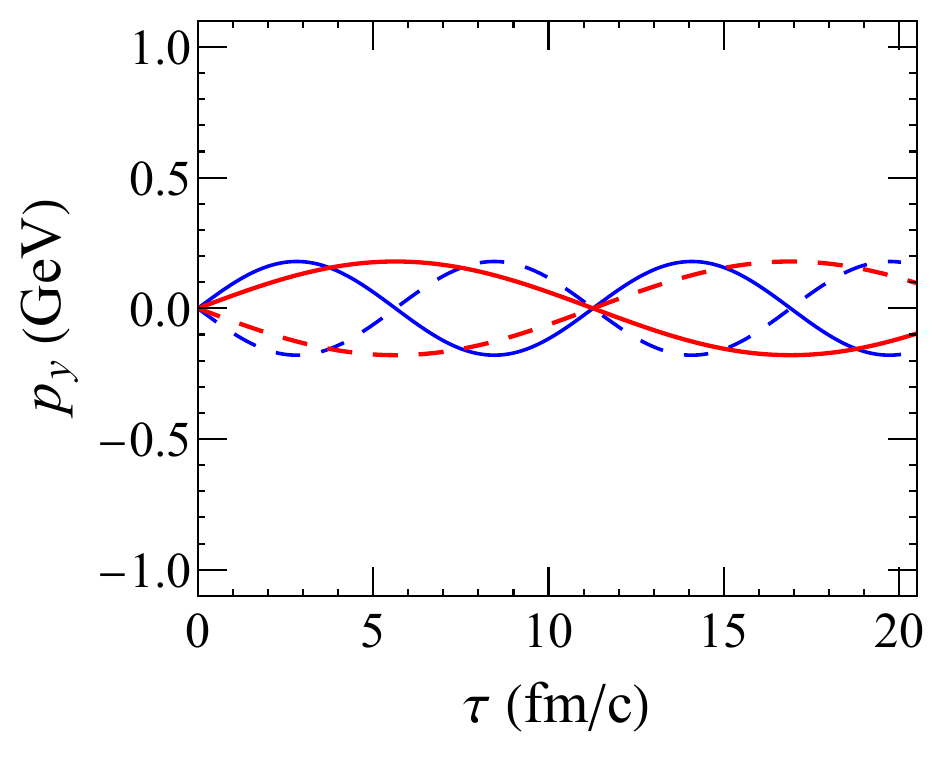}\\
\includegraphics[width=0.23\textwidth]{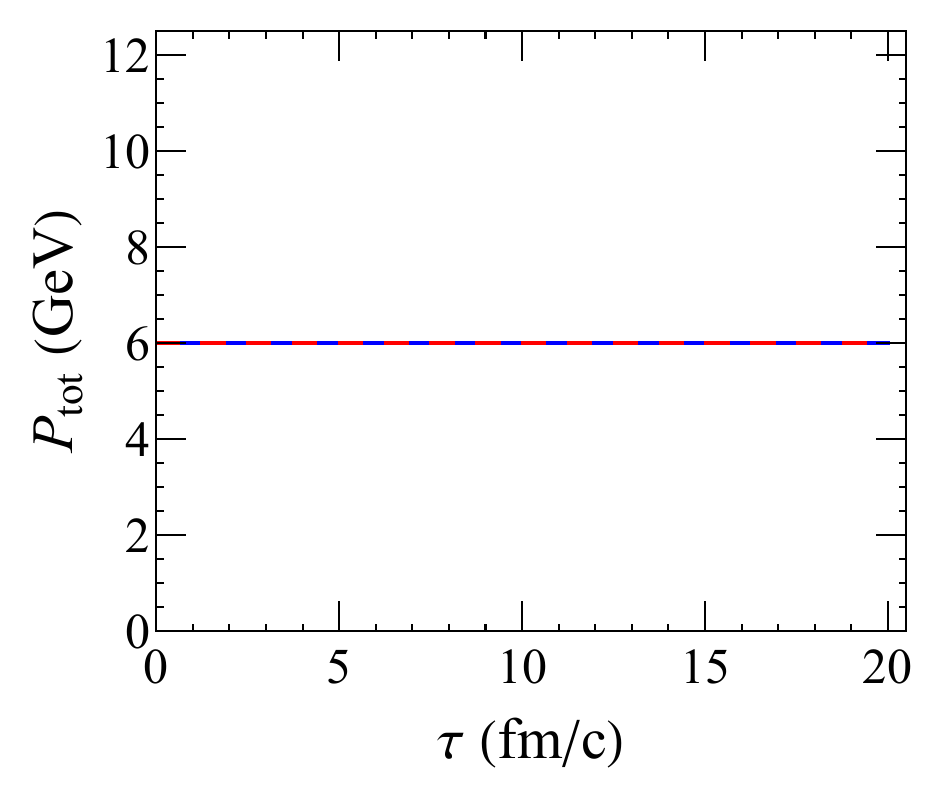}
\includegraphics[width=0.23\textwidth]{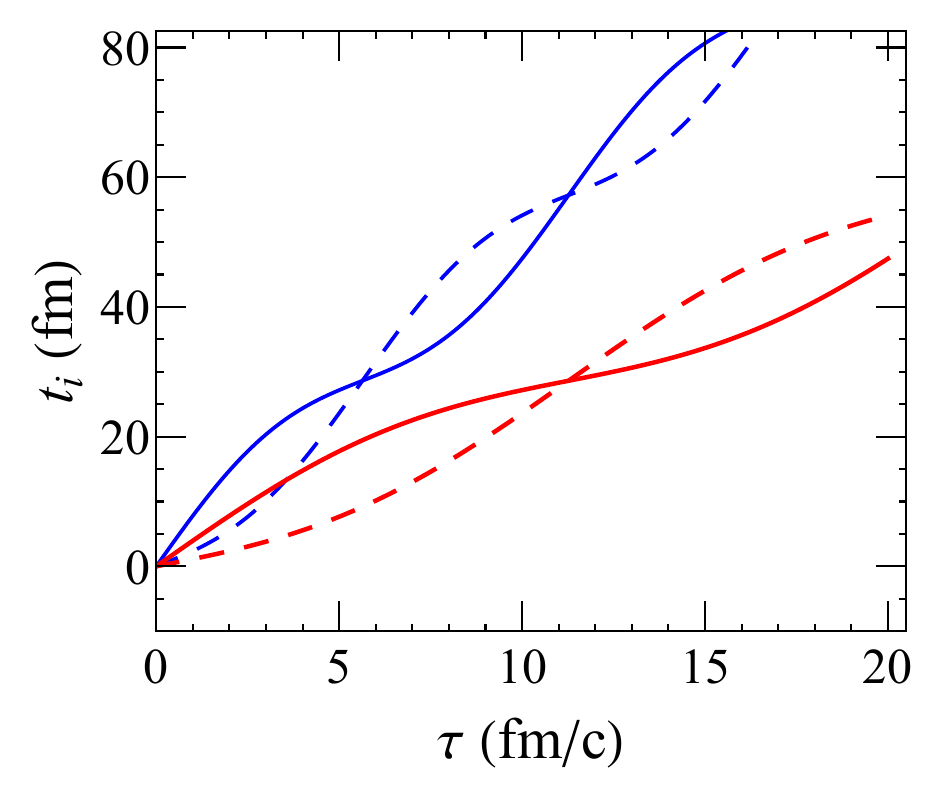}\\
\includegraphics[width=0.23\textwidth]{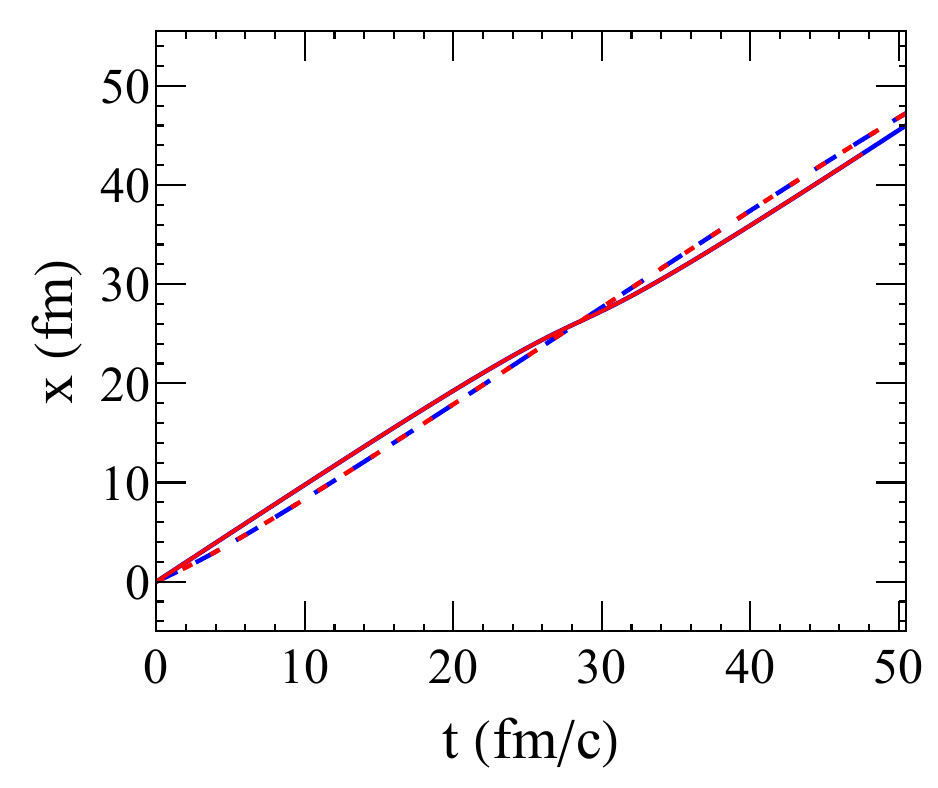}\includegraphics[width=0.23\textwidth]{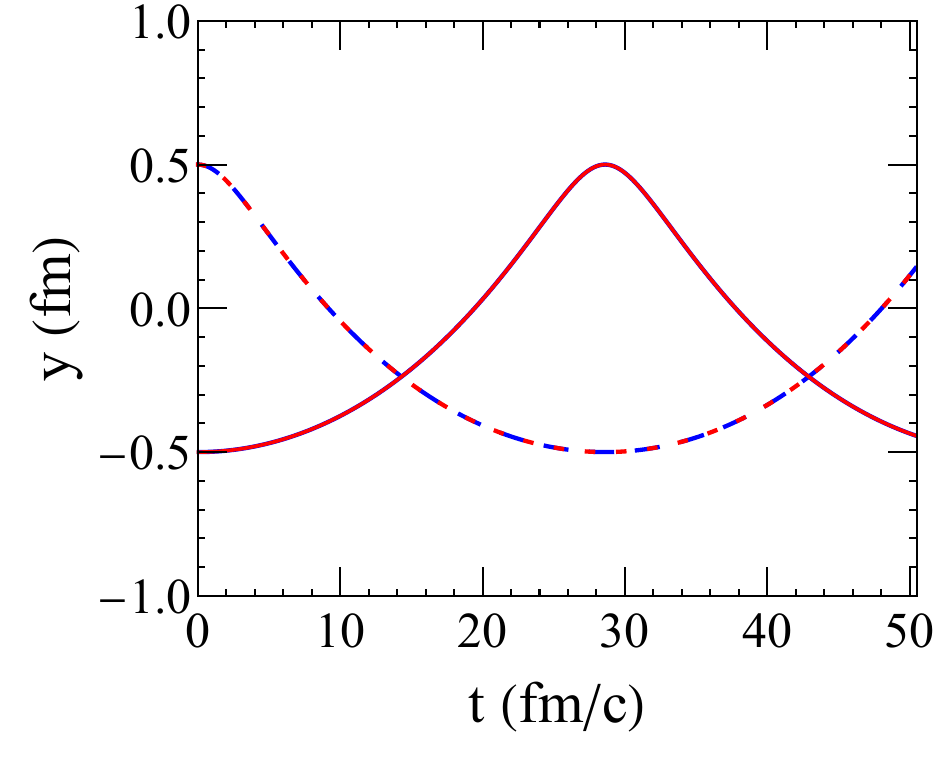}\\
\includegraphics[width=0.23\textwidth]{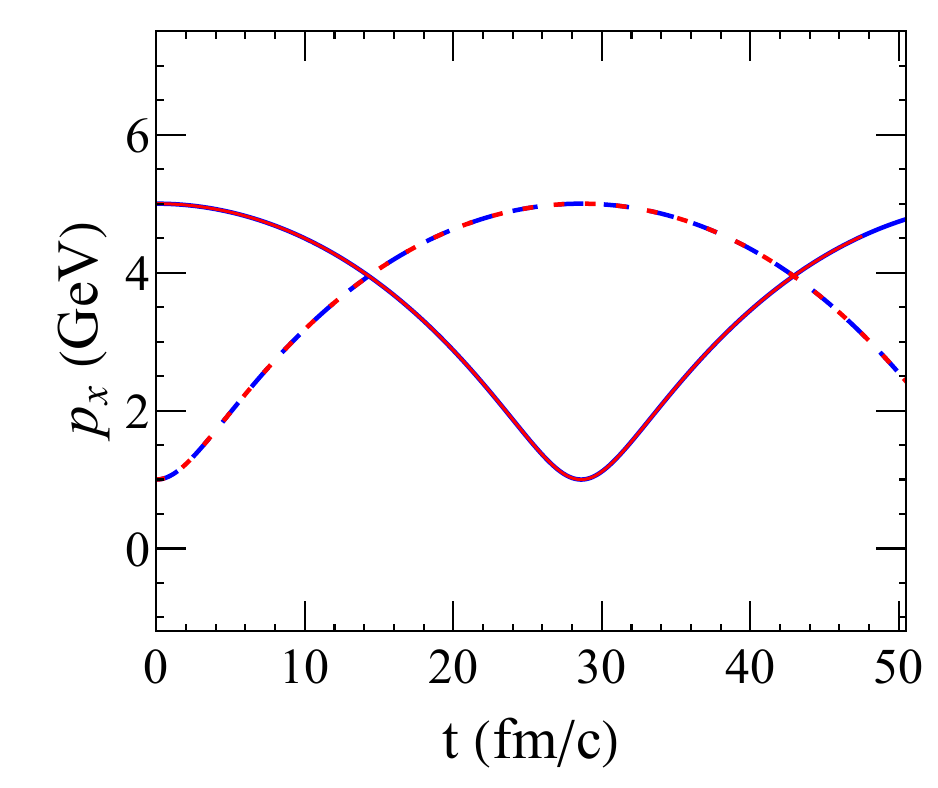}\includegraphics[width=0.23\textwidth]{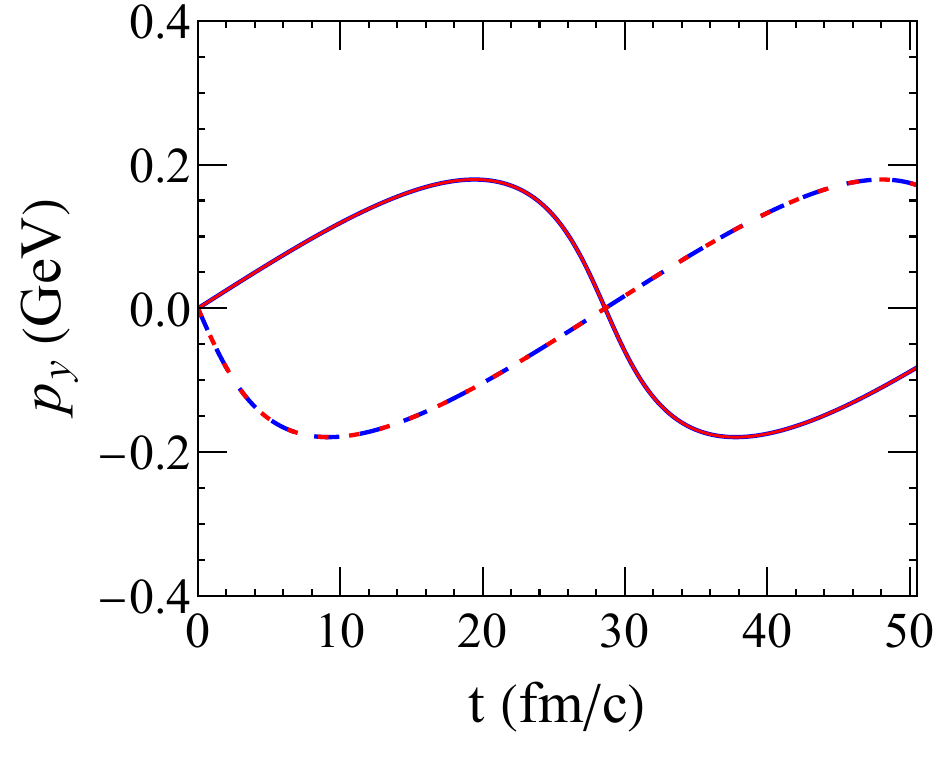}
\caption{The coordinates and momenta of particles 1 (solid lines) and 2 (dashed lines). The red and blue lines represent the results for time constraints 1 (Eq.~\eqref{eq.const2}) and time constraints 2 (Eq.~\eqref{eq.timec2}), respectively. Here we take an attractive interaction between the two particles ($c=-1$). The initial condition I is chosen.} 
\label{fig.relanewtc}
\end{figure}
\begin{figure}[!htb]
\centering
\includegraphics[width=0.23\textwidth]{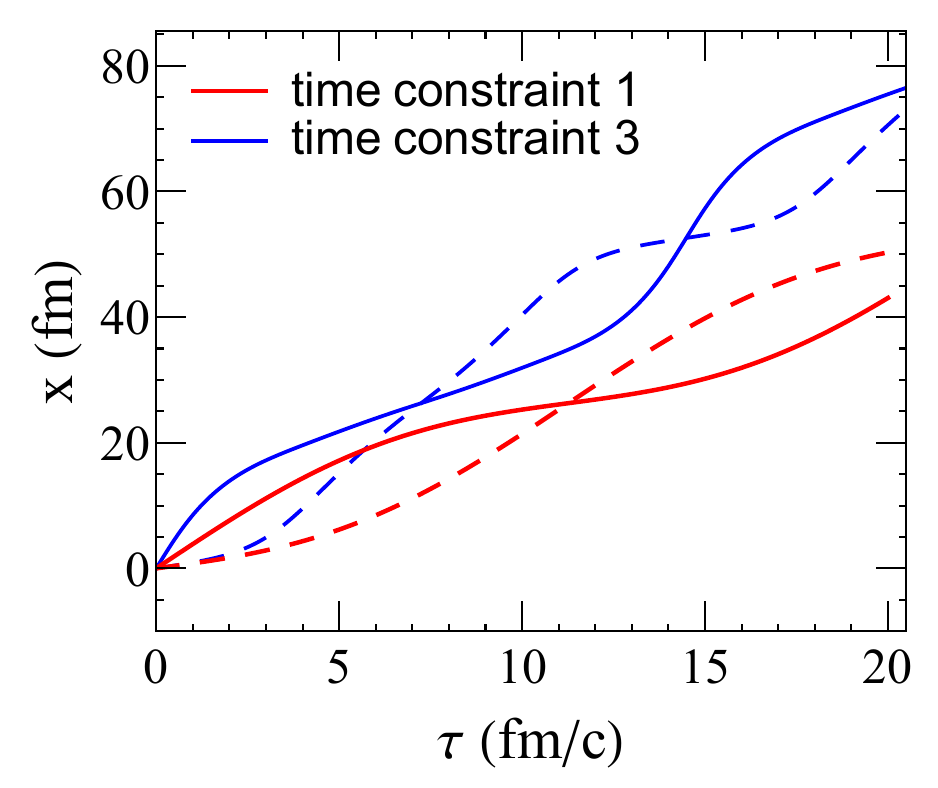}\includegraphics[width=0.23\textwidth]{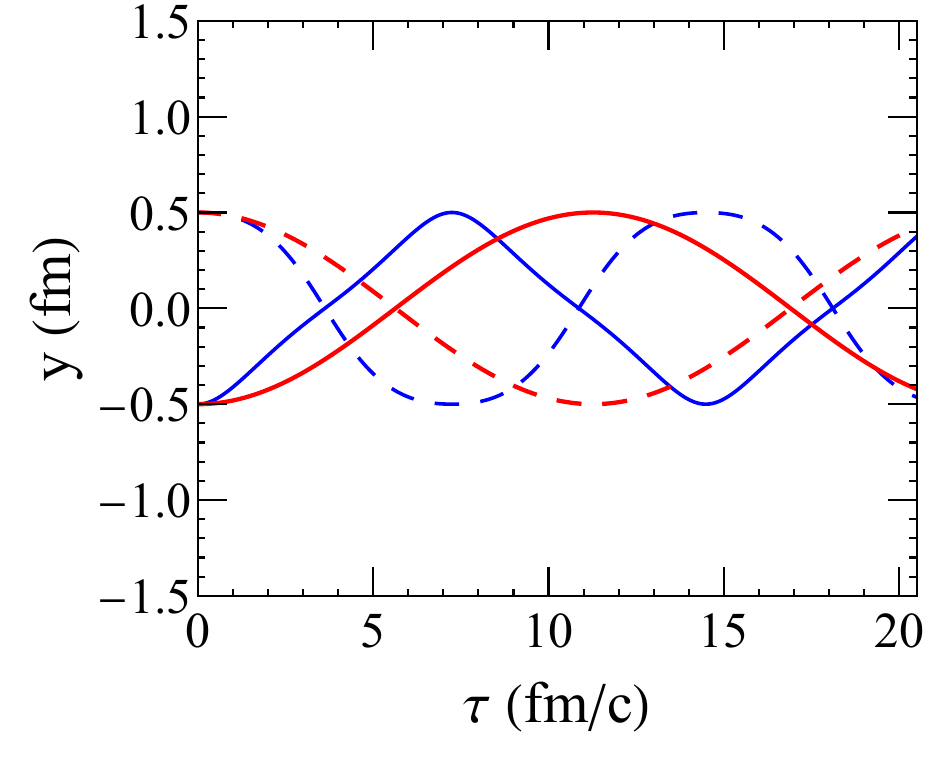}\\
\includegraphics[width=0.23\textwidth]{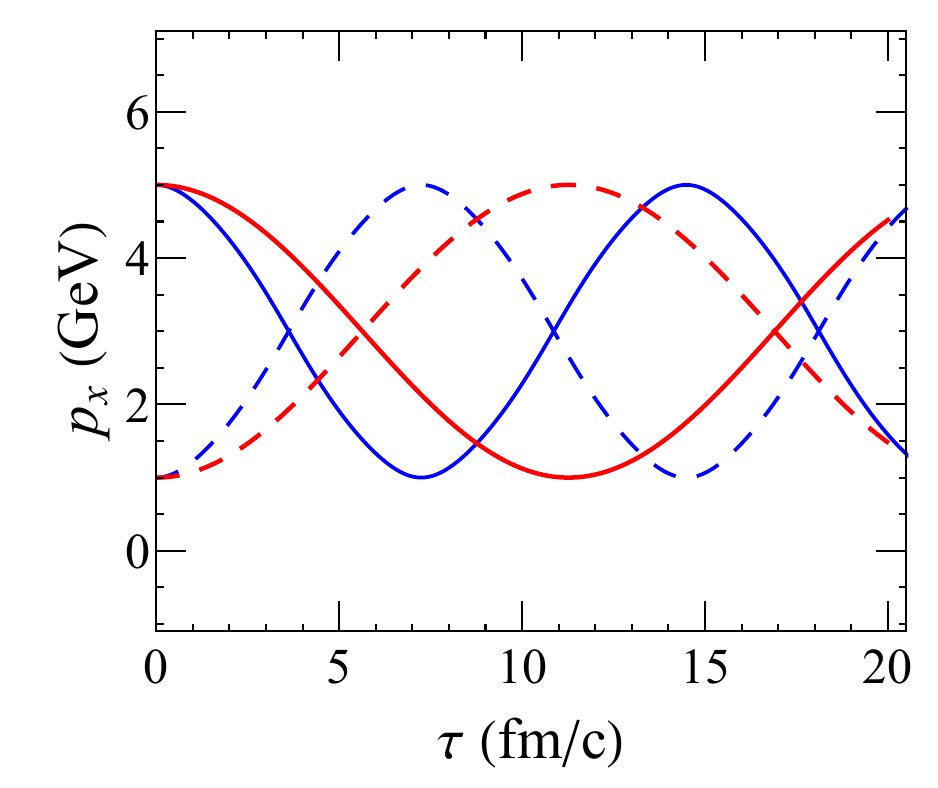}\includegraphics[width=0.23\textwidth]{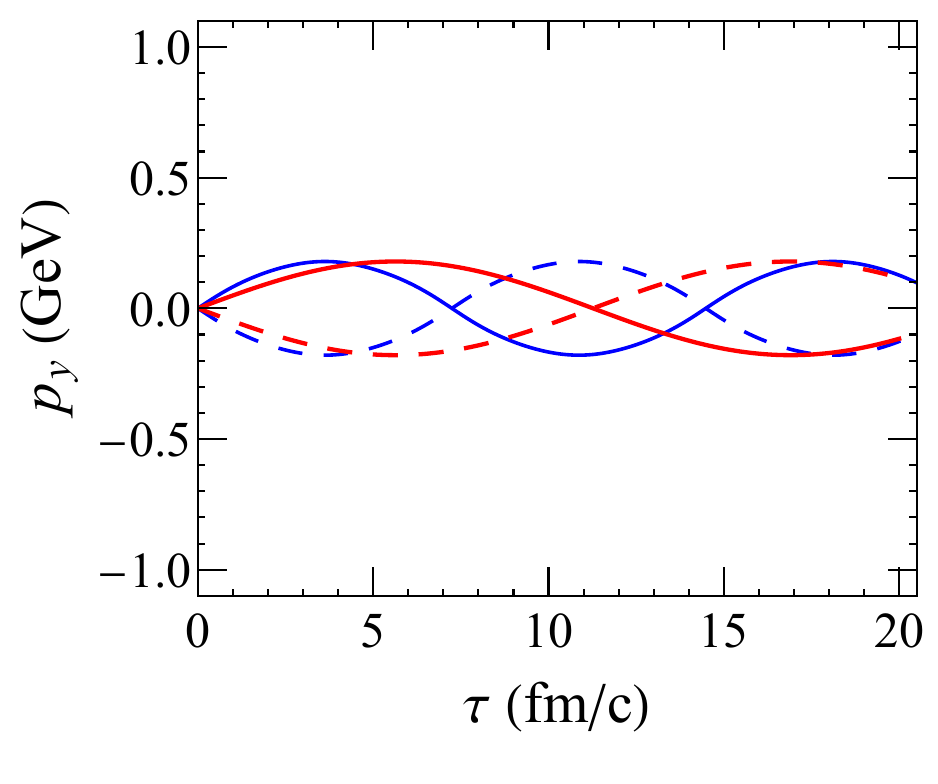}\\
\includegraphics[width=0.23\textwidth]{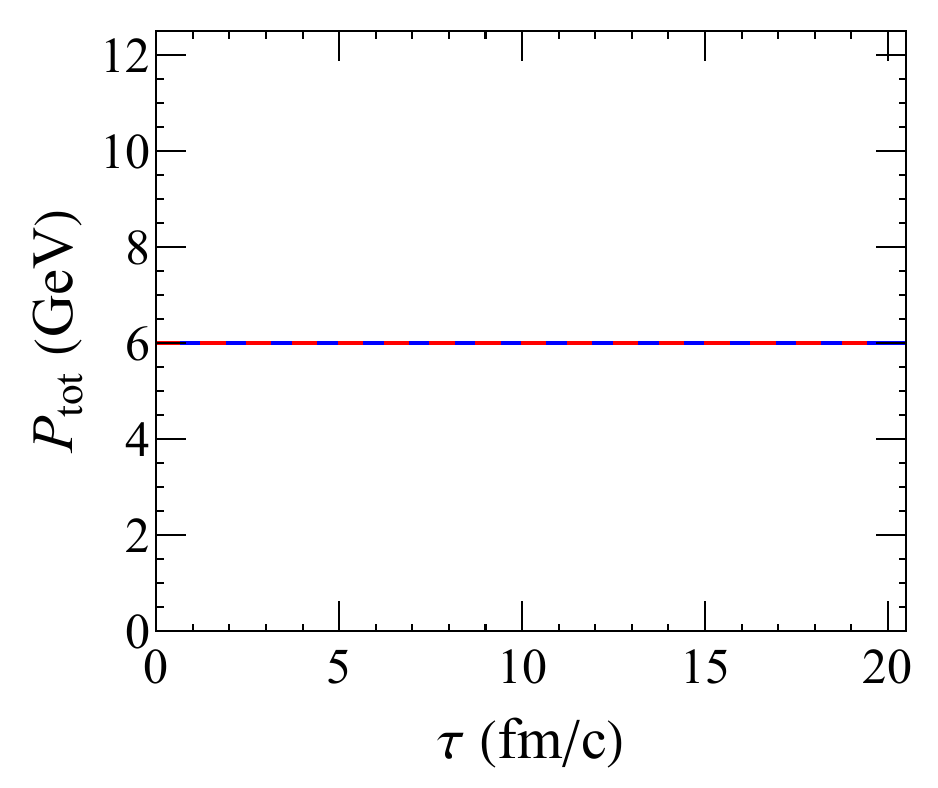}
\includegraphics[width=0.23\textwidth]{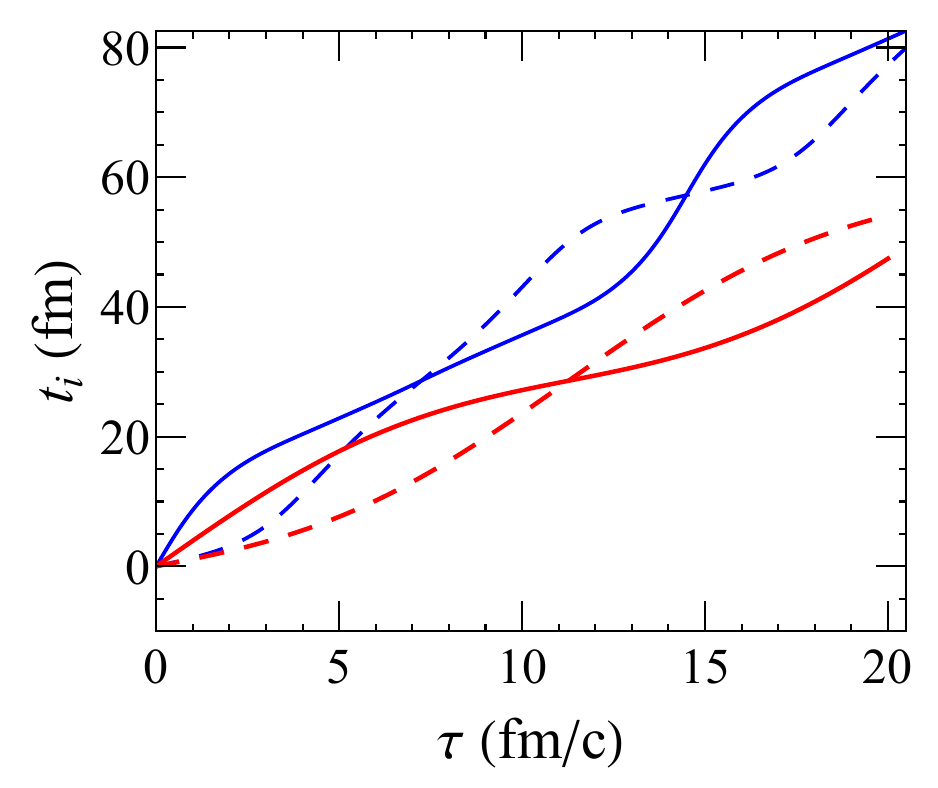}\\
\includegraphics[width=0.23\textwidth]{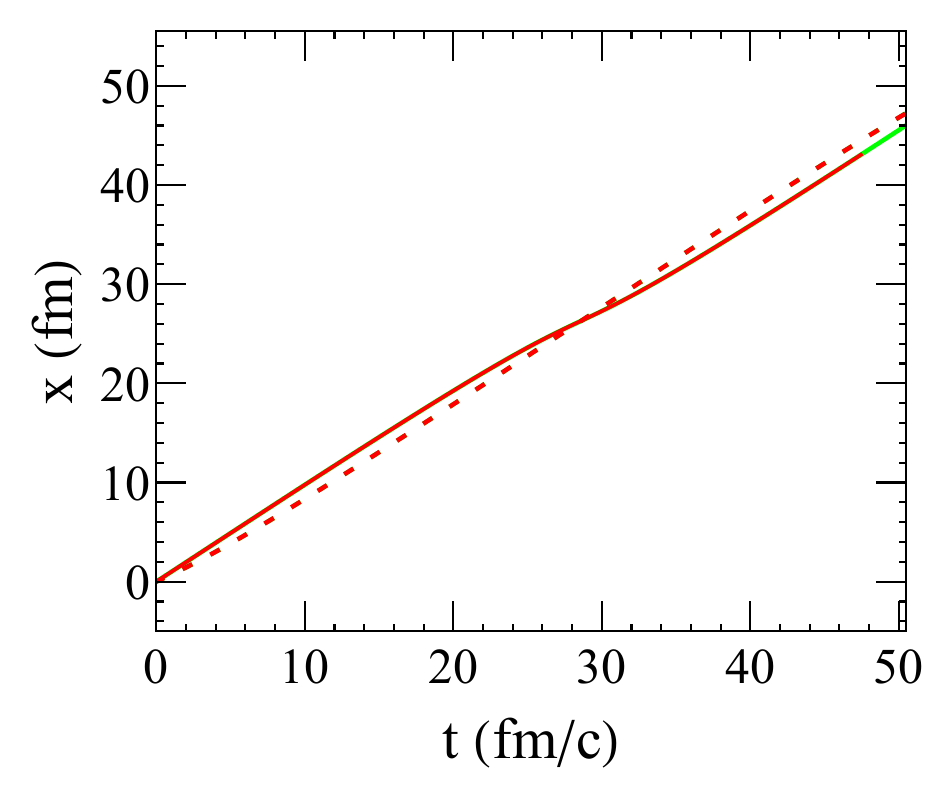}\includegraphics[width=0.23\textwidth]{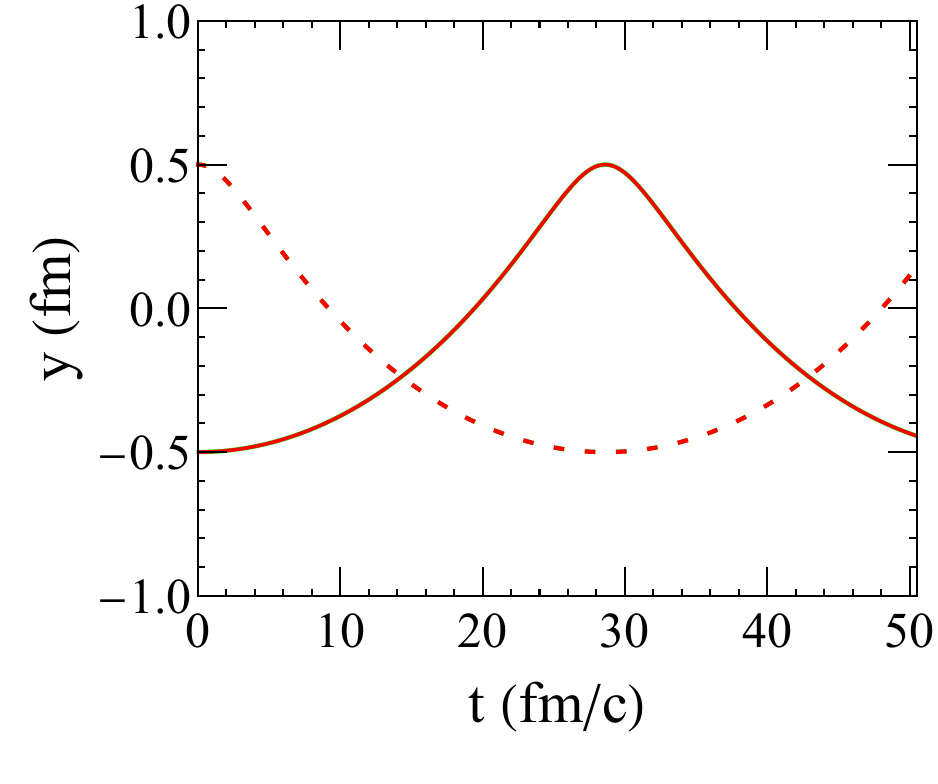}\\
\includegraphics[width=0.23\textwidth]{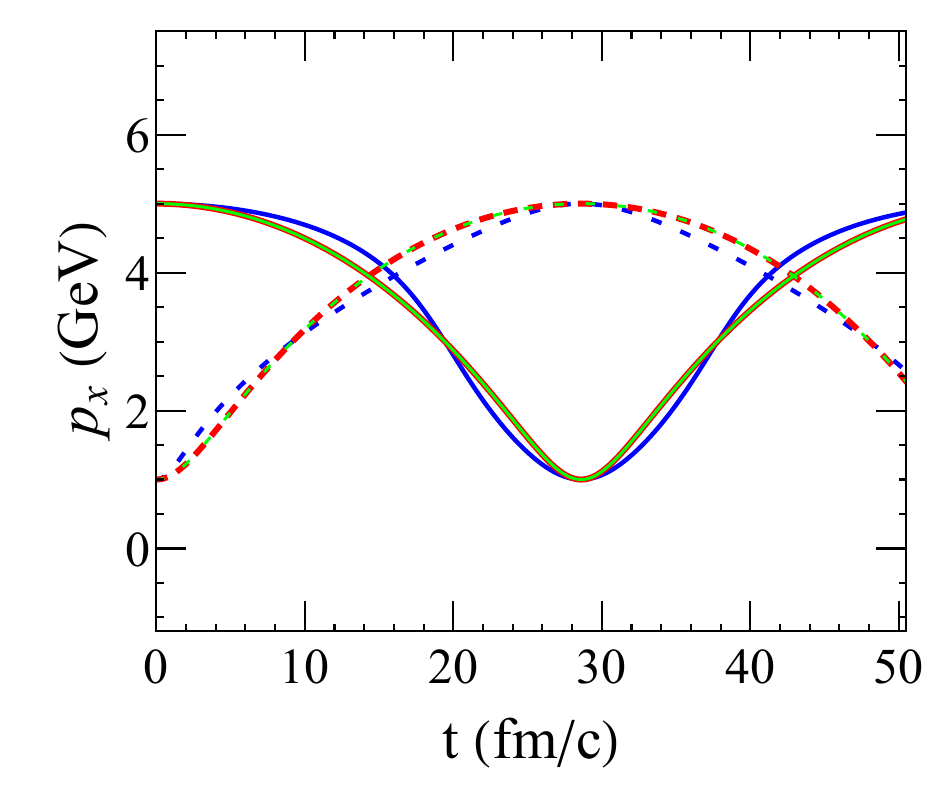}\includegraphics[width=0.23\textwidth]{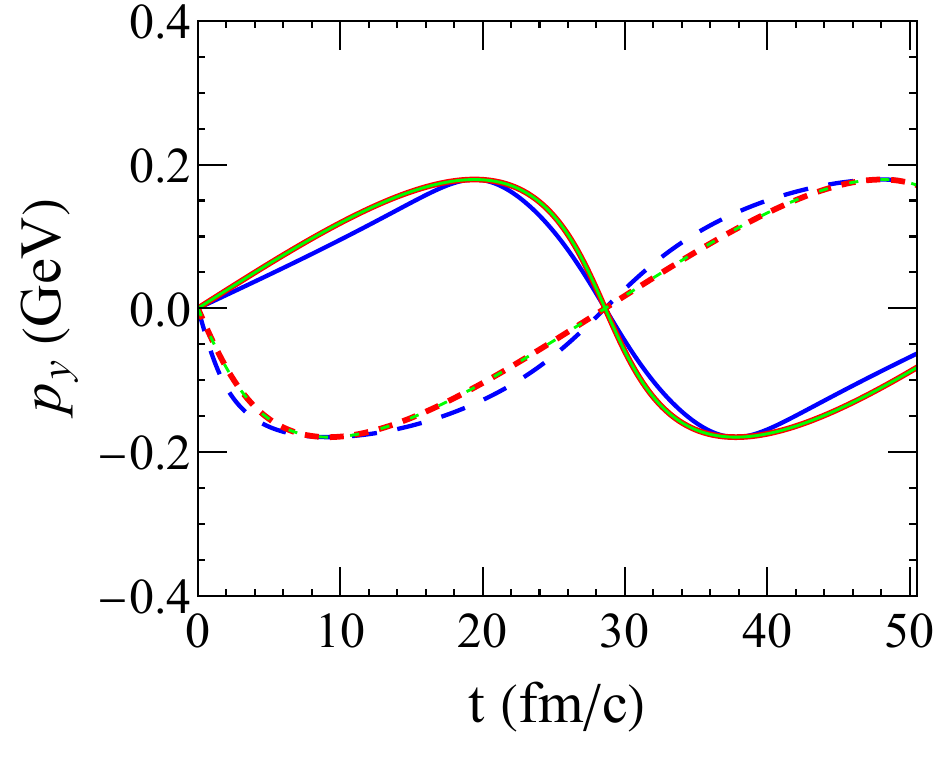}
\caption{The coordinates and momenta of particles 1 (solid lines) and 2 (dashed lines). The red and blue lines represent the results for time constraints 1 (Eq.~\eqref{eq.const2}) and time constraints 3 (Eq.~\eqref{eq.timec3}), respectively. Here we take an attractive interaction between the two particles ($c=-1$). The initial condition I is chosen. The green curves in the bottom panel is the momentum after transformation.} 
\label{fig.relanewtc3}
\end{figure}

In this case, although the equations of motion (EoM) cannot be directly reduced to the non-relativistic form, the particle trajectories are identical. This is ensured by the World line invariance condition. The results obtained, using the previous time constraint, Eq.~\eqref{eq.const2}, and the new time constraint, Eq.~\eqref{eq.timec2}, are shown in Fig.~\ref{fig.relanewtc}.
As we can see it is really not visible you have to present the curves in a different form, the trajectories plotted as a function of the computational time $\tau$ appear different for the two time constraints.  
However, when we take into account the relation between the computational time $\tau$ and the times $t_i$ of the particles, we can calculate the trajectories  ${\bf x}_i(t_i)$. After performing this transformation, we find that the physical particle trajectories are indeed identical under both time constraints, as shown in the bottom two panels. In this case, despite differences in parametrization with respect to $\tau$, the actual space-time evolution of the system remains unchanged, consistent with the requirement of world line invariance, as discussed in section~\ref{sec.preparation}.

We now consider a qualitatively different and asymmetric choice of time constraints, in which particle 1 defines the synchronization hypersurface,
\begin{eqnarray}
\chi_1&=&{1\over 2m}(q_1^\mu-q_2^\mu) p_{1,\mu} =0 , \nonumber\\
\chi_2&=&{1\over 2m}q_1^\mu p_{1,\mu}-\tau =0.
\label{eq.timec3}
\end{eqnarray} 
With this constraints, $q^\mu P_\mu \neq 0$ with $q^\mu= q_1^\mu-q_2^\mu$ as defined before, the momentum derivative terms exist, 
\begin{eqnarray}
{\partial \Phi \over \partial p_{i,\mu}}=-2{q^\nu P_\nu \over P^{2}}q_{T}^{\mu}{\partial \Phi \over \partial q_T^2}.
\end{eqnarray}
$\Phi$ is the scalar potential which has been defined in Eq.~\eqref{eq.phi12}.  
Then,
\begin{eqnarray}
\{\chi_1,H_1\}&=&{p_{1,\mu}p_1^\mu \over m}-{q^\mu\over 2m}{\partial \Phi\over \partial q_1^\mu}, \nonumber\\
\{\chi_1,H_2\}&=&-{p_{1,\mu}p_2^\mu \over m}-{q^\mu\over 2m}{\partial \Phi\over \partial q_1^\mu},\nonumber\\
\{\chi_2,H_1\}&=&{p_{1,\mu}p_1^\mu \over m}+{p_{1,\mu} \over 2m}{\partial \Phi \over \partial p_{1,\mu}}-{q_1^\mu\over 2m}{\partial \Phi\over \partial q_1^\mu},\nonumber\\
\{\chi_2,H_2\}&=&{p_{1,\mu} \over 2m}{\partial \Phi \over \partial p_{1,\mu}}-{q_1^\mu\over 2m}{\partial \Phi\over \partial q_1^\mu}.
\end{eqnarray} 
The related $\lambda_i$ can be calculated as
\begin{eqnarray}
\lambda_1&=&{m[2p_{1,\mu}p_2^\mu+q^\mu{\partial \Phi\over \partial q_1^\mu}]\over \mathcal {M}_2}, \nonumber\\
\lambda_2&=&{m[2p_{1,\mu}p_1^\mu-q^\mu{\partial \Phi\over \partial q_1^\mu}]\over \mathcal {M}_2},
\end{eqnarray} 
where
\begin{eqnarray}
\mathcal {M}_2&\equiv&2(p_{1,\nu}p_2^\nu)(p_{1,\nu}p_1^\nu)-[(p_{1,\nu}p_1^\nu+p_{1,\nu}p_2^\nu)q_1^\mu \\
&-& (p_{1,\nu}p_1^\nu)q^\mu] {\partial \Phi\over \partial q_1^\mu}+[(p_{1,\nu}p_1^\nu+p_{1,\nu}p_2^\nu)p_{1,\mu}]{\partial \Phi\over \partial p_{1,\mu}}.\nonumber
\end{eqnarray} 
Then, the equations of motion become,
\begin{eqnarray}
{dq_i^\mu \over d\tau}&=&\lambda_i 2p_i^\mu+\sum_{k=1}^2\lambda_k {\partial \Phi \over \partial p_{i,\mu}},\nonumber\\
{dp_i^\mu \over d\tau}&=&-\sum_{k=1}^2\lambda_k {\partial \Phi \over \partial q_{i,\mu}}.
\end{eqnarray} 
In this case, the times $t_1$ and $t_2$ can be calculated via the time constraints,
\begin{eqnarray}
t_1&=&{2m\tau + {\bf p}_1\cdot {\bf r}_1\over E_1},\nonumber\\ 
t_2&=&{2m\tau + {\bf p}_1\cdot {\bf r}_2\over E_1}.
\label{eq.t1t2tc3}
\end{eqnarray} 
We take the same initial condition I for the evolution and the results are shown in Fig.~\ref{fig.relanewtc3}.

We can see that the trajectories plotted as a function of the computational time $\tau$ appear different for the two time constraints. However, when we take into account the relation between the computational time $\tau$ and the times $t_i$ of the particles as shown in Eq~\eqref{eq.t1t2tc3}, we can calculate the trajectories ${\bf x}_i(t_i)$. After performing this transformation, we find that the physical particle trajectories are still different under both time constraints, as shown in the bottom two panels by comparing the red lines with blue lines. 

As discussed above, the mass-shell constraints alone do not uniquely determine the evolution of the system. Instead, they define a family of physically equivalent trajectories in the constrained phase space, which differ in their parametrization of the individual particle worldlines. The time constraints then specify how the particle time coordinates are synchronized and thereby select a particular parametrization of the evolution. Different admissible time constraints therefore correspond to different synchronization prescriptions rather than to different underlying physical dynamics.

For a constrained Hamiltonian two-body system, the infinitesimal gauge transformation of any phase-space variable $A(q,p)$ is,
\begin{eqnarray}
\delta A = \{A, \mathcal G_H\}
\end{eqnarray}
where $\mathcal G_H=\epsilon_1 H_1+\epsilon_2 H_2$, or as an another form, $\mathcal G_H=\epsilon_+(H_1+H_2)+\epsilon_- (H_1-H_2)$. 
$\epsilon_i$ is completely arbitrary infinitesimal parameter. 
From the Poisson brackets of $\{q^\mu, \mathcal G_H\}$, 
one finds that the combination $H_1+H_2$ mainly generates the
collective evolution of the two-body system. Due to 
\begin{eqnarray}
\{q^\mu,H_1-H_2\}= 2P^\mu,
\end{eqnarray}
and hence
\begin{eqnarray}
\delta q^\mu= 2\epsilon_+(p_1^\mu-p_2^\mu)+2\epsilon_-P^\mu.
\label{eq.deltaqmu}
\end{eqnarray}
Therefore the $H_1-H_2$ transformation changes only the component of
$q^\mu$ parallel to the total four-momentum $P^\mu$, while leaving the
transverse relative coordinate $q_T^\mu$
unchanged. This transformation can thus be interpreted as a change of
the synchronization of time of the two particles.

Although the phase-space trajectories obtained with Time Constraints 1 and 3 appear different when expressed in their respective time parametrizations, this difference does not imply different underlying physical dynamics. Rather, the two solutions employ different synchronization prescriptions for the particle worldlines. Therefore, a direct point-by-point comparison of the two trajectories at the same value of the evolution parameter is not meaningful. To compare them consistently, the solution obtained with Time Constraint 3, Eq.~\eqref{eq.timec3}, must first be transformed to satisfy the synchronization condition defined by Time Constraint 1, Eq.~\eqref{eq.const2}. In the following, we explicitly construct the corresponding finite transformation and demonstrate the relation between the two descriptions. For this purpose, we consider two time constraints, which can be written as,
\begin{eqnarray}
\chi_1^{(1)}
&=&q^\mu P_\mu=0,\\
\chi_1^{(3)}
&=&q^\mu p_{1,\mu}=0,
\end{eqnarray}
The first constraint synchronizes the two particles on the hyperplane
orthogonal to the total four-momentum,
corresponding to simultaneity in the pair center-of-mass frame.
The second constraint instead synchronizes the particles on the
hyperplane orthogonal to the four-momentum of particle~1.
We want to transform the $q^\mu$ to $q'{}^{\mu}=q^{\mu}+2\epsilon_- P^\mu$ (based on Eq.~\eqref{eq.deltaqmu}) to make sure $q'{}^{\mu} P_\mu=0$. This means,
\begin{eqnarray}
(q^{\mu}+2\epsilon_- P^\mu) P_\mu=0.
\end{eqnarray}
It gives $\epsilon_-=\frac{q^\nu P_\nu}{2P^2}$.
So, all transformation are
\begin{eqnarray}
q_1'{}^\mu &=& q_1^\mu - \frac{q^\nu P_\nu}{P^2}p_1^\mu, \nonumber\\
q_2'{}^\mu &=& q_2^\mu + \frac{q^\nu P_\nu}{P^2}p_2^\mu, \nonumber\\
p_i'{}^\mu &=& p_i^\mu.
\end{eqnarray}
The transformed relative coordinate is
\begin{eqnarray}
q'^\mu = q^\mu-\frac{q^\nu P_\nu}{P^2}P^\mu
= q_T^\mu.
\end{eqnarray}

Since the transformation acts on the complete four-coordinates
$q_i^\mu=(t_i,\mathbf{x}_i)$, both the spatial coordinates and the corresponding particle times are transformed simultaneously, whereas the four-momenta remain unchanged. Consequently, the transformed trajectory should be expressed in terms of the transformed particle time as
$\mathbf{x}_i'(t_i')$ and
$\mathbf{p}_i'(t_i')$.
As shown by the green curves in Fig.~\ref{fig.relanewtc3}, after this transformation, the trajectory obtained with Time Constraint~3 coincides with that obtained directly with Time Constraint~1 (red curves). This demonstrates that the apparent differences between the trajectories before the transformation arise from the different synchronization prescriptions used to relate the particle time coordinates to the common evolution parameter. Once the two solutions are expressed using the same synchronization prescription, they yield the same physical particle motion.

It should be emphasized that the initial conditions used for the two
time constraints must represent the same physical state. If two calculations start from different physical
initial conditions rather than different gauge representatives of the
same state, no subsequent gauge transformation can make the resulting
trajectories identical.

It is worth emphasizing that not every choice of time constraint requires
an additional gauge transformation. For example, Time Constraints~1 and~2
lead to identical particle trajectories when expressed in terms of the
particle coordinate time $t_i$, without applying any gauge mapping.
This is because the two constraints define the same synchronization
surface in the constrained phase space, differing only in the
parameterization of the evolution. In other words, the evolution
parameter $\tau$ is related by a simple reparameterization, while the
selected phase-space representative on each gauge orbit remains
unchanged. Consequently, the canonical coordinates
$q_i^\mu(\tau)$ differ only through the choice of evolution parameter,
and the trajectories become identical once they are expressed as
$\mathbf{x}_i(t_i)$ and $\mathbf{p}_i(t_i)$.

The above discussion naturally raises the question of whether the
particle trajectory depends on the choice of time constraint.
In Dirac's constrained Hamiltonian theory,
the canonical coordinates $q_i^\mu$ are gauge-dependent quantities and
therefore cannot, by themselves, be regarded as physical observables.
Consequently, the coordinate trajectory
$\mathbf{x}_i(t_i)$ obtained in a particular synchronization gauge is,
in general, gauge dependent. Physical observables are invariant under gauge transformations generated
by the first-class constraints. In Dirac's constrained Hamiltonian
theory, an observable $O$ satisfies
\begin{eqnarray}
\{O,H_i\}=0, \qquad i=1,2,
\end{eqnarray}
and is therefore called a Dirac observable.
Examples include the total four-momentum $P^\mu$, the invariant
transverse relative distance
$q_T^2$, the interaction potential
$\Phi(q_T^2)$, and other quantities constructed solely from
gauge-invariant combinations of the canonical variables.

\subsection{Numerical study of a two-nucleon system with scalar and vector interactions}
\label{subsec.nucleon}
The dynamics of relativistic heavy-ion collisions is characterized by the collisions among hadrons as well as by their potential interactions. Observables have been identified, which are sensitive to these potential interactions. It is one of the primary goals of heavy ion physics to identify the density and momentum dependence of the potential interaction. Applying these potentials to matter in equilibrium allows to identify the equation of state of strongly interacting matter, the energy as a function of baryon density (the zero component of the baryon current in the rest frame) and temperature. This quantity, which can presently not calculated theoretically, is of importance for other fields of physics like that of neutron stars and gravitational waves. To explore the equation of state, experimental results of heavy-ion collisions are compared with that of transport approaches. In these transport approaches usually a so called Schr\"odinger equivalent potential (SEP) is used~\cite{Cooper:1993nx}. Sometimes the momentum dependence of this potential is even completely neglected. 
The SEP arises from the non-relativistic reduction of the Dirac equation with relativistic scalar and vector mean fields. It provides an effective central potential that can be constrained, for densities below nuclear saturation, through analyses of elastic proton–nucleus scattering data.
However, this approximation scheme breaks down at beam energies in the GeV-per-nucleon regime. The reason is that the reduction to a single SEP conceals the very different dynamical roles of scalar and vector fields: the scalar potential modifies the effective mass of the particle, while the vector potential shifts the particle 4- momentum and acts differently in the equations of motion. At low energies, where the difference between scalar density and vector density (the zero component of the baryon 4-current in the local rest frame) is small, the simplification  in the SEP to identify both with the scalar density is justified, as it has been done in static Skyrme-type potentials, which are frequently used in low energy calculations. At higher beam energies this approximation becomes inadequate, because  the distinct roles of scalar and vector fields become crucial. In a fully relativistic EoM the scalar and vector fields enter the EoMs differently, as shown in Eq.~\eqref{enconst} and~\eqref{eq.rela2as}, and will influence the time evolution of particle trajectories in qualitatively different ways.

When proceeding to the study of covariant time-evolution equations for a two-body system, interacting via scalar and vector potentials, we adopt forms of these potentials that are particularly suited for future investigations of the EoS.
\begin{eqnarray}
A_i^\mu=\sum_{j\neq i}^Np_j^{*\mu}\rho_{ij}(q_T),\quad
S_i=\sum_{j\neq i}^N m_j \rho_{ij}(q_T),
\label{eq.potentialvandsN}
\end{eqnarray}
is on the one side of a form, which can be used to study an interacting system of nucleons ~\cite{Aichelin:1986wa,Hartnack:1989sd,Aichelin:1991xy}
and offers on the other side the possibility to study the different influence of scalar and vector potentials on nuclear trajectories.

For a two-body system we employ
\begin{eqnarray}
&&A_1^\mu=p_2^{*\mu}\rho_{12}(q_T),\nonumber\\
&&A_2^\mu=p_1^{*\mu}\rho_{12}(q_T),\nonumber\\
&&\Phi_1=\Phi_2=2m_\mu S= 2m_\mu m\rho_{12}(q_T),
\label{eq.potentialvands}
\end{eqnarray}
where $m_\mu=m_1m_2/(m_1+m_2)$ is the reduced mass. $p_i^{*\mu}=p_i^\mu-A_i^\mu$ is the mechanical momentum. In the numerical calculation, we take, of not otherwise stated,  $m_1=m_2=m=1\rm \ GeV$.
The definition of the density $\rho_{12}$ is~\cite{Nara:2020ztb} 
\begin{eqnarray}
\rho_{12}(q_T)={\kappa\over (2\pi L)^{3/2}}\exp(q_{T}^2/(2L)),
\label{eq:dens}
\end{eqnarray}
where the Gaussian width is taken as $L=1.5~\rm fm^2$.  $\kappa$ is a parameter with the dimension of $\rm fm^3$, which is introduced to ensure that $\rho_{12}$ is dimensionless and to control the strength of the interaction. We develop the formalism for $\rho_{12}<1$ because 
we intend to apply the formalism in an upcoming publication in a region where this condition is fulfilled.

From the above definition, we can show that the $A_i^\mu$ can be expressed as
\begin{eqnarray}
&&A_1^\mu={p_2^\mu \rho_{12} -p_1^\mu \rho_{12}^2 \over 1-\rho_{12}^2},\nonumber\\
&&A_2^\mu={p_1^\mu \rho_{12} -p_2^\mu \rho_{12}^2 \over 1-\rho_{12}^2}.
\label{eq.a1a22body}
\end{eqnarray}
In this case, we can see that the vector potential is momentum-dependent. The Poisson bracket gives, 
\begin{eqnarray}
\{p_1^{*\mu},p_2^{*\mu}\}&=& {(p_1^\mu+p_2^\mu)\over (1+\rho_{12})^2(1-\rho_{12})}{\partial \rho_{12}\over \partial q_1^\mu}\nonumber\\
&=&{2P^{*\mu} q_{T,\mu} \over (1-\rho_{12})^2 L}{\partial \rho_{12}\over \partial q_T^2}=0.
\end{eqnarray}
Therefore, it fulfills the compatibility condition, $\{H_1,H_2\}=0$ together with the scalar potential.

The differentiation of the mechanical momentum yields the following:
\begin{eqnarray}
&&{\partial p_{1,\nu}^*\over \partial p_{1,\mu}}={1\over 1-\rho_{12}^2}\delta_{\nu}^{\mu}, \nonumber\\
&&{\partial p_{1,\nu}^* \over \partial p_{2,\mu}}={-\rho_{12} \over 1-\rho_{12}^2}\delta_{\nu}^{\mu},\nonumber\\
&& {\partial p_{2,\nu}^*\over \partial p_{1,\mu}}={-\rho_{12}\over 1-\rho_{12}^2}\delta_{\nu}^{\mu}, \nonumber\\
&&{\partial p_{2,\nu}^*\over \partial p_{2,\mu}}={1 \over 1-\rho_{12}^2}\delta_{\nu}^{\mu},\nonumber\\
&&{\partial \Phi\over \partial p_{i,\mu}}=0,
\end{eqnarray}
and,
\begin{eqnarray}
&&{\partial p_{1,\nu}^*\over \partial q_{i,\mu}}=-{p_{2,\nu}-2p_{1,\nu}\rho_{12}+p_{2,\nu}\rho_{12}^2 \over (1-\rho_{12}^2)^2}{\partial \rho_{12}\over \partial q_{i,\mu}}, \quad i=1,2\nonumber\\
&& {\partial p_{2,\nu}^*\over \partial q_{i,\mu}}=-{p_{1,\nu}-2p_{2,\nu}\rho_{12}+p_{1,\nu}\rho_{12}^2 \over (1-\rho_{12}^2)^2}{\partial \rho_{12}\over \partial q_{i,\mu}}, \quad i=1,2\nonumber\\
&&{\partial \Phi\over \partial q_{i,\mu}}=2m_\mu m {\partial \rho_{12}\over \partial q_{i,\mu}},\quad i=1,2.
\end{eqnarray}

Following the similar procedure as shown at the beginning of Sec.~\ref{sec.2body}, we find 
\begin{eqnarray}
\lambda_1&=&{1+\rho_{12}\over 2U_\mu p_1^{*\mu}}={(1+\rho_{12})^2(1-\rho_{12}) \over 2(p_1^\mu-p_2^\mu\rho_{12}) U_\mu }, \nonumber\\
\lambda_2&=&{1+\rho_{12}\over 2U_\mu p_2^{*\mu}}={(1+\rho_{12})^2(1-\rho_{12}) \over 2(p_2^\mu-p_1^\mu\rho_{12}) U_\mu }. 
\end{eqnarray}
Finally, we obtain the explicit form of the EoMs from Eq.~\eqref{eq.rela2as}.
\begin{widetext}
\begin{eqnarray}
{dq_1^\mu \over d\tau}&=&{p_1^\mu-p_2^\mu\rho_{12} \over (p_1^\mu-p_2^\mu\rho_{12}) U_\mu}{1 \over 1-\rho_{12}}- {p_2^\mu-p_1^\mu\rho_{12} \over (p_2^\mu-p_1^\mu \rho_{12})U_\mu}{\rho_{12}\over 1-\rho_{12}},\nonumber\\
{dq_2^\mu \over d\tau}&=&{p_1^\mu-p_2^\mu\rho_{12} \over (p_1^\mu-p_2^\mu\rho_{12}) U_\mu}{-\rho_{12} \over 1-\rho_{12}}+ {p_2^\mu-p_1^\mu\rho_{12} \over (p_2^\mu-p_1^\mu \rho_{12})U_\mu}{1\over 1-\rho_{12}},\nonumber\\
{dp_1^\mu \over d\tau}&=&\left[{(p_1^\nu-p_2^\nu \rho_{12})(p_{2,\nu}-2 p_{1,\nu}\rho_{12}+ p_{2,\nu}\rho_{12}^2) \over  (p_1^\mu-p_2^\mu \rho_{12})U_\mu} {1 \over (1+\rho_{12})(1-\rho_{12})^2} \right]{\partial \rho_{12} \over \partial q_{1,\mu}}\nonumber\\
&+&\left[{(p_2^\nu-p_1^\nu \rho_{12})(p_{1,\nu}-2 p_{2,\nu}\rho_{12}+ p_{1,\nu}\rho_{12}^2 )\over (p_2^\mu-p_1^\mu \rho_{12})U_\mu} {1 \over (1+\rho_{12})(1-\rho_{12})^2}\right]{\partial \rho_{12} \over \partial q_{1,\mu}}\nonumber\\
&-&{(1+\rho_{12})^2(1-\rho_{12})m_\mu m \over (p_1^\mu-p_2^\mu\rho_{12})U_\mu}{\partial \rho_{12} \over \partial q_{1,\mu}}-{(1+\rho_{12})^2(1-\rho_{12})m_\mu m \over (p_2^\mu-p_1^\mu\rho_{12})U_\mu}{\partial \rho_{12} \over \partial q_{1,\mu}},\nonumber\\
{dp_2^\mu \over d\tau}&=&\left[{(p_1^\nu-p_2^\nu \rho_{12})(p_{2,\nu}-2 p_{1,\nu}\rho_{12}+ p_{2,\nu}\rho_{12}^2) \over  (p_1^\mu-p_2^\mu \rho_{12})U_\mu} {1 \over (1+\rho_{12})(1-\rho_{12})^2} \right]{\partial \rho_{12} \over \partial q_{2,\mu}}\nonumber\\
&+&\left[{(p_2^\nu-p_1^\nu \rho_{12})(p_{1,\nu}-2 p_{2,\nu}\rho_{12}+ p_{1,\nu}\rho_{12}^2 )\over (p_2^\mu-p_1^\mu \rho_{12})U_\mu} {1 \over (1+\rho_{12})(1-\rho_{12})^2}\right]{\partial \rho_{12} \over \partial q_{2,\mu}}\nonumber\\
&-&{(1+\rho_{12})^2(1-\rho_{12})m_\mu m \over (p_1^\mu-p_2^\mu\rho_{12})U_\mu}{\partial \rho_{12} \over \partial q_{2,\mu}}-{(1+\rho_{12})^2(1-\rho_{12})m_\mu m \over (p_2^\mu-p_1^\mu\rho_{12})U_\mu}{\partial \rho_{12} \over \partial q_{2,\mu}}.
\label{eq.eomrela}
\end{eqnarray} 
Non-relativisitic EoMs (Eq.~\eqref{eq.nonrela2as}) can be written as,
\begin{eqnarray}
{d{\bf q}_1\over dt}&=&{{\bf p}_1-2{\bf p}_2 \rho_{12}+{\bf p}_1\rho_{12}^2 \over m(1-\rho_{12}^2)^2}+{{\bf p}_1\rho_{12}\over m(1+\rho_{12})},\nonumber\\
{d{\bf q}_2\over dt}&=&{{\bf p}_2-2{\bf p}_1 \rho_{12}+{\bf p}_2\rho_{12}^2 \over m(1-\rho_{12}^2)^2}+{{\bf p}_2\rho_{12}\over m(1+\rho_{12})},\nonumber\\
{d{\bf p}_1\over dt}&=&\left[{2{\bf p}_1\cdot {\bf p}_2-3({\bf p}_1^2+{\bf p}_2^2)\rho_{12}+6{\bf p}_1\cdot {\bf p}_2\rho_{12}^2-({\bf p}_1^2+{\bf p}_2^2)\rho_{12}^3 \over m(1-\rho_{12}^2)^3}-{E_1+E_2\over (1+\rho_{12})^2}\right]{\partial \rho_{12} \over \partial {\bf q}_1}-{\partial V \over \partial {\bf q}_1},\nonumber\\
{d{\bf p}_2\over dt}&=&\left[{2{\bf p}_1\cdot {\bf p}_2-3({\bf p}_1^2+{\bf p}_2^2)\rho_{12}+6{\bf p}_1\cdot {\bf p}_2\rho_{12}^2-({\bf p}_1^2+{\bf p}_2^2)\rho_{12}^3 \over m(1-\rho_{12}^2)^3}-{E_1+E_2\over (1+\rho_{12})^2}\right] {\partial \rho_{12} \over \partial {\bf q}_2}-{\partial V \over \partial {\bf q}_2},
\label{eq.eomnonrela}
\end{eqnarray} 
\end{widetext}
where $E_i=m+p_i^2/(2m)$.
In the non-relativistic limit, which requires $|{\bf p}^*|\ll m$, consequently, $\rho_{12}\ll 1$, we see that the relativistic equations of motion (EoMs) become non-relativistic one.

\begin{figure}[!htb]
\centering
\includegraphics[width=0.23\textwidth]{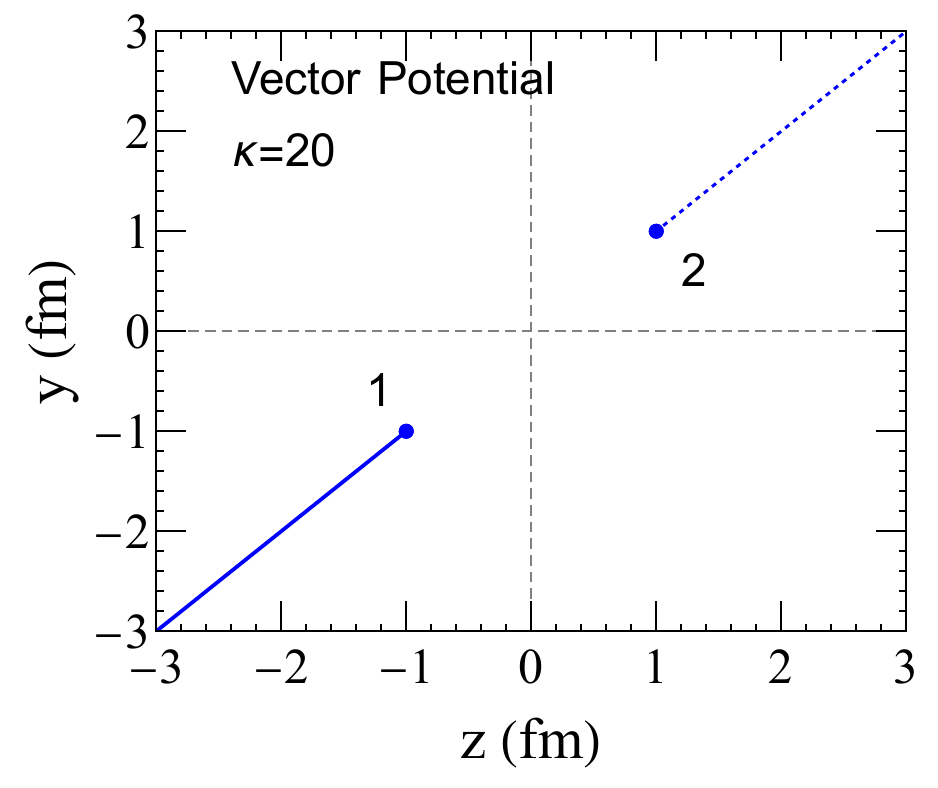}
\includegraphics[width=0.23\textwidth]{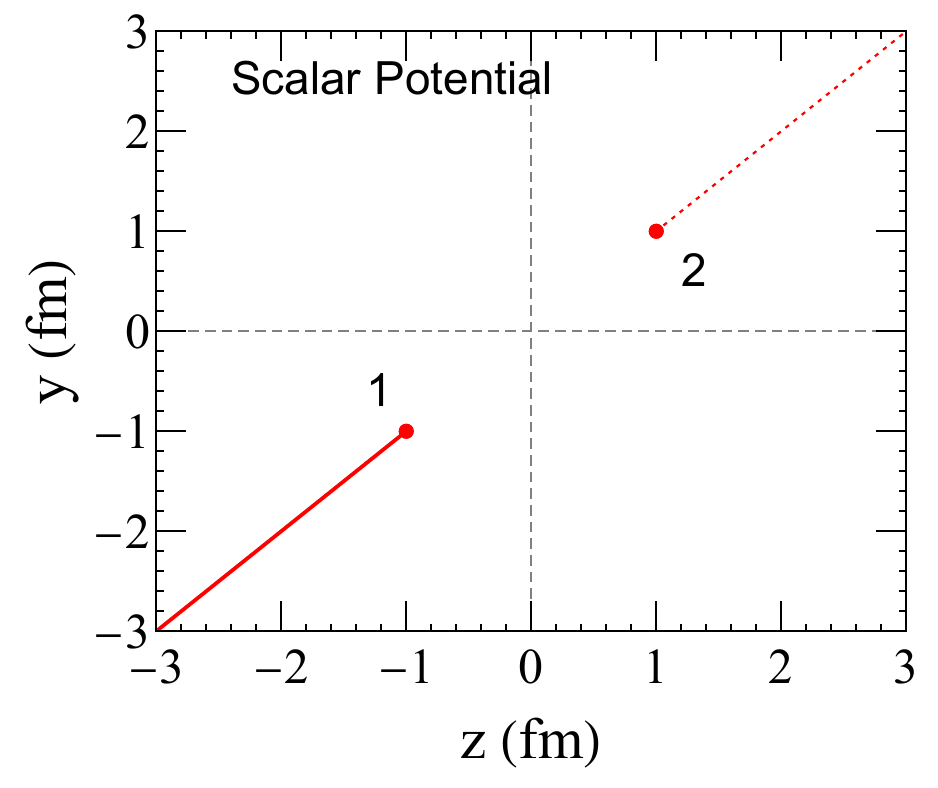}\\
\includegraphics[width=0.23\textwidth]{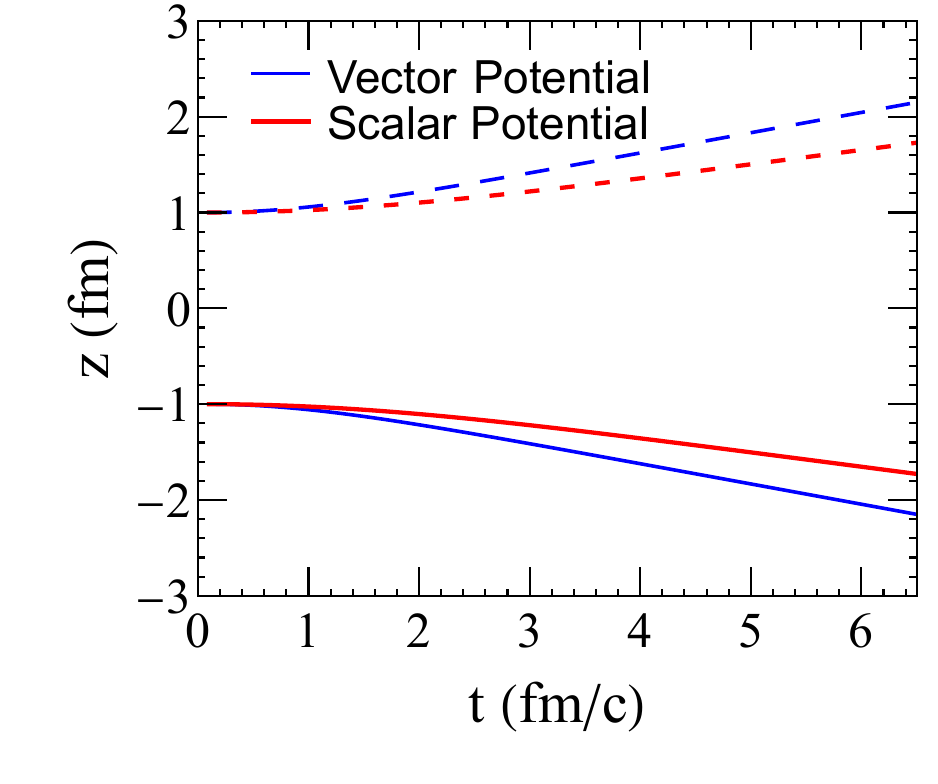}
\includegraphics[width=0.23\textwidth]{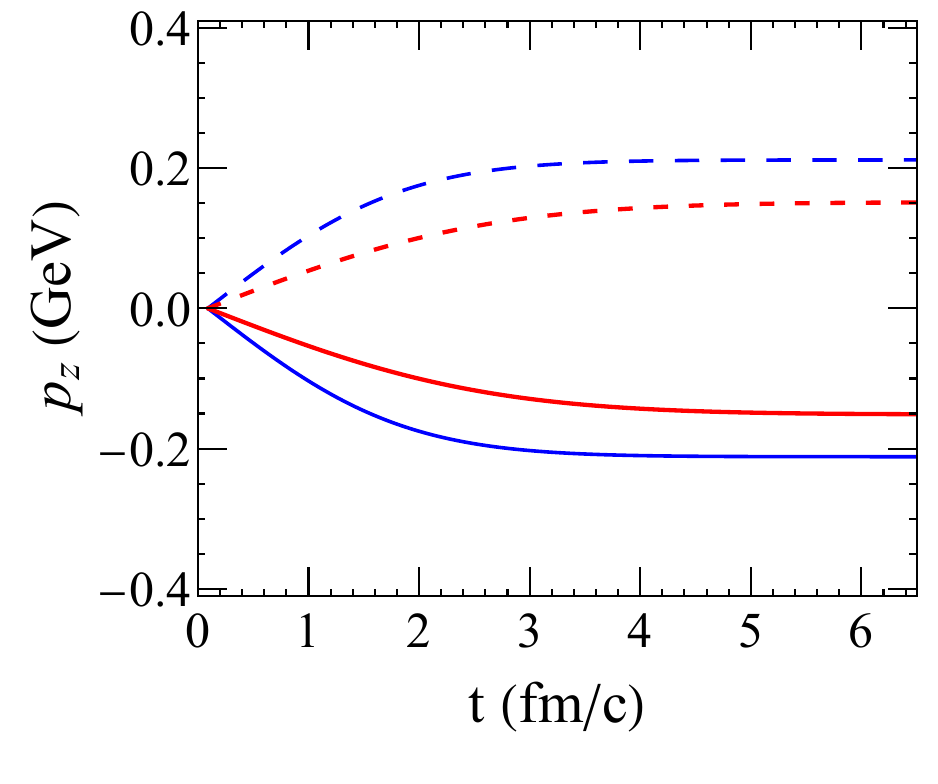}
\caption{Time evolution of the positions and momenta of particle 1 (solid lines) and 2 (dashed lines), applying the relativistic time evolution equations. The blue
line represents the time evolution with the vector potential only, the red line that with a scalar potential only.
The initial condition III is chosen and with $\kappa=20$ in the density. At $t=0$, $A_i^0=S_i$ is imposed.} 
\label{fig.vands}
\end{figure}

\begin{figure*}[!htb]
\centering
\includegraphics[width=0.32\textwidth]{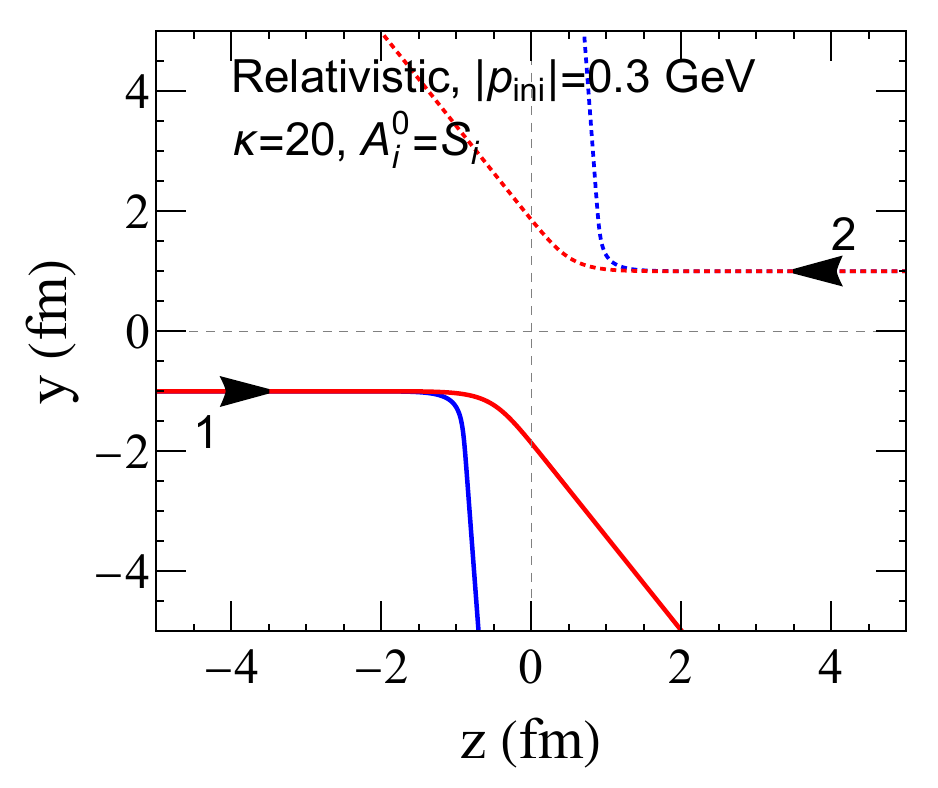}
\includegraphics[width=0.32\textwidth]{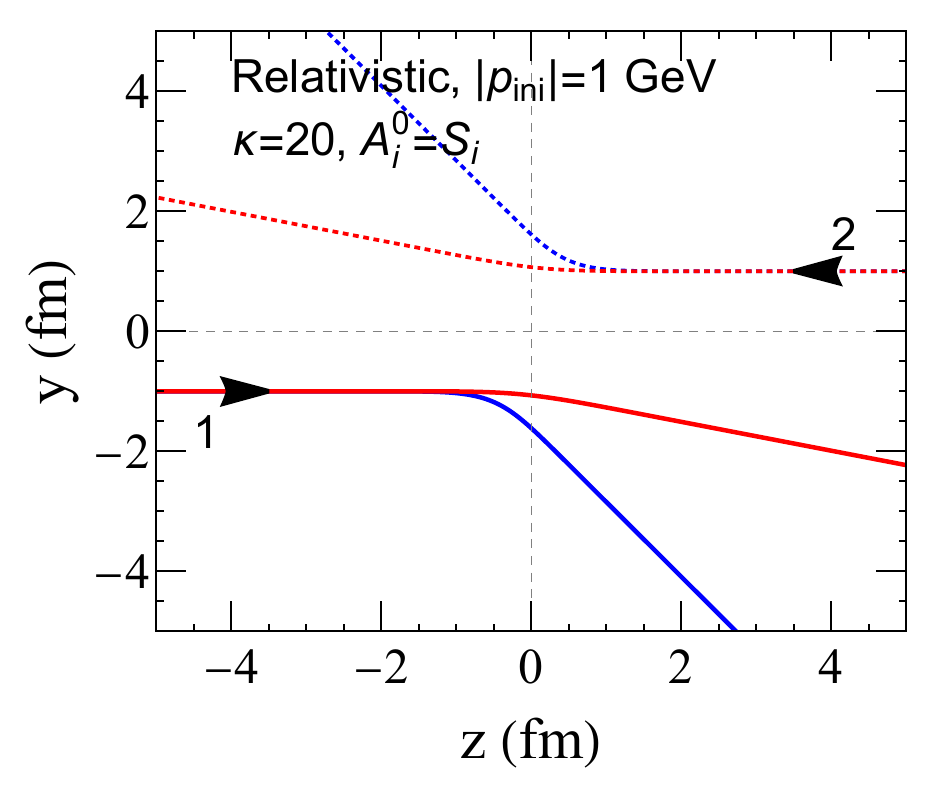}
\includegraphics[width=0.32\textwidth]{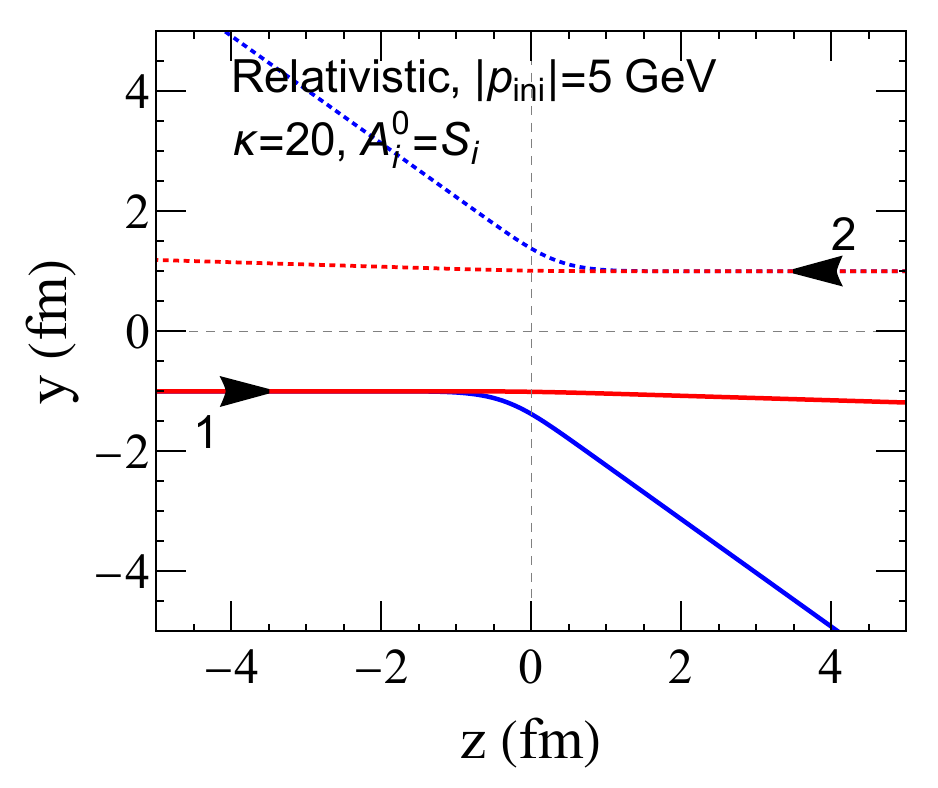}\\
\includegraphics[width=0.32\textwidth]{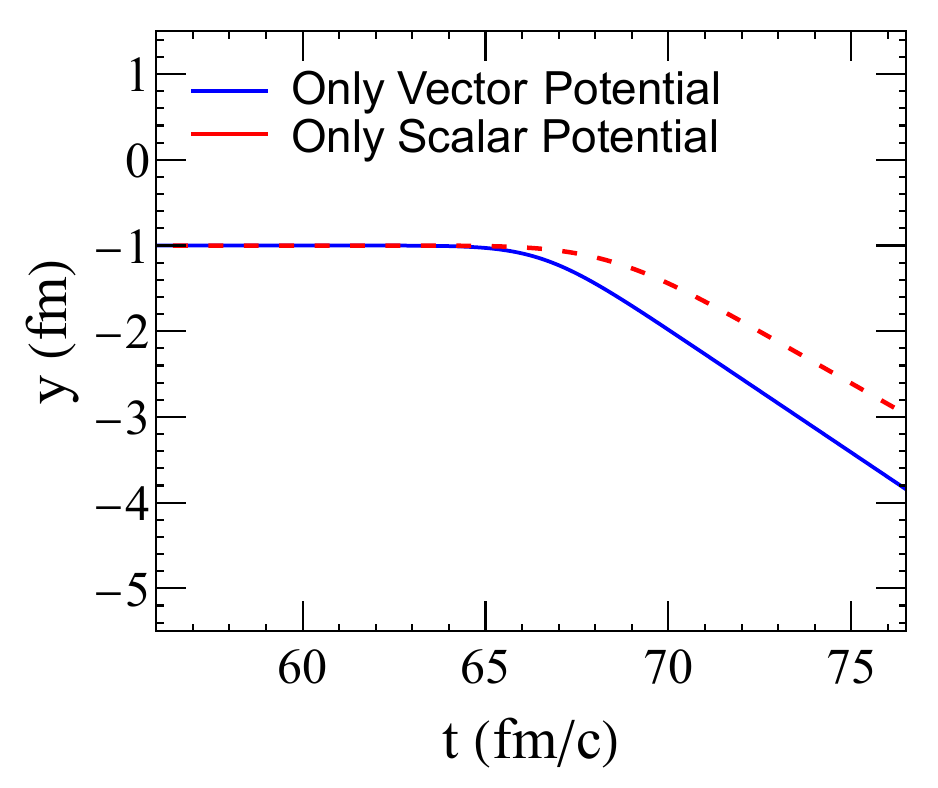}
\includegraphics[width=0.32\textwidth]{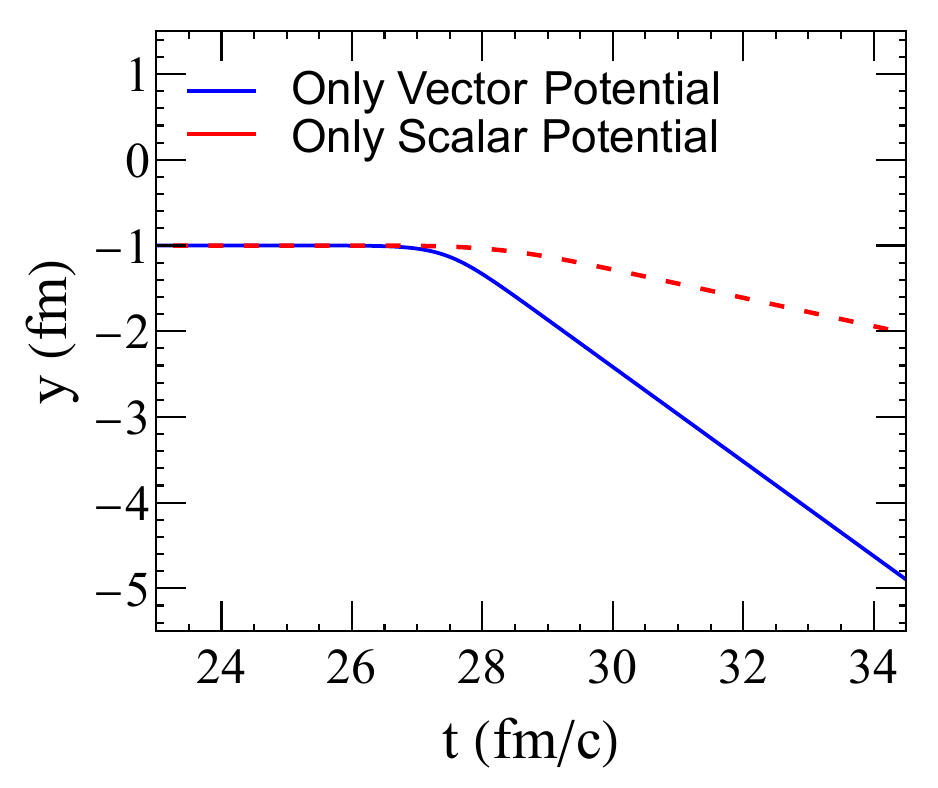}
\includegraphics[width=0.32\textwidth]{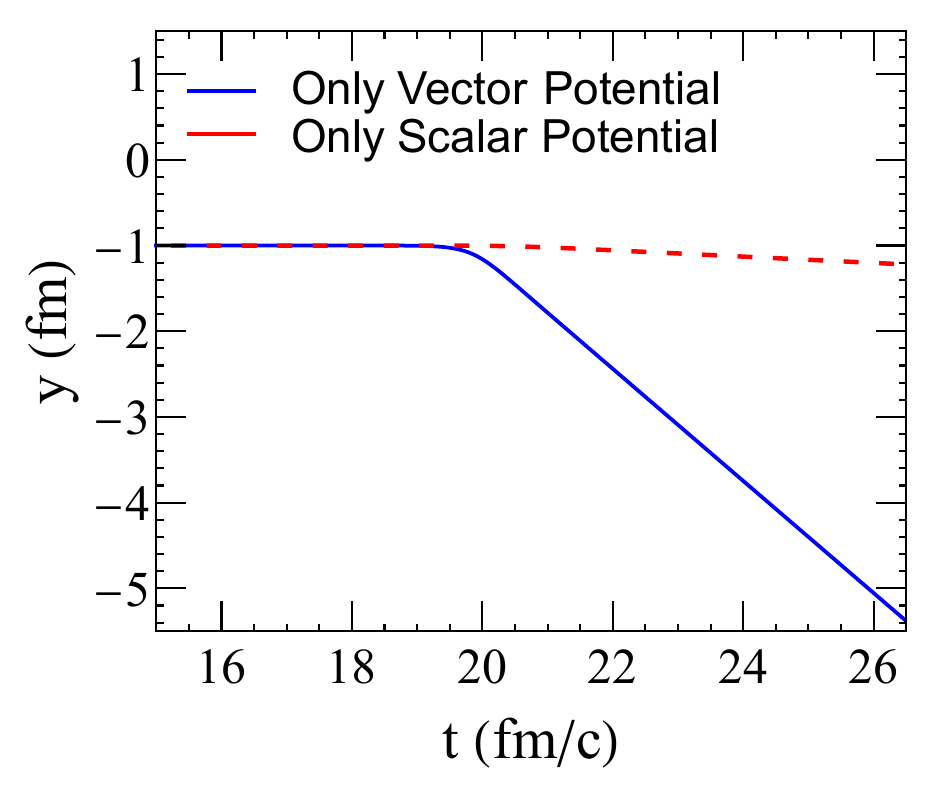}\\
\includegraphics[width=0.32\textwidth]{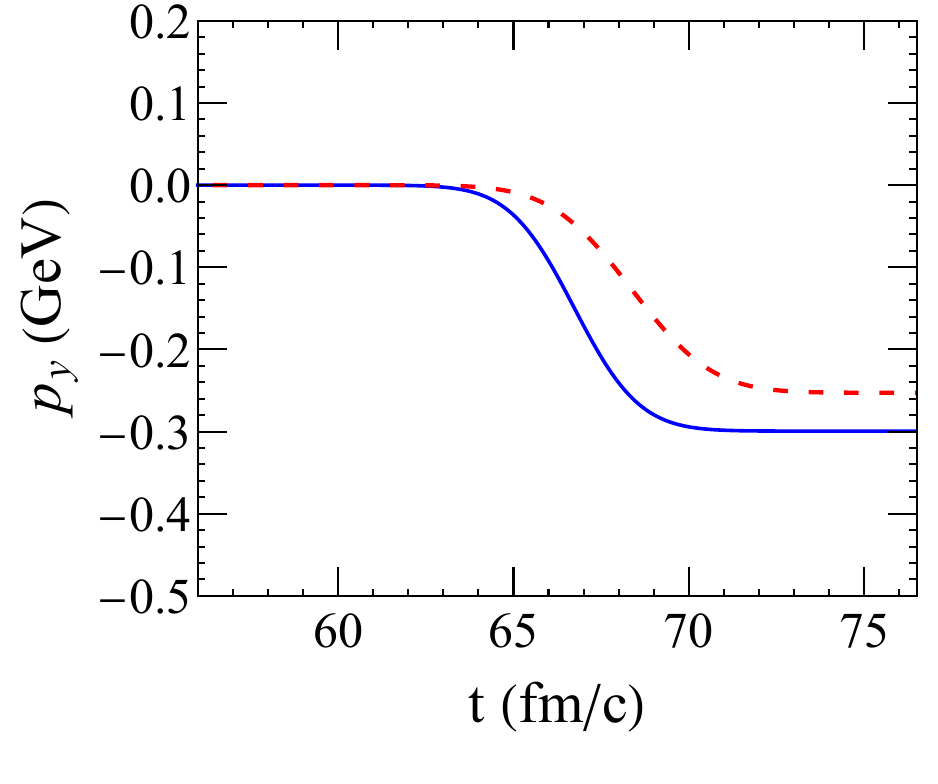}
\includegraphics[width=0.32\textwidth]{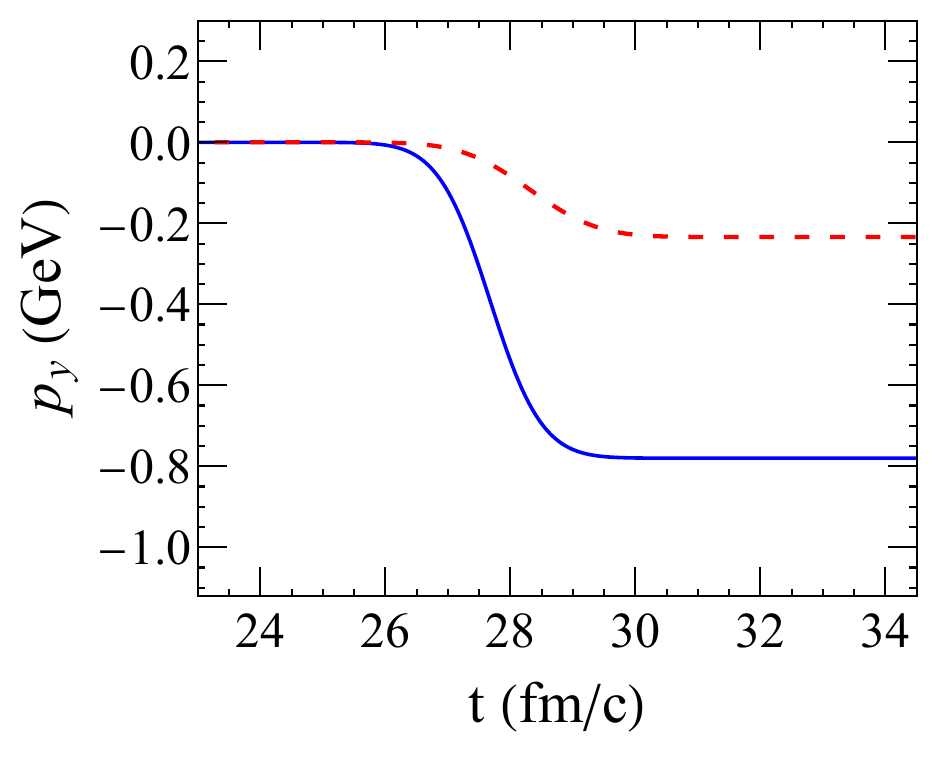}
\includegraphics[width=0.32\textwidth]{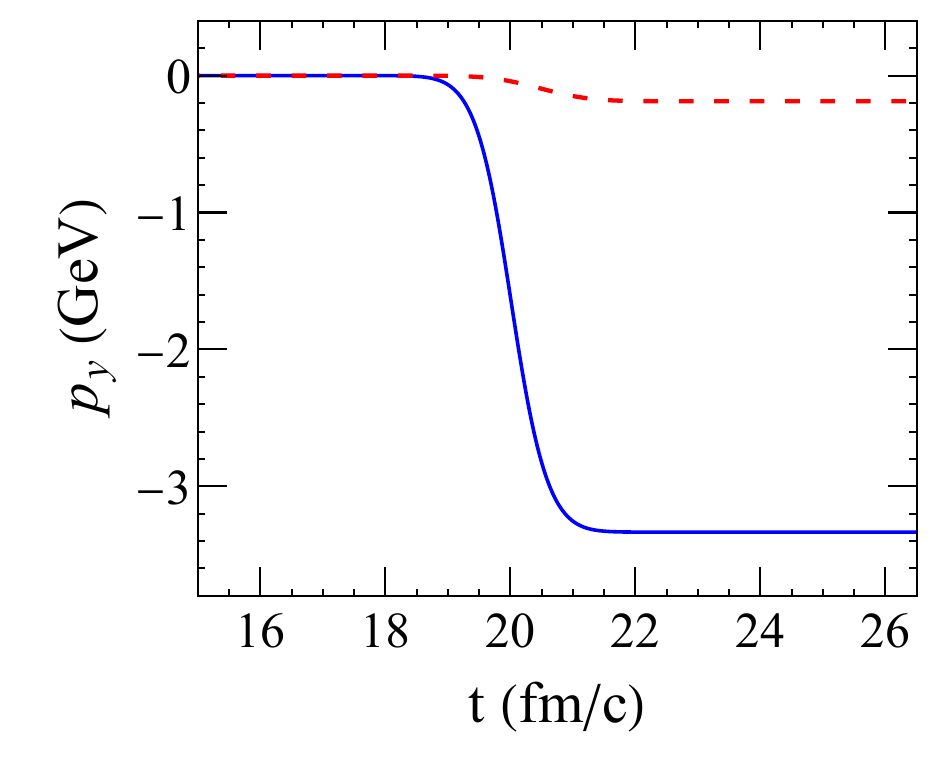}\\
\includegraphics[width=0.32\textwidth]{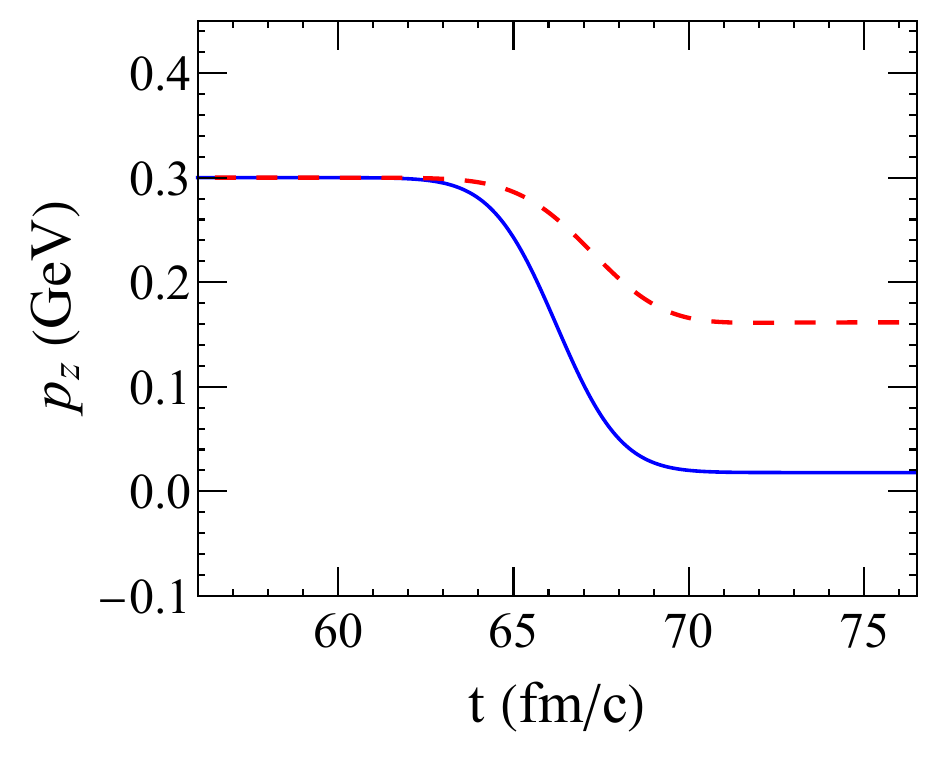}
\includegraphics[width=0.32\textwidth]{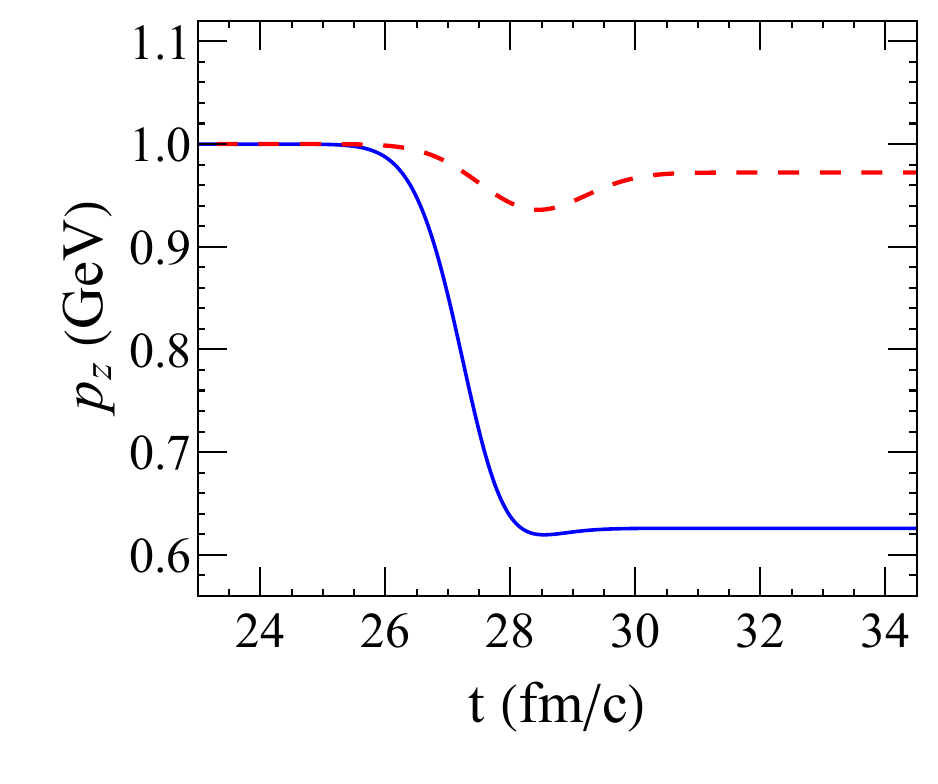}
\includegraphics[width=0.32\textwidth]{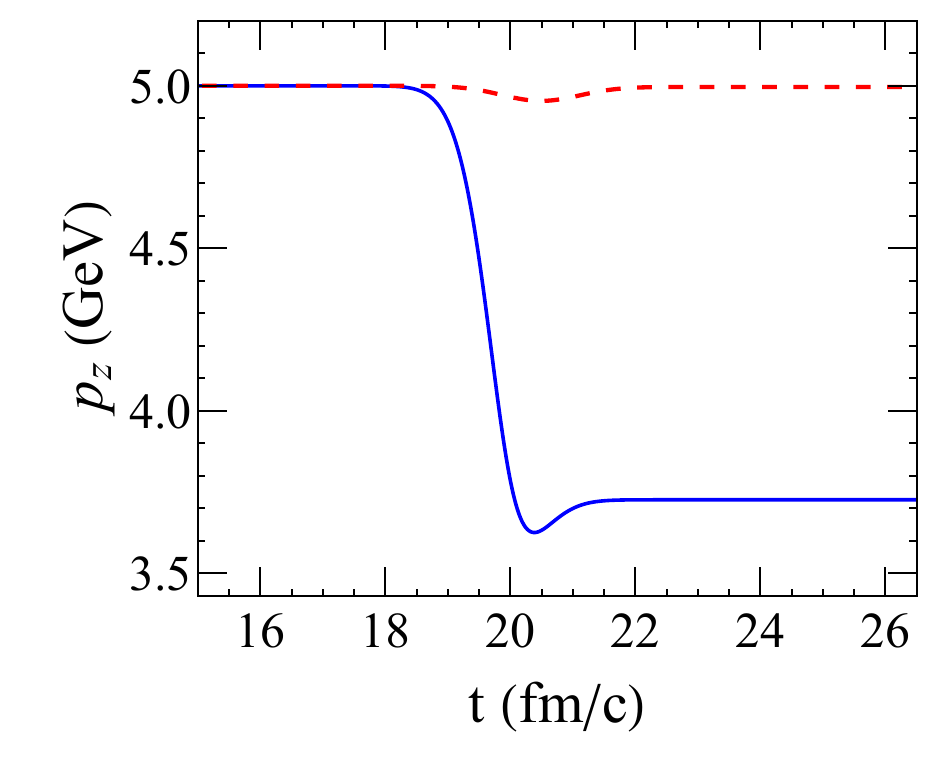}\\
\caption{
Comparison of the relativistic trajectories with only vector (solid color lines) and only scalar potential (dashed color lines) (see Eq.~\eqref{eq.potentialvands}). $A_i^0=S_i$ is imposed. For the initial positions the initial condition IV is chosen and $\kappa=20$, with an initial momentum $|{\bf p}|=0.3$~GeV (left), $|{\bf p}|=1$~GeV (middle), and $|{\bf p}|=5$~GeV (right). } 
\label{fig.vandsit3}
\end{figure*}

\subsubsection{The influence of the vector and scalar potentials}

To see the influence of the vector and scalar potential, we solve numerically the equations of motion for a two-body system with two simple initial conditions:
\begin{enumerate}[label=\Roman*., start=3]
\item ${\bf x}_1=(0,-1,-1)~\rm fm$, 
${\bf x}_2=(0,1,1)~\rm fm$, 
${\bf p}_1=(0,0,0)~\rm GeV$, 
${\bf p}_2=(0,0,0)~\rm GeV$;
\item ${\bf x}_1=(0,-1,-20)~\rm fm$, ${\bf x}_2=(0,1,20)~\rm fm$, ${\bf p}_1=(1,0,0)~\rm GeV$, ${\bf p}_2=(-1,0,0)~\rm GeV$.
\end{enumerate}
The initial condition III is the static case in which two particles are placed at positions 1 and 2, but with zero momentum.
Condition IV is the initial condition for two-particle scattering, where the particles start far away with opposite momentum.

We investigate now how vector and scalar potentials influence the particle trajectories. 
For this purpose we study the trajectories of two particles, separately for only a vector potential or only a scalar potential and impose initially $A_i^0=S_i$, where e.g. $A_1^0=(E_2\rho_{12}-E_1\rho_{12}^2)/(1-\rho_{12}^2)$ is the zero-component of the vector potential $A_1^\mu$ as shown in Eq.~\ref{eq.a1a22body}.

We start out with the initial condition III, where the particles are at rest and initially  $A_i^0\neq 0$, $S_i\neq 0$. The initial condition $A_i^0=S_i$ has the consequence that $H_i^A \neq H_i^S$, where $H_i^A=(p_{i,\mu}-A_{i,\mu})(p_i^\mu-A_i^\mu)-m_i^2$ and $H_i^S=p_{i,\mu}p_i^\mu-m_i^2+\Phi_i$ as shown in Eq.~\eqref{enconst}. To set $A_i^0=S_i$, we multiply $S_i$ by a factor that changes its value to $A_i^0$. An additional -1 is multiplied to make this the scalar potential repulsive. Therefore the asymptotic momentum differs for the two cases. To clearly see the difference, we take a large value of $\kappa$, $\kappa = 20$. This gives strong vector and scalar potentials but $\rho_{12}$ still smaller than 1. By multiplying by a constant factor to scalar potential, we obtain $A_i^0 = S_i = 45.83$~MeV.
The different
trajectories are shown in Fig.~\ref{fig.vands}, top row. In the lower panel we display for the two potentials the time evolution of the position ($z$ which is due to the initial condition = $y$)  and the momentum $p_z$ (= $p_y$) of both particles. 
Initially $p_z$ is zero. At the beginning, the trajectories for the scalar and vector interaction are very similar, a difference emerge only when the system evolves. Particles, accelerated by the repulsive force, acquire velocity, which in turn generates finite vector components of $A^\mu$. The final momentum is determined by energy conservation $H(t\to \infty)=H(t=0)$ and is larger for a vector potential than for a scalar potential. This means that for this initial condition  and $A_i^0(t=0)=S_i(t=0)$ initially more potential energy is stored for a vector potential than for a scalar potential, as can be also seen from Eq.~\eqref{enconst}.

\begin{figure*}[!htb]
\centering
\includegraphics[width=0.32\textwidth]{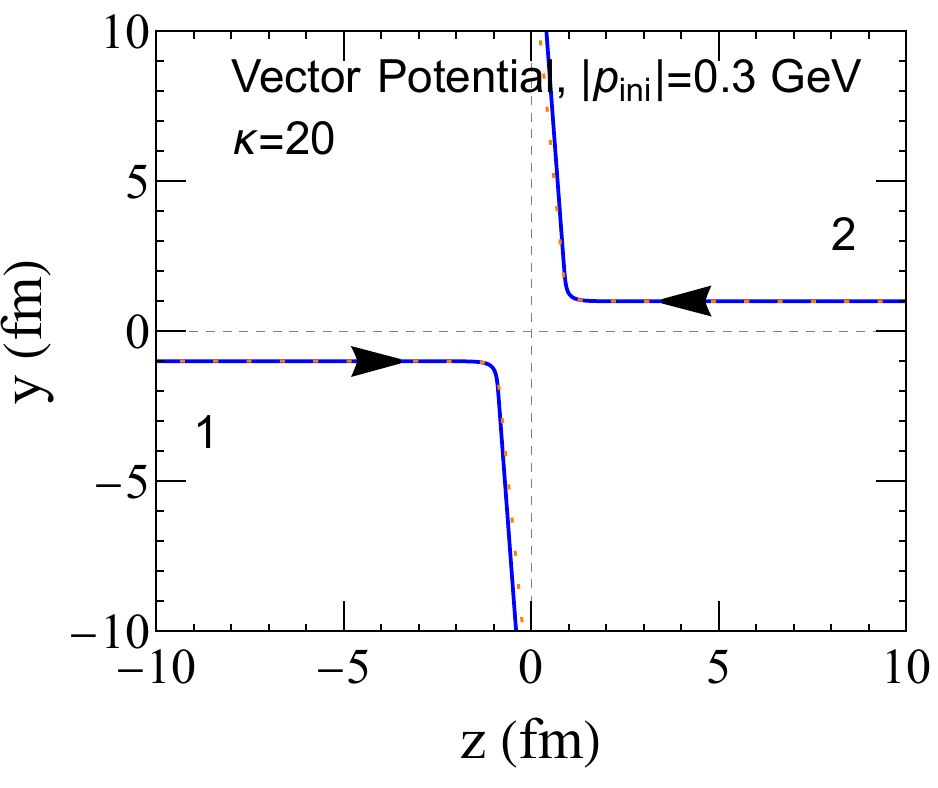}
\includegraphics[width=0.32\textwidth]{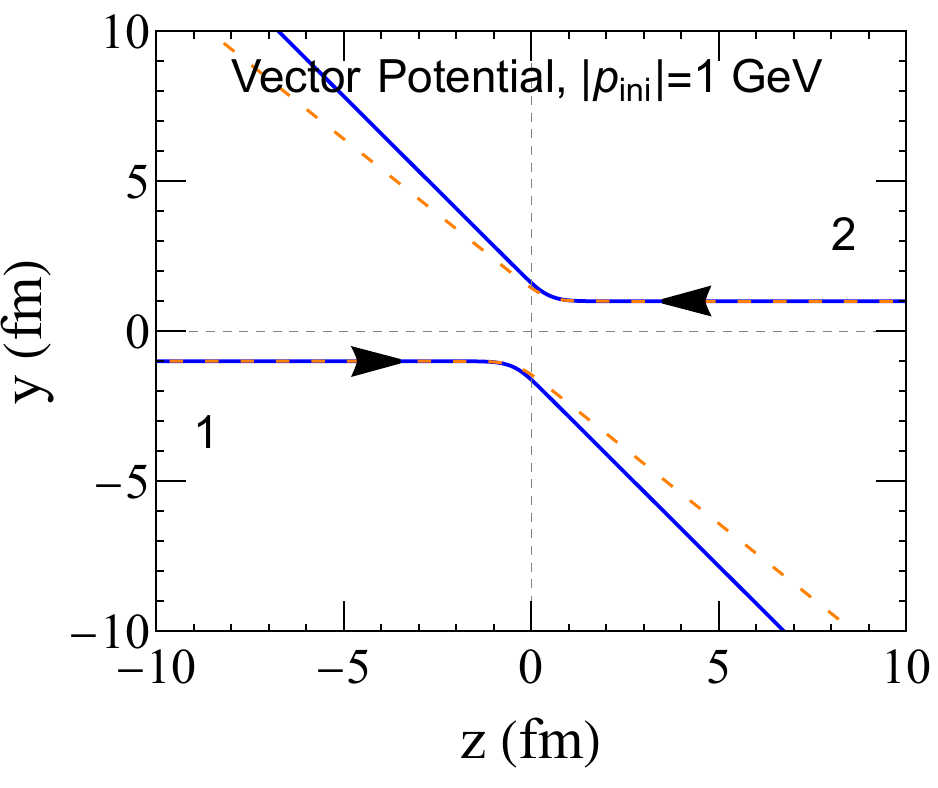}
\includegraphics[width=0.32\textwidth]{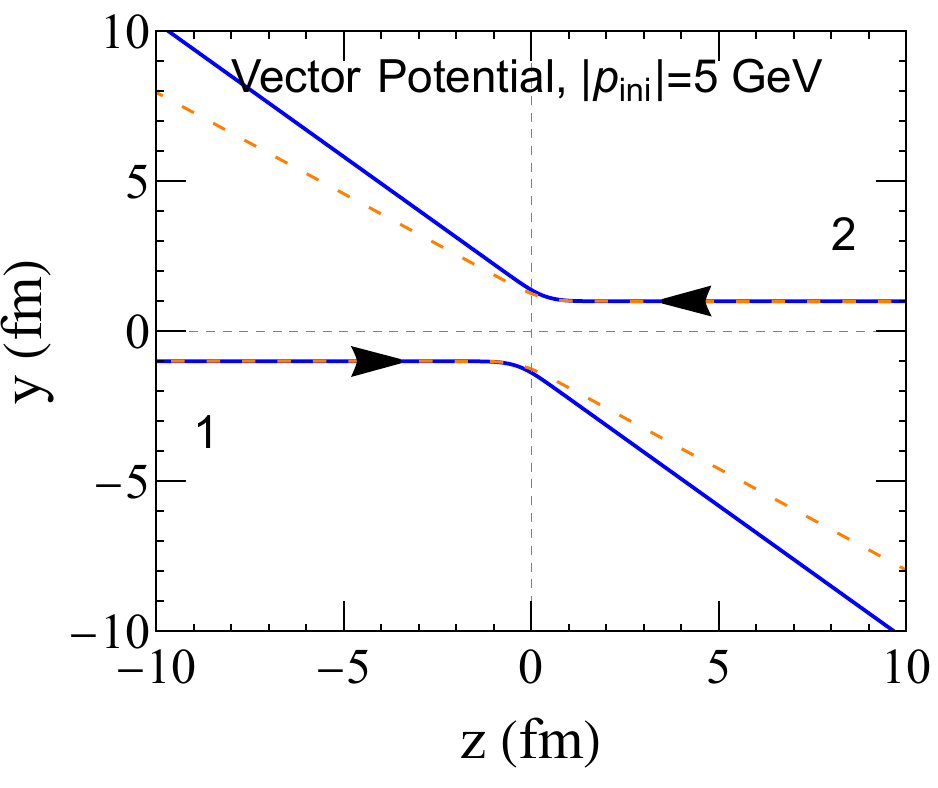}
\\
\includegraphics[width=0.32\textwidth]{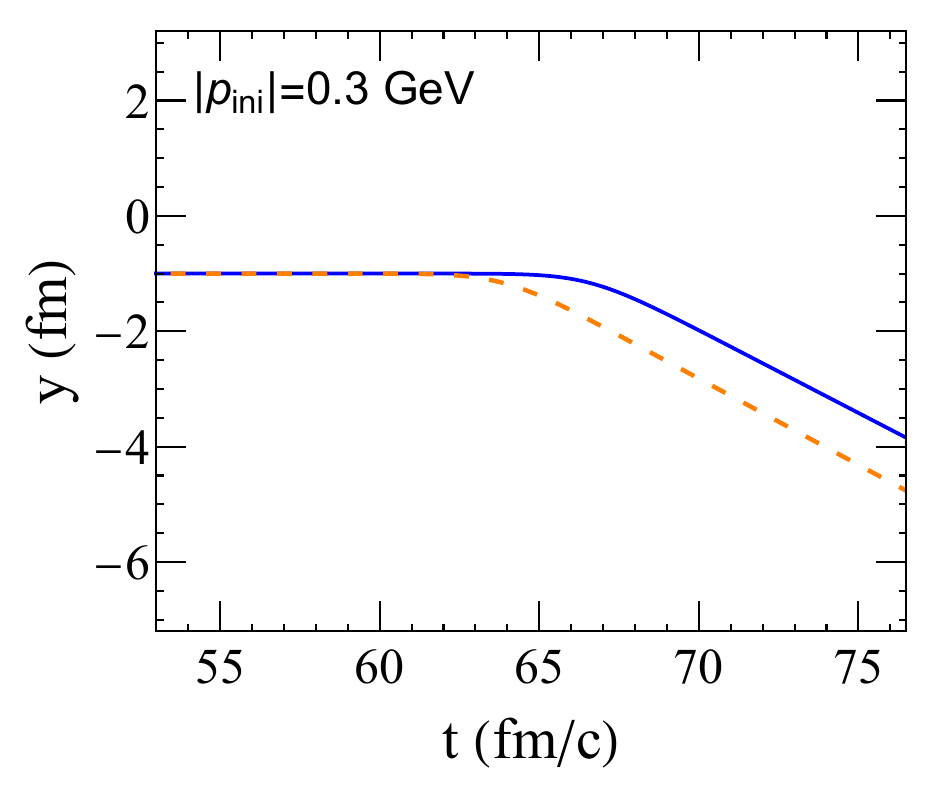}
\includegraphics[width=0.32\textwidth]{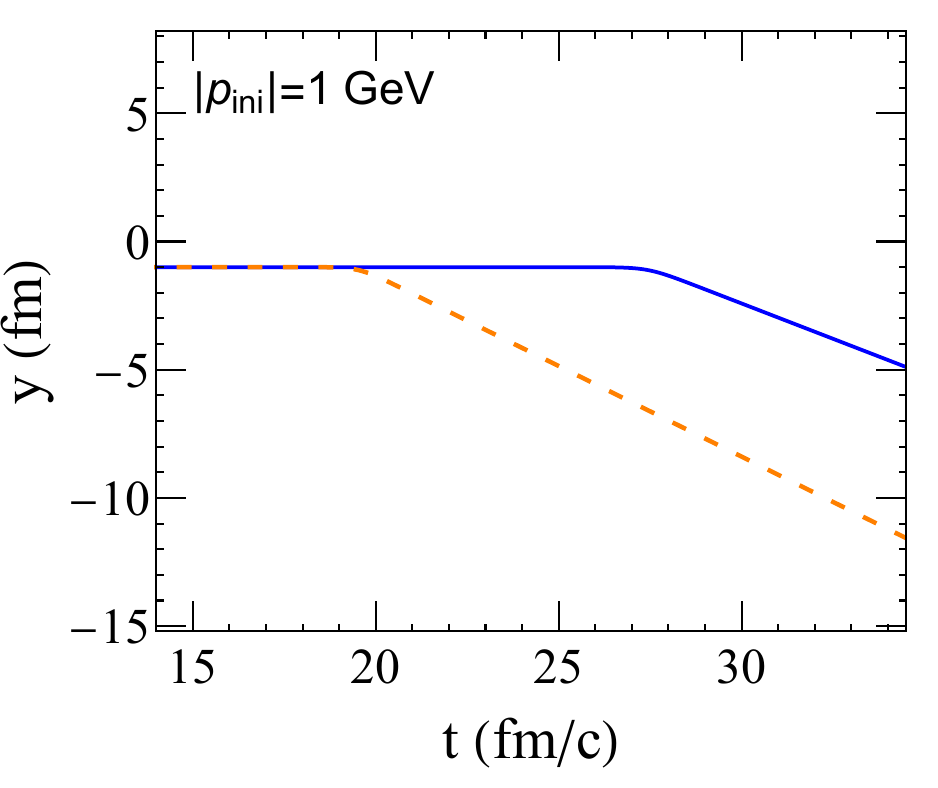}
\includegraphics[width=0.32\textwidth]{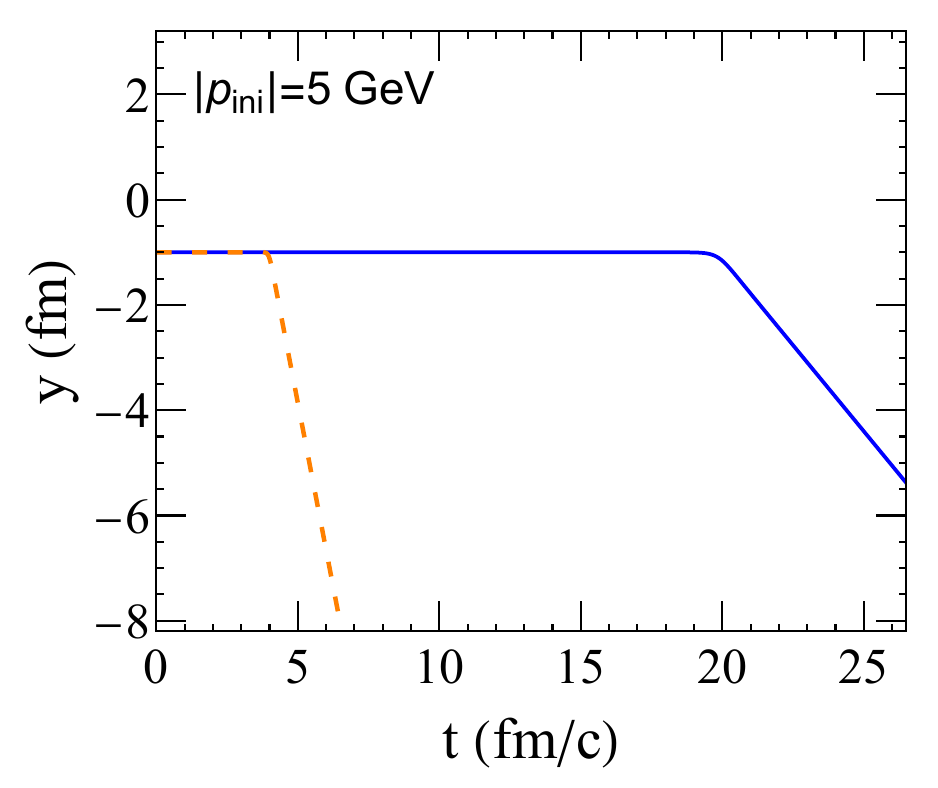}
\\
\includegraphics[width=0.32\textwidth]{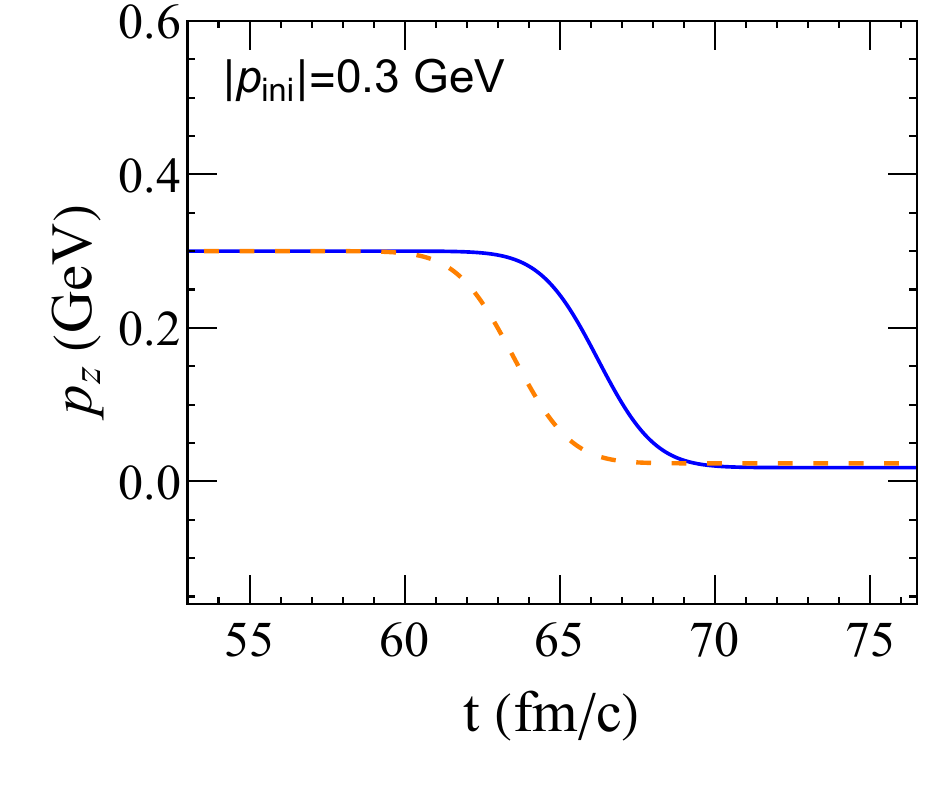}
\includegraphics[width=0.32\textwidth]{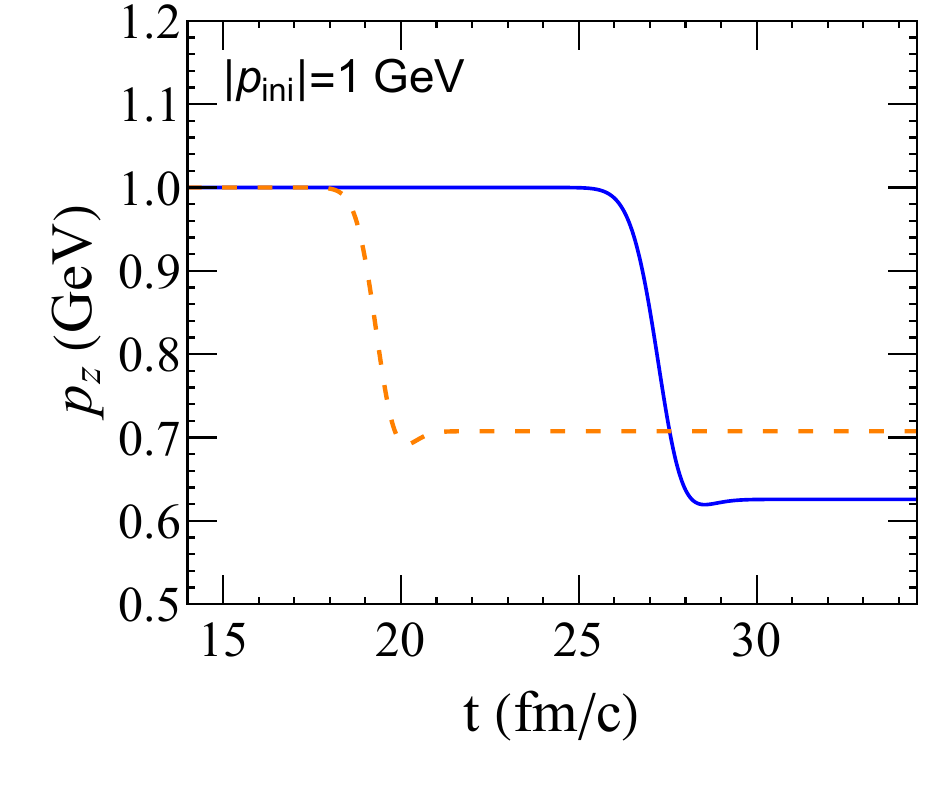}
\includegraphics[width=0.32\textwidth]{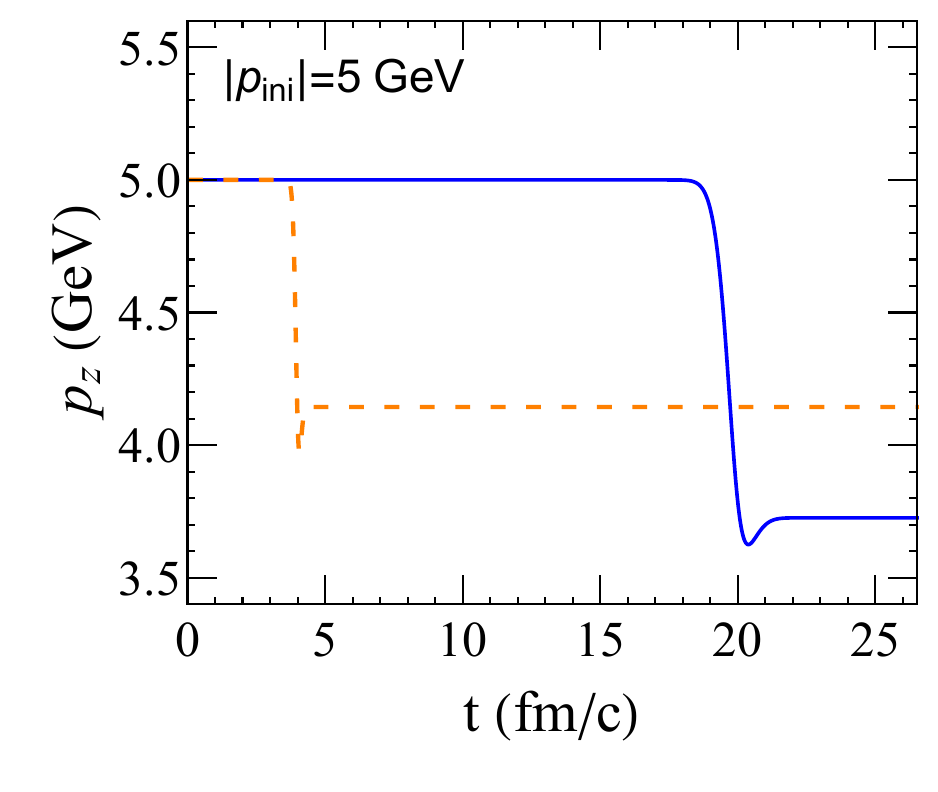}
\\
\includegraphics[width=0.32\textwidth]{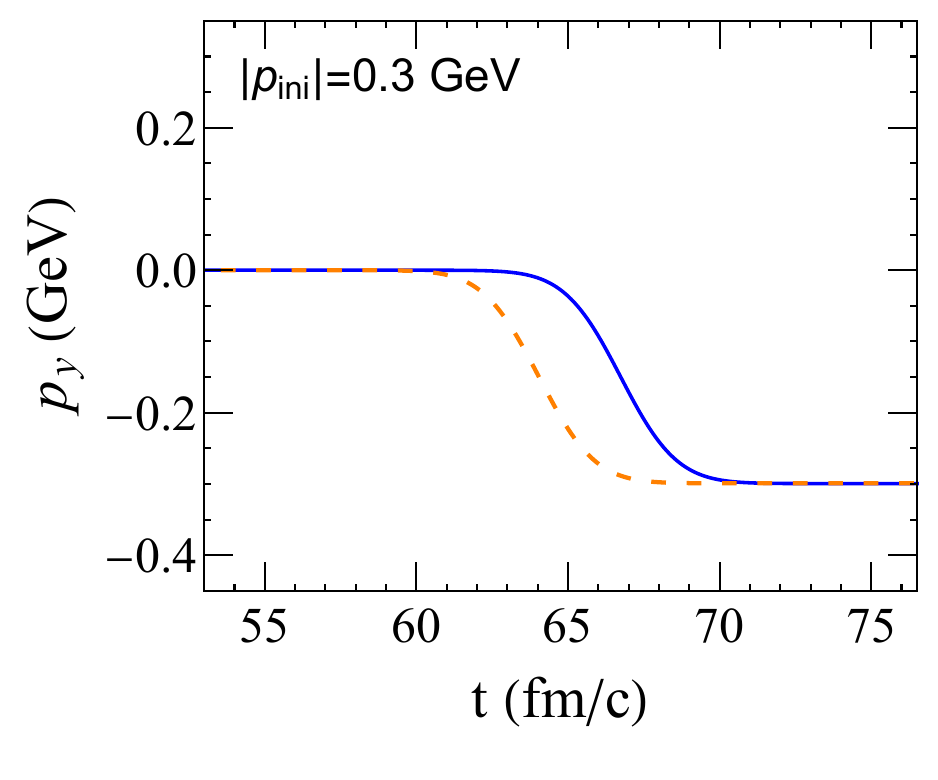}
\includegraphics[width=0.32\textwidth]{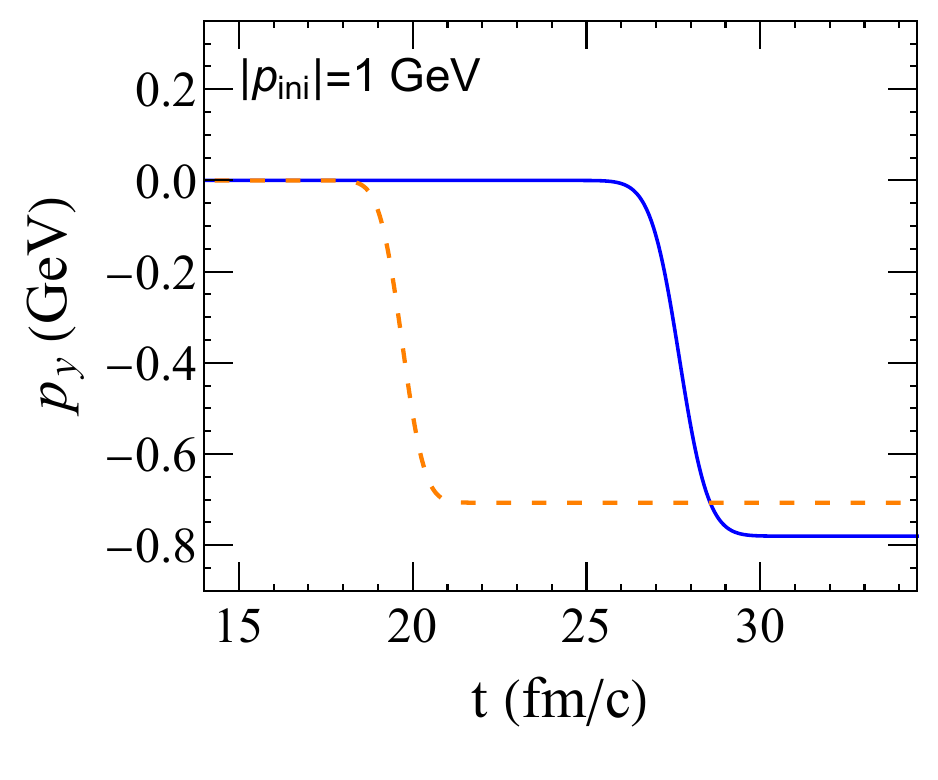}
\includegraphics[width=0.32\textwidth]{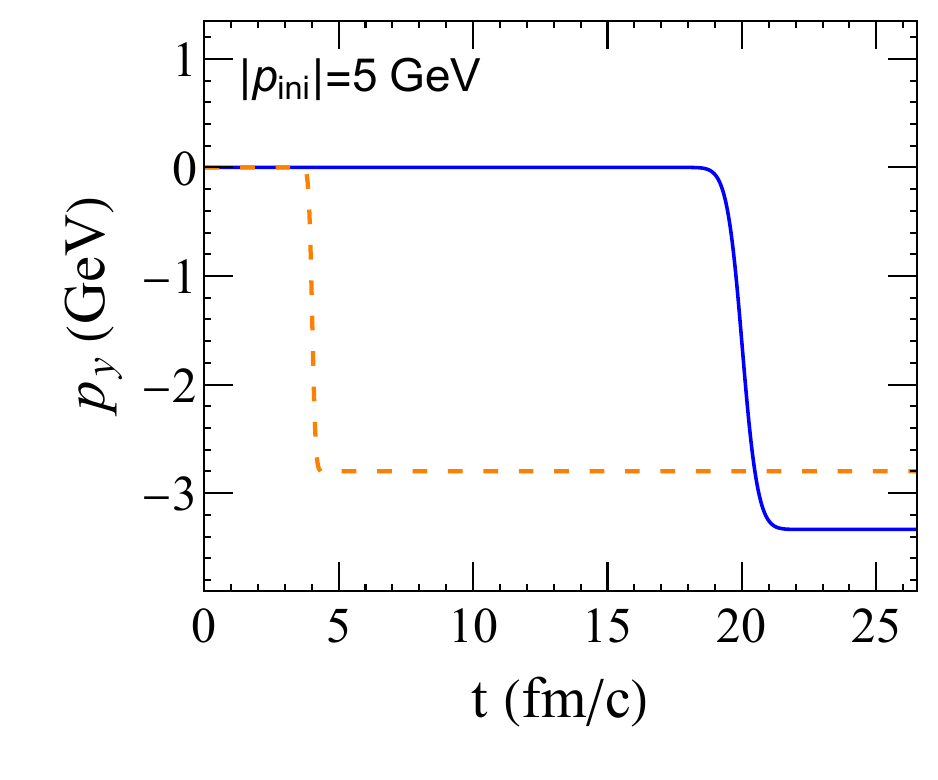}
\caption{Comparison of the relativistic trajectories (solid color lines) to the non-relativistic trajectories (dashed color lines) with only the vector potential of Eq.~\eqref{eq.potentialvands}. The initial condition IV is applied and $\kappa=20$.  The initial momentum is $|{\bf p}|=0.3$~GeV (left), $|{\bf p}|=1$~GeV (middle), and $|{\bf p}|=5$~GeV (right). The bottom three panels show $y$, $p_z$, and $p_y$ of particle 1 as a function of time applying relativistic and non-relativistic evolution equations.
}
\label{fig.vandsit3comp}
\end{figure*}

\begin{figure*}[!htb]
\centering
\includegraphics[width=0.32\textwidth]{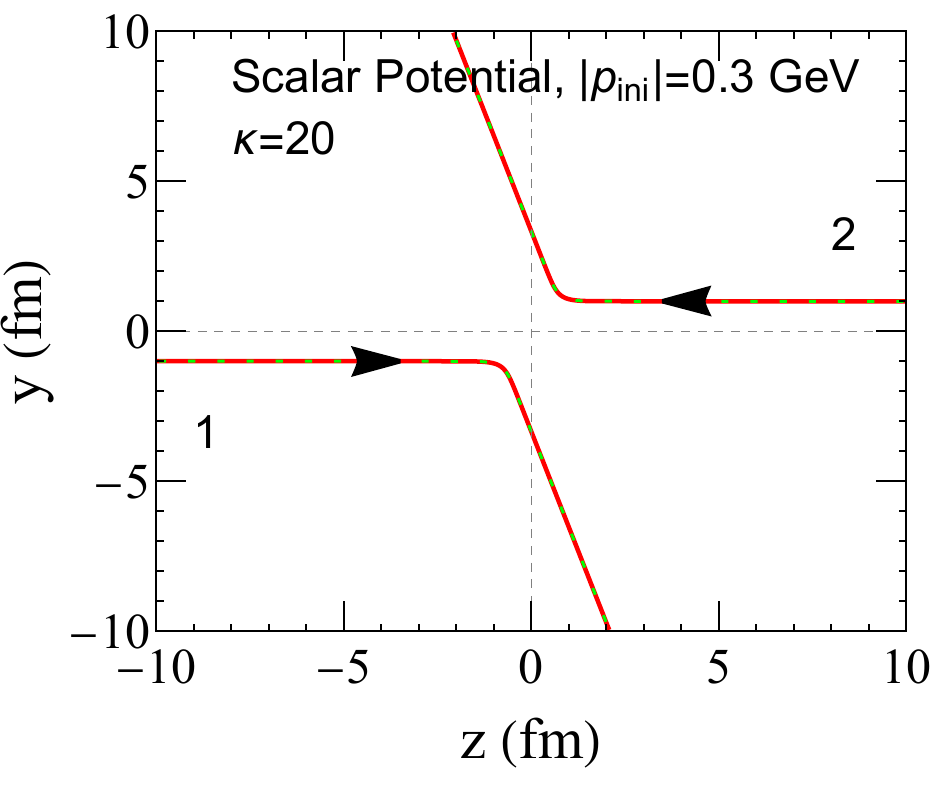}
\includegraphics[width=0.32\textwidth]{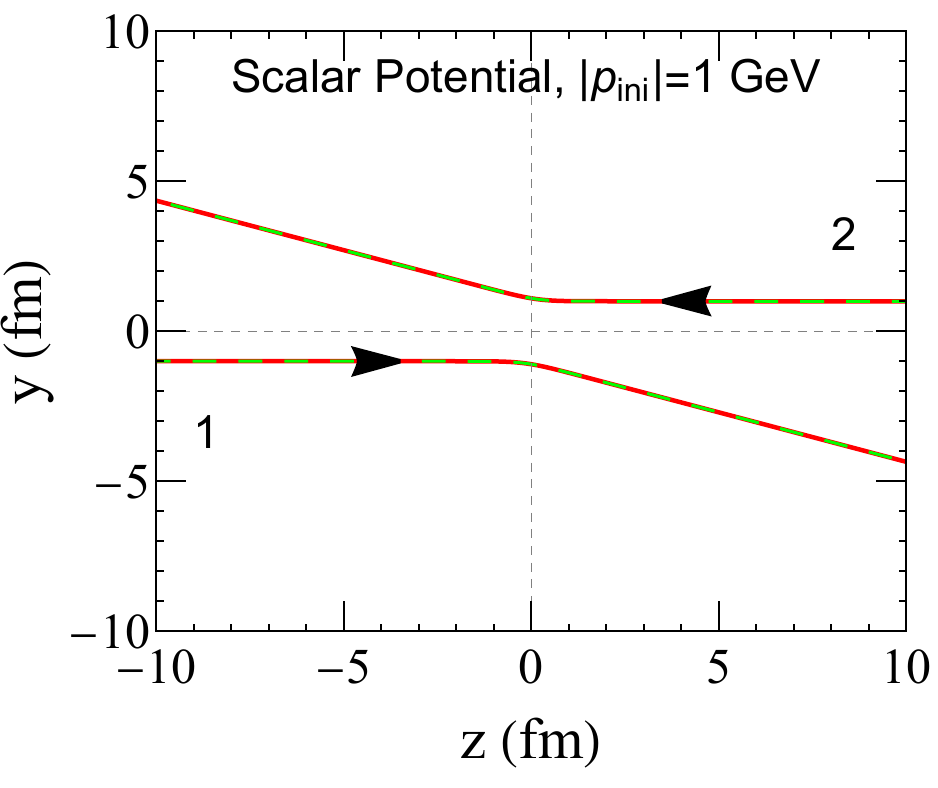}
\includegraphics[width=0.32\textwidth]{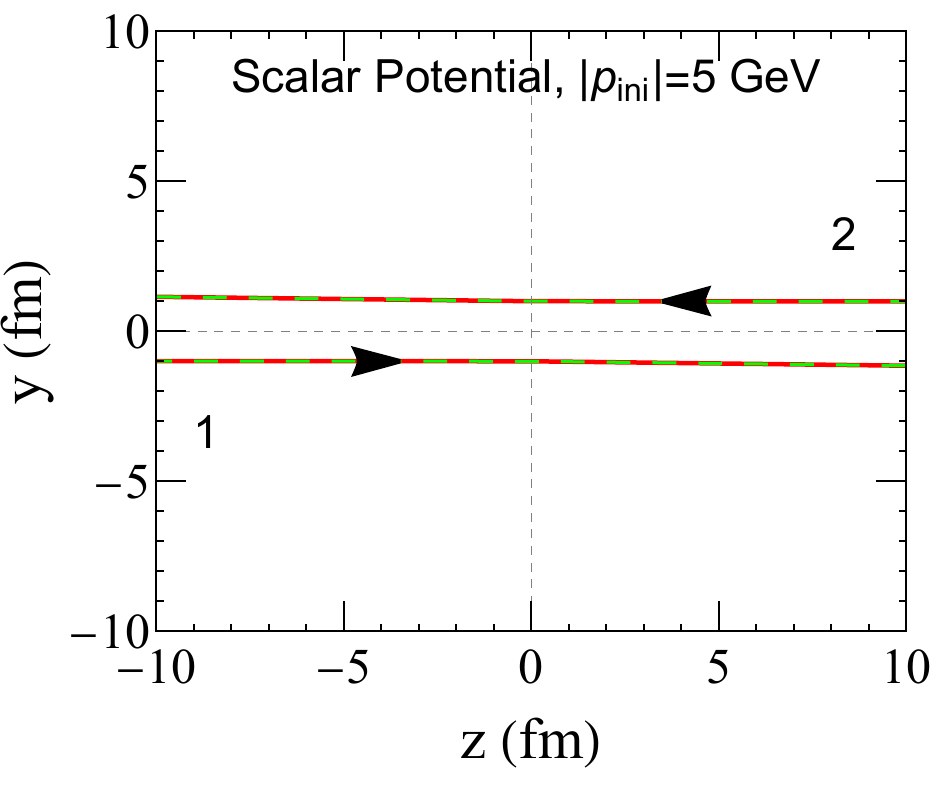}
\\
\includegraphics[width=0.32\textwidth]{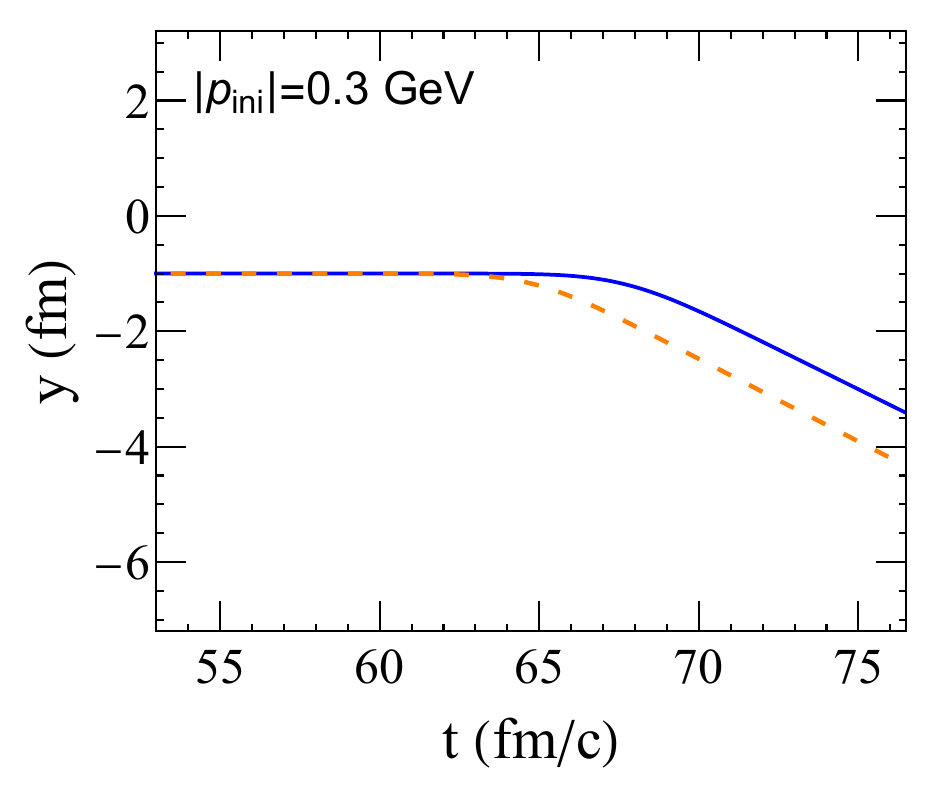}
\includegraphics[width=0.32\textwidth]{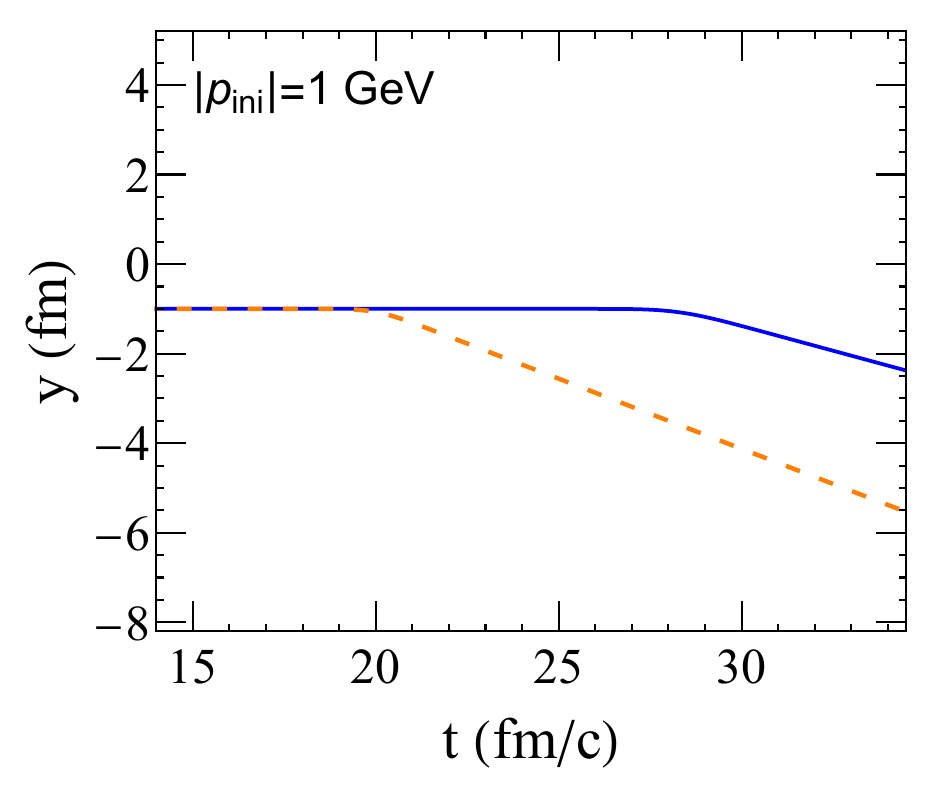}
\includegraphics[width=0.32\textwidth]{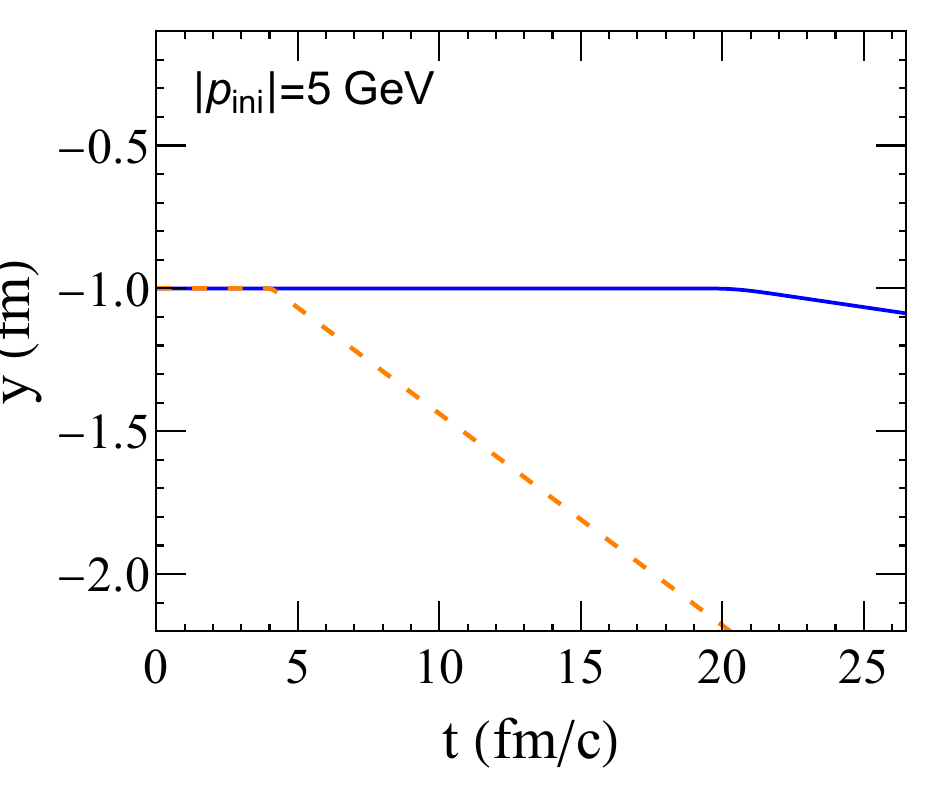}
\\
\includegraphics[width=0.32\textwidth]{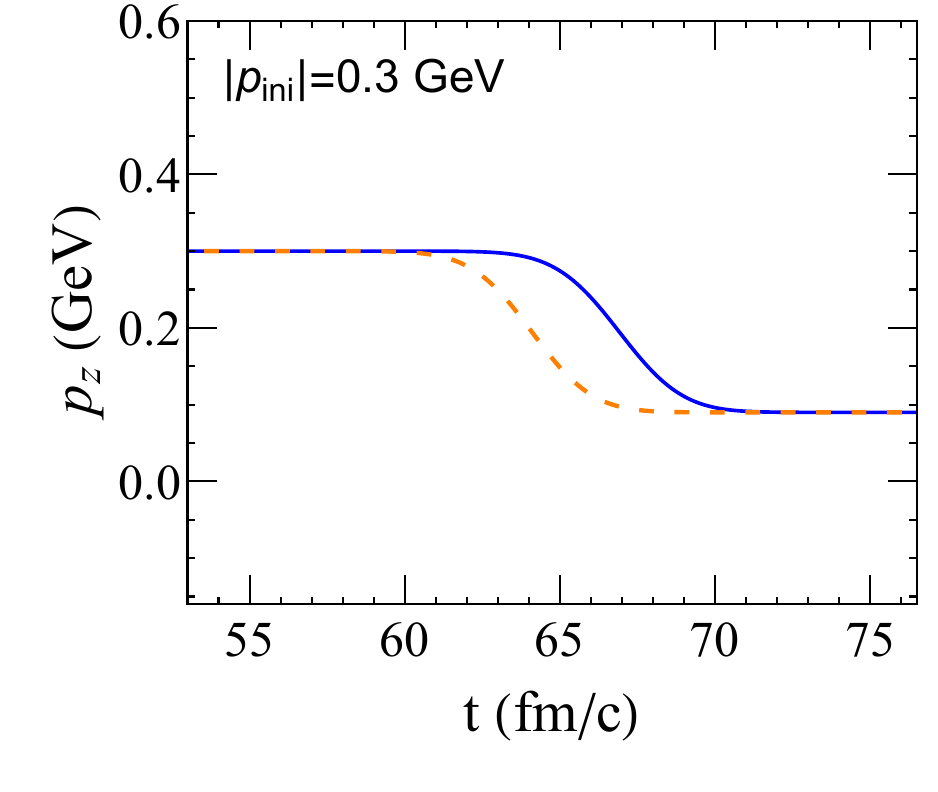}
\includegraphics[width=0.32\textwidth]{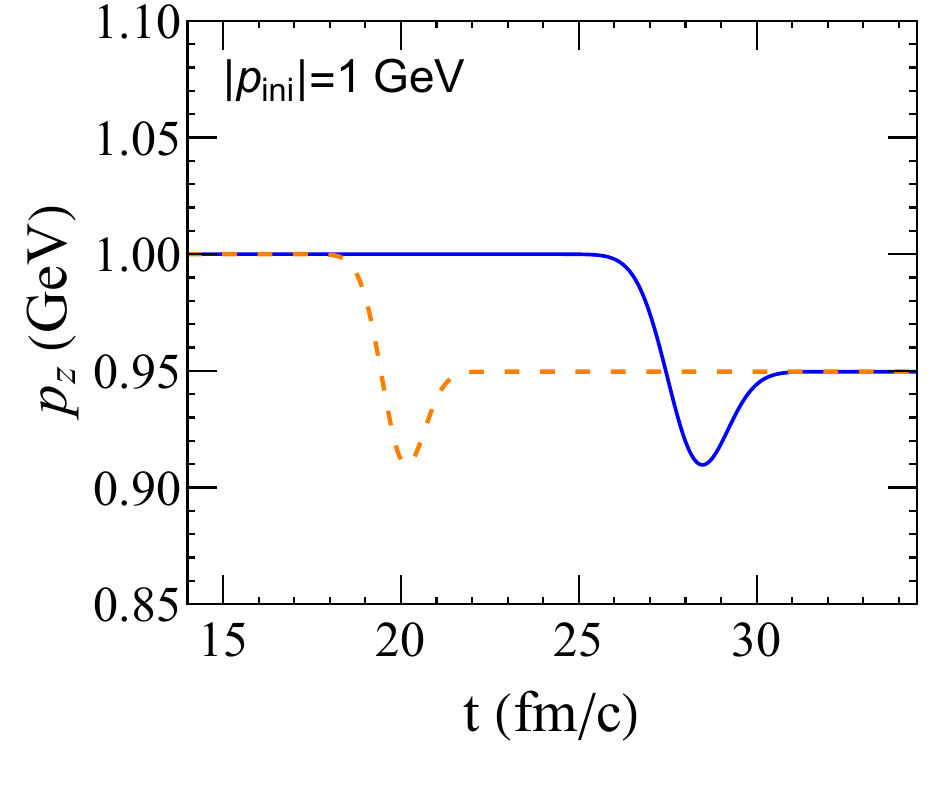}
\includegraphics[width=0.32\textwidth]{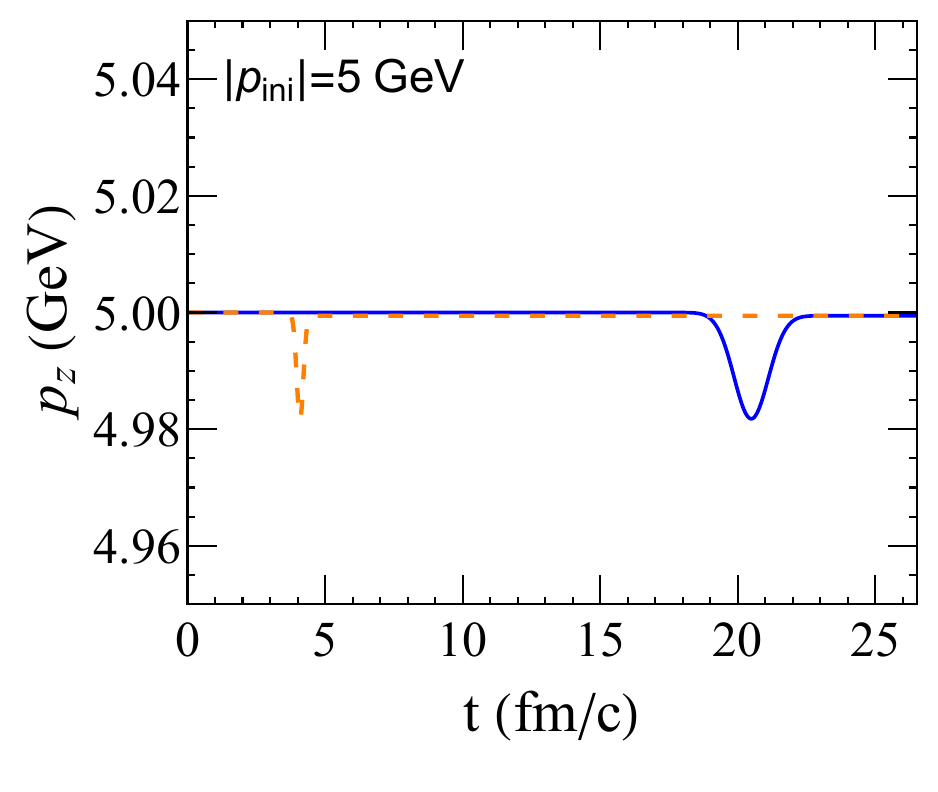}
\\
\includegraphics[width=0.32\textwidth]{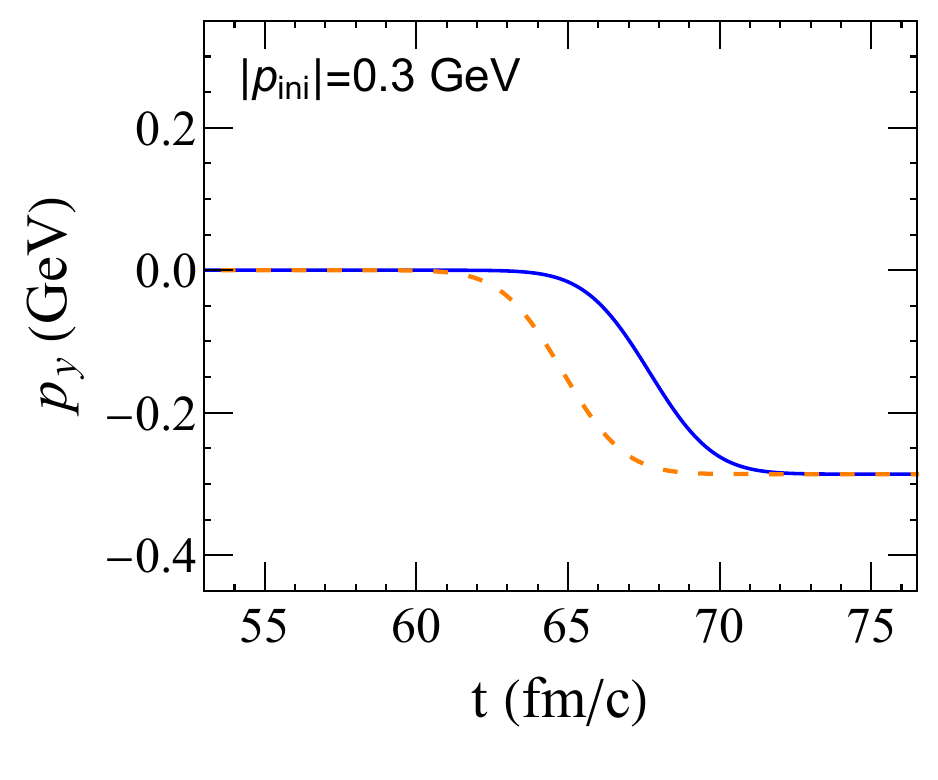}
\includegraphics[width=0.32\textwidth]{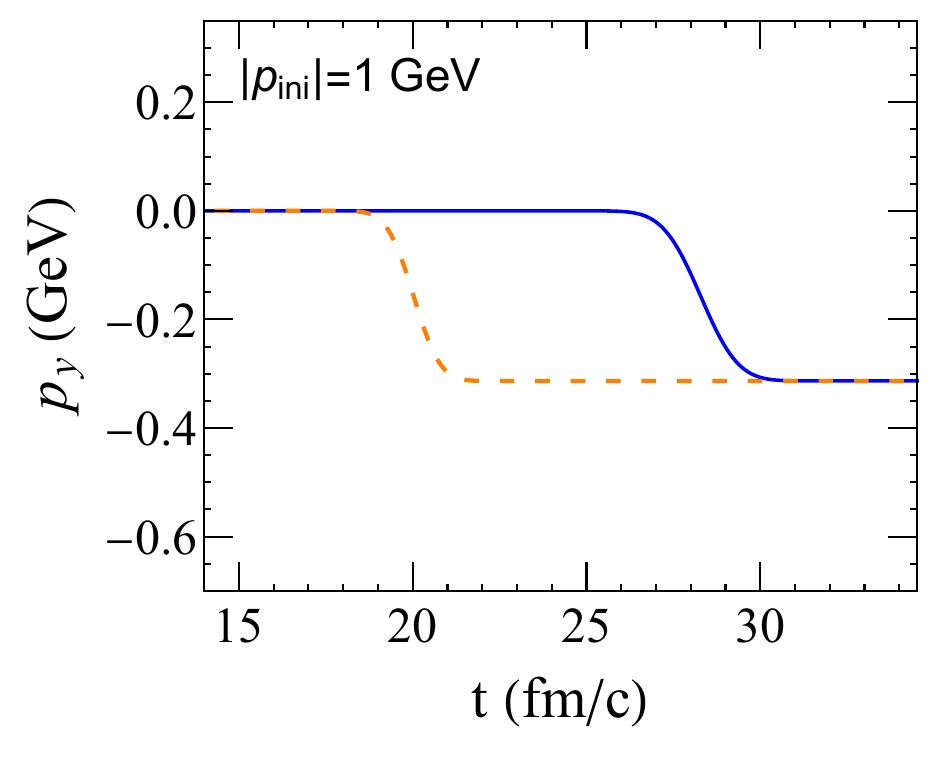}
\includegraphics[width=0.32\textwidth]{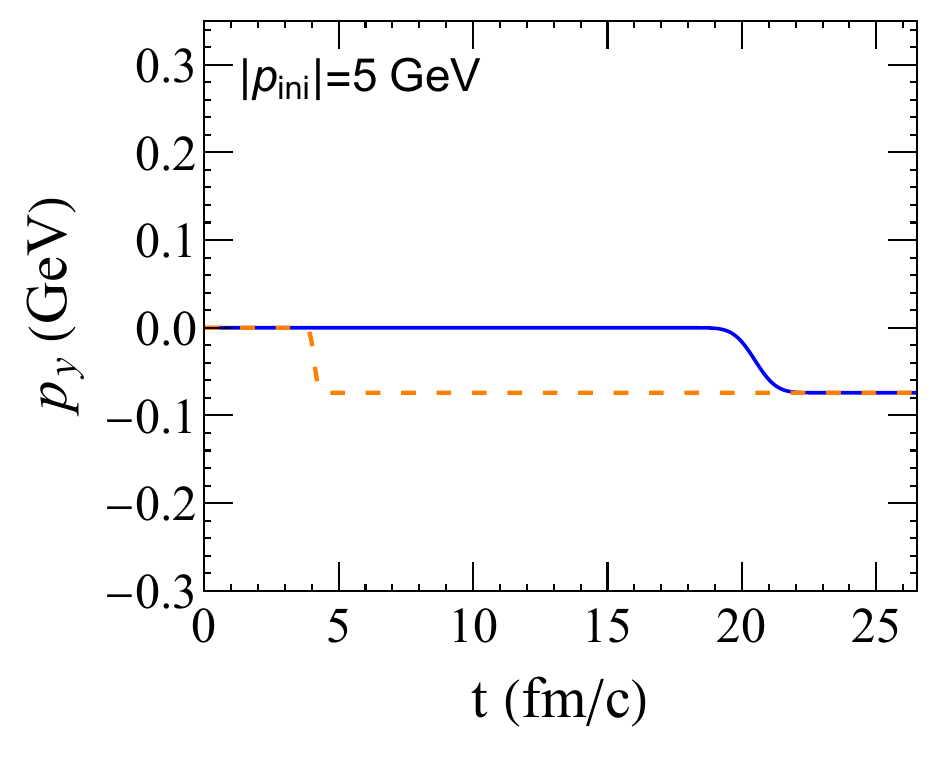}
\caption{Same as Fig.~\ref{fig.vandsit3comp} but with only the scalar potential $S_i$.
}
\label{fig.vandsit4comp}
\end{figure*}

We study now the trajectories if the particles have initially a finite momentum. We apply the initial condition IV, which implies that the particles are initially not interacting and the trajectories obey $H_i^A=H_i^S=0$. 
In this case, we assume that $A_i^0=E\rho_{12}/(1+\rho_{12})$ and $S_i=m\rho_{12}$ are identical in a medium of density $\rho_{12}=1$, where $E=E_1=E_2$ denotes the initial particle energy. This requires multiplying a constant factor of $E/2$ by the scalar potential to ensure that $A_i^0=S_i$. An additional -1 is multiplied to make this the scalar potential repulsive. 
This choice is made in view of the possible application of our approach to proton - heavy-ion collisions, where the analysis of the experimental data yields $A_i^0 \approx S_i$~\cite{Cooper:1993nx}.
Here we choose $\kappa = 20$ in the interaction density, Eq.~\eqref{eq:dens}.

In Fig.~\ref{fig.vandsit3}  we display the results for an initial momentum $p=0.3$~GeV (left), $p=1$~GeV (middle) and $p=5$~GeV (right). The first row shows the trajectories in the $z-y$ plane, the next three rows  display $y(t)$, $p_y(t)$, and $p_z(t)$.  For all initial momenta the vector potential gives a larger final $p_y$ than the  scalar potential and consequently, due to energy conservation, a smaller $p_z$.  Already for the non-relativistic energy ($p_z = 0.3$~GeV, $m=1$~GeV) the trajectories differ substantially because a vector potential creates a Lorentz force whereas for a scalar potential the force is given by the density gradient.
The vector potential deflects the particle to almost 90° with a very small $p_z$ left,
at $p=1(5)$~GeV the scattering angle is 51(41°), thus it decreases with increasing energy.

For the scalar potential the scattering angle decreases from
73° over 18° to 0.9° if the momentum increases from 0.3 to 5 GeV  due to the increasingly shorter interaction time. Hence the difference of the scattering angles and the trajectories between a scalar and a vector interaction increases with increasing initial momentum.  This is due to the different mechanism which create a transverse momentum. For a scalar potential it is the gradient of the potential whereas the vector potential acts in a similar way as a Lorentz force, which is proportional to the particle velocity. By comparing the scattering angle, it decreases much faster for a scalar potential than for a vector potential.

\subsubsection{Relativistic vs. non-relativistic trajectories}
In this section, we compare the trajectories from the relativistic vs. non-relativistic EoMs, as shown in Eq.~\eqref{eq.eomrela} and Eq.~\eqref{eq.eomnonrela}. The results are shown in Fig.~\ref{fig.vandsit3comp} for the vector potential and in Fig.~\ref{fig.vandsit4comp} for the scalar potential. Again we employ the initial condition IV with $\kappa = 20$ but don't require $A_i^0=S_i$. Same as the previous figure we perform these calculations for the initial momenta $p=0.3$, 1 and 5 GeV. The non-relativistic quantities are given as dashed lines whereas the relativistic ones as solid lines. Since we describe an elastic scattering process the initial and final momentum $\sqrt{p_z^2+p_y^2}$ is identical, for the relativistic as well as for the non-relativistic case.
 
At $p=0.3$~GeV, we see a very small difference between the relativistic and non-relativistic trajectories, as expected because $p/m=0.3$. The main difference in the time evolution comes from the different velocities, $p/m$ and $p/E$, which makes the non-relativistic particles faster. In the case of a scalar potential, the asymptotic values of the momentum for a relativistic and non-relativistic calculation are also identical at final time because the only difference in the EoMs is the velocity (or the pre-factor $\lambda$)(see Eq.~\eqref{eq.eomrela} and Eq.~\eqref{eq.eomnonrela}). For the vector potential, on the contrary, already for beam momenta of 1 GeV the momentum transfer in non-relativistic calculation is considerably smaller than in the relativistic counterpart, leading consequently to very different scattering angles.

We can conclude at already at $p=1$~GeV for a vector potential the non-relativistic trajectories deviate substantially from their relativistic counterparts, whereas for a scalar potential there is very little difference in the scattering angle. If we exchange the vector potential by a scalar potential the trajectories and the transverse momentum transfer changes considerably and shows  a completely different systematics as a function of the beam momentum. This is of importance for the non-relativistic reduction of the relativistic potentials, the so called Schr\"odinger equivalent potential, in which the scalar density (on which depends $S$) and the vector density (on which depends $A^0$) are considered as identical.

\subsubsection{The frame-independence of the trajectories}

\begin{figure}[!htb]
\centering
\includegraphics[width=0.23\textwidth]{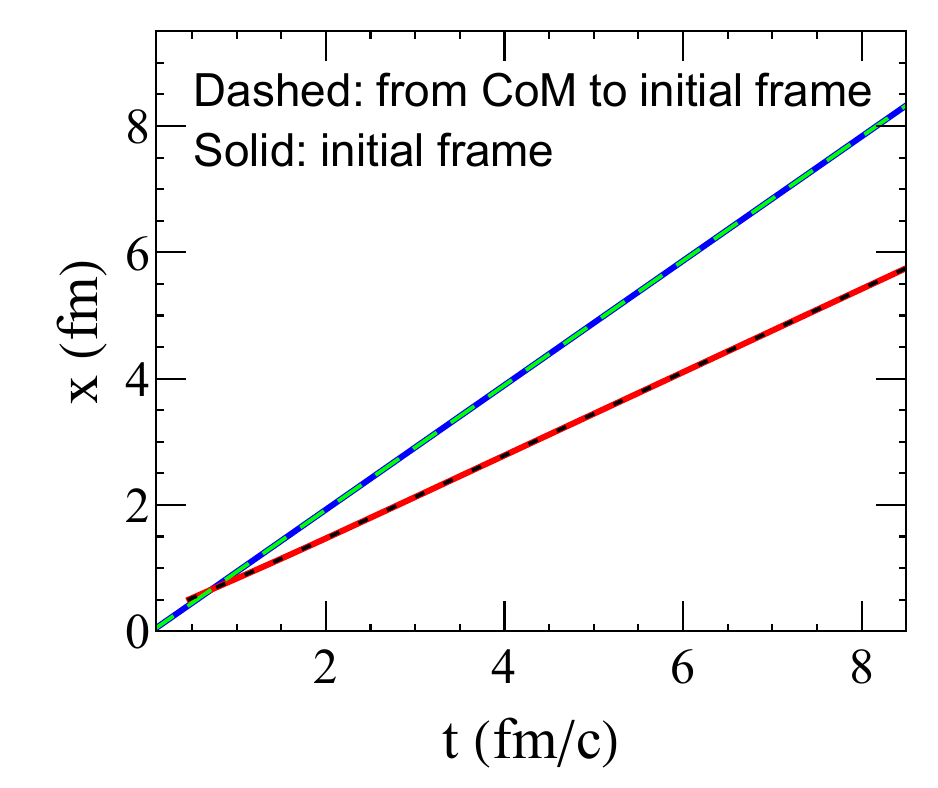}
\includegraphics[width=0.23\textwidth]{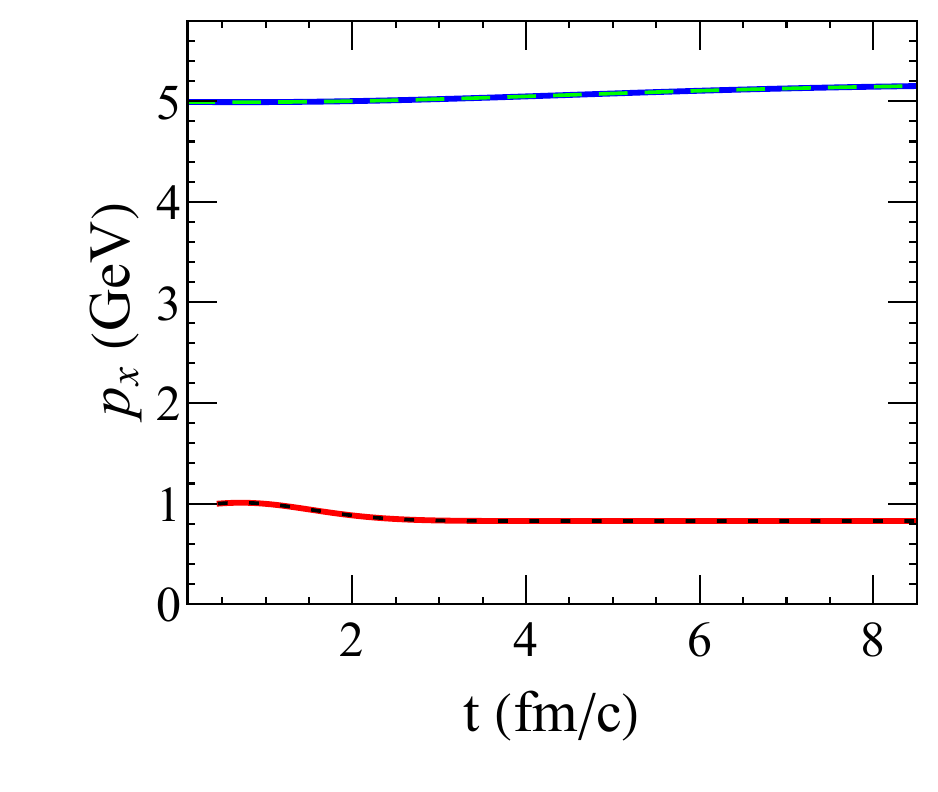}\\
\caption{The positions and momenta of particle 1 (blue-solid and green-dashed lines) and 2 (red-solid and black-dashed lines) as a function of the time $t$ applying the relativistic evolution equations. The dashed lines show the trajectory calculated first in the center-of-mass frame and then boosted to the initial frame, while the solid line shows the trajectory calculated directly in the initial frame. The initial condition I is taken and $\kappa=1$. 
} 
\label{fig.vc}
\end{figure}
It is essential to verify also numerically that the trajectories of the particles are independent of the reference system, in which the EoMs are calculated.
To demonstrate this, we compute the particle trajectories using the equations of motion~\eqref{eq.eomrela} with both vector and scalar potential (Eq.~\eqref{eq.potentialvands} in the frame in which the initial condition is given. In the second calculation, we transform the initial condition to the center-of-mass (CoM) frame of the two-particle system and calculate the EoMs~\eqref{eq.eomrela} in this frame. Finally, we transform the results from the CoM frame to the initial frame and compare the trajectories. 

The results are presented in Fig.~\ref{fig.vc}.
As shown in the figure, the trajectories are independent of the frame in which the time evolution has been calculated, demonstrating that the particle evolution is frame-independent, as expected from  our relativistic equations. This confirms the consistency and Lorentz invariance of our numerical implementation. The upper panel shows the position and the middle panel show the momentum of both particles as a function of the time $t$. The lower panel confirms that the total momentum is conserved during the time evolution.


\section{Relativistic N-body evolution}
\label{sec.nbody}
Now, we extend our approach to N-body systems. The mass shell condition yields N first-class constraints.
For four interacting particles the energy constraints are
\begin{eqnarray}
H_1&=&p^*_{1,\mu}p^{*\mu}_{1}-m_1^2+\Phi_1 ,\nonumber\\
H_2&=&p^*_{2,\mu}p^{*\mu}_{2} -m_2^2+\Phi_2 ,\nonumber\\
&...&\nonumber\\
H_N&=&p^*_{N,\mu}p^{*\mu}_{N} -m_N^2+\Phi_N.
\end{eqnarray} 
The total Dirac Hamiltonian is $H=\sum_i^N\lambda_i H_i$. 
The conservation of each Dirac Hamiltonian $H_i$ in time requires that the commutation relations are fulfilled, $\{H_i,H_j\}=0$ for any $i,j=1,...,N$.
In practice, however, this condition cannot generally be fulfilled for
interacting systems containing more than two particles, even when only
scalar interactions are considered. It is trivially satisfied in the
cascade limit, where the particles propagate freely between binary
collisions and no mean-field interaction is present.

For scalar interactions, one generally finds
$\{H_i,H_j\}\neq 0$ for an individual pair of constraints. Nevertheless,
a weaker relation, $
\{H_i,\sum_{j\neq i}H_j\}=0$, 
can still be satisfied~\cite{Marty:2012vs}. This implies that the
nonvanishing contributions from the individual Poisson brackets may
cancel when summed over all other particles. Such a weaker condition,
however, does not guarantee the pairwise compatibility
$\{H_i,H_j\}=0$ and therefore does not ensure the separate conservation
of each individual Dirac Hamiltonian during the evolution. Thus, it
does not provide a complete solution to the constraint-compatibility
problem for an interacting many-body system.

Alternative implementations of relativistic constrained dynamics,
designed to preserve the Dirac Hamiltonian of each particle throughout
the evolution, have been proposed in RQMD by Sorge~\cite{Sorge:1989dy}
and later adopted in JAM by Nara~\cite{Nara:2021fuu}. These approaches,
however, employ different sets of second-class constraints. While they
provide practical prescriptions for maintaining the mass-shell
conditions during the dynamical evolution, a general analytical proof
of the mutual compatibility of all individual mass-shell constraints
for an interacting many-body system is not provided.

Introduce multipliers for the time constraints as well:
\begin{eqnarray}
H_E=\sum_{i=1}^N \lambda_i H_i +\sum_{a=1}^N \mu_a \chi_a.
\end{eqnarray} 
There are $2N$ constraint-preservation equations. and unknown multipliers, $(\lambda_i,...\lambda_N,\mu_1,...,\mu_N)$.
For any observables, its time evolution satisfies,
\begin{eqnarray}
{d \mathcal{O} \over d\tau}=\{\mathcal{O}, H_E\}+{\partial \mathcal{O} \over \partial \tau}.
\end{eqnarray} 
Now the time evolution equations for coordinate and momentum can be expressed as,
\begin{eqnarray}
{dq_i^\mu \over d\tau}&=&\sum_{j=1}^N \lambda_j \{q_i^\mu, H_j\}+\sum_{a=1}^N \mu_a \{q_i^\mu, \chi_a\}, \nonumber\\
{dp_i^\mu \over d\tau}&=&\sum_{j=1}^N \lambda_j \{p_i^\mu, H_j\}+\sum_{a=1}^N \mu_a \{p_i^\mu, \chi_a\}.
\end{eqnarray} 
The preservation of $H_i$, requires $\{H_i, H_E\}=0$. Moreover, preservation of the time constraints gives $\{\chi_a, H_E\}=0$. They can be written explicitly as, 
\begin{eqnarray}
\sum_{j=1}^N \lambda_j \{H_i,H_j\}+\sum_{a=1}^N \mu_a \{H_i, \chi_a\}&=&0,\nonumber\\
{\partial \chi_a \over \partial \tau}+ \sum_{j=1}^N \lambda_j \{\chi_a,H_j\}+\sum_{a=1}^N \mu_a \{\chi_a, \chi_a\}&=&0.
\end{eqnarray}
It can be expressed as a matrix form,
\begin{equation}
\left(\begin{array}{cc} 
{\bm C} & -{\bm A}^T \\
{\bm A}  & {\bm D}  \\
 \end{array}\right)\left(\begin{array}{c} 
{\bm \lambda}\\
{\bm \mu} \\
 \end{array}\right)=\left(\begin{array}{c} 
0\\
{\bm b} \\
 \end{array}\right),
\label{eq.nbodymart}
\end{equation}
where $A_{ai}=\{\chi_a, H_i\}$, $C_{ij}=\{H_i, H_j\}$, $D_{ab}=\{\chi_a, \chi_b\}$, and $b_a=-{\partial \chi_a \over \partial \tau}$. If we take the previous time constraints. Then ${\bm b}=(0,...,1)^T$ and ${\bm D}=0$.

For two-body case, the Poisson bracket $\{H_1,H_2\}=0$, Then the first block equation becomes. $-{\bm A}^T{\bm \mu}=0$. In general, ${\rm det} {\bm A}\neq 0$ and therefore ${\bm \mu}=0$. The second block equation becomes ${\bm A} {\bm \lambda}={\bm b}$. This is the same as Eq.~\eqref{eq.2bodychiH}, as shown in  section~\ref{sec.2body}.
For N-body case, due to $\{H_i,H_j\}\neq 0$, we need to solve the matrix equation~\eqref{eq.nbodymart} exactly. 

There are N second-class constraints that can be extended from the previous two-body cases. The last constraint is the only one that includes the computational time $\tau$,
\begin{eqnarray}
\chi_1&=&{1\over N}(q_1^\mu-q_2^\mu) U_\mu =0,\nonumber\\ 
\chi_2&=&{1\over N}(q_2^\mu-q_3^\mu) U_\mu =0,\nonumber\\ 
&...&,\nonumber\\
\chi_{n}&=&{1\over N}(q_{n}^\mu-q_{n+1}^\mu) U_\mu =0,\nonumber\\
&...&,\nonumber\\
\chi_N&=&{1\over N}(q_1^\mu+q_2^\mu+...+q_N^\mu) U_\mu-\tau =0,
\label{eq.nobodychia}
\end{eqnarray} 
where $U_\mu=P^{*}_{\mu}/\sqrt{P^{*2}}$ is the four velocity of the system.
This means that the uniform evolution time $\tau$  in the center-of-mass frame is given by $\tau=\sum_i t_i^{com}/N$.
The particle times $t_i$ can be reconstructed,
\begin{eqnarray}
t_i={1\over U_0}(\tau+x_iU_x+y_iU_y+z_iU_z).\quad i=1,...,N.
\end{eqnarray} 
With these second-class constraints, one can easily prove $\{\chi_a,\chi_b\}=0$, that means ${\bm D}=0$. Therefore, one can obtain ${\bm \lambda}$ the from the second block ${\bm \lambda}={\bm A}^{-1} {\bm b}$. 

With the above time constraints, $A_{ai}$ can be expressed analytically,
\begin{eqnarray}
\{\chi_1,H_1\}&=&{2\over N}U_\mu p_1^{*\nu}\left({\partial p^*_{1,\nu}\over \partial p_{1,\mu}}-{\partial p^*_{1,\nu}\over \partial p_{2,\mu}}\right)\\
&+& {U_\mu\over N} \left({\partial \Phi_1\over \partial p_{1,\mu}}-{\partial \Phi_1\over \partial p_{2,\mu}}\right),\nonumber\\
\{\chi_1,H_2\}&=&{2\over N}U_\mu p_1^{*\nu}\left({\partial p^*_{2,\nu}\over \partial p_{1,\mu}}-{\partial p^*_{2,\nu}\over \partial p_{2,\mu}}\right)\nonumber\\
&+& {U_\mu\over N} \left({\partial \Phi_2\over \partial p_{1,\mu}}-{\partial \Phi_2\over \partial p_{2,\mu}}\right), \nonumber\\
&...&, \nonumber\\
\{\chi_1,H_N\}&=&{2\over N}U_\mu p_N^{*\nu}\left({\partial p^*_{N,\nu}\over \partial p_{1,\mu}}-{\partial p^*_{N,\nu}\over \partial p_{2,\mu}}\right)\nonumber\\
&+& {U_\mu\over N} \left({\partial \Phi_N\over \partial p_{1,\mu}}-{\partial \Phi_N\over \partial p_{2,\mu}}\right), \nonumber\\
&...&, \nonumber\\
\{\chi_n,H_m\}&=&{2\over N}U_\mu p_m^{*\nu}\left({\partial p^*_{m,\nu}\over \partial p_{n,\mu}}-{\partial p^*_{m,\nu}\over \partial p_{n+1,\mu}}\right)\nonumber\\
&+& {U_\mu\over N} \left({\partial \Phi_m\over \partial p_{n,\mu}}-{\partial \Phi_m\over \partial p_{n+1,\mu}}\right), \nonumber\\
&...&, \nonumber\\
\{\chi_N,H_1\}&=&{2\over N}U_\mu p_1^{*\nu} \left(\sum_{i=1}^N{\partial p^*_{1,\nu} \over \partial p_{i,\mu}}\right)+ {U_\mu\over N}\sum_{i=1}^N {\partial \Phi_i\over \partial p_{i,\mu}},\nonumber\\
&...&, \nonumber\\
\{\chi_N,H_N\}&=&{2\over N}U_\mu p_N^{*\nu} \left(\sum_{i=1}^N{\partial p^*_{N,\nu} \over \partial p_{i,\mu}}\right)+ {U_\mu\over N}\sum_{i=1}^N {\partial \Phi_i\over \partial p_{i,\mu}}.\nonumber
\end{eqnarray} 
Solving the eigen equation, we obtain the analytical form of the ${\bm \lambda}$, 
\begin{eqnarray}
\lambda_1&=&{T_1\over  2~{\rm det} \left[M\right]}, \nonumber\\
&...&,\nonumber\\
\lambda_N&=&{T_N \over 2~{\rm det} \left[M\right]},
\label{eq.nobodylambda}
\end{eqnarray}
where $M$ is a $N\times N$ matrix with the element,
\begin{eqnarray}
M_{ij}=U_\mu p_i^{*\nu}{\partial p^*_{i,\nu} \over \partial p_{j,\mu}}, \quad i,j=1,...,N .
\end{eqnarray}
The numerator of $\lambda_m$ can be expressed as
\begin{eqnarray}
T_{m}=\sum_{n=1}^N C_{mn},
\end{eqnarray}
where $C_{mn}$ is the cofactor matrix of matrix $M$. The details of the algorithm used to compute the above derivatives are shown in the Appendix.~\ref{app.a}. 

With this, ${\bm u}={\bm A}^{-T} {\bm C}{\bm A}^{-1}{\bm b}$ from the first block equation. 
The explicit form of element of ${\bm C}$ is 
\begin{eqnarray}
C_{ij}&=&\{H_i,H_j\}\nonumber\\
&=&\sum_{k=1}^N\sum_{\mu=0}^3g_{\mu\mu}\left({\partial H_i\over\partial q_k^\mu}{\partial H_j\over\partial p_k^\mu}-{\partial H_i\over\partial p_k^\mu}{\partial H_j\over\partial q_k^\mu}\right).
\end{eqnarray} 

The final EoM for the N-body system can be expressed as,
\[
\boxed{
\begin{gathered}
\frac{dq_i^\mu}{d\tau}
 = \sum_{k=1}^N \lambda_k
 \left(2 p_k^{*\nu}\frac{\partial p^*_{k,\nu}}{\partial p_{i,\mu}} + \frac{\partial\Phi_k}{\partial p_{i,\mu}}\right)+\sum_{a=1}^N\mu_a{\partial \chi_a\over \partial p_{i,\mu}},
\\[6pt]
\frac{dp_i^\mu}{d\tau}
 = -\sum_{k=1}^N \lambda_k
 \left(2 p_k^{*\nu}\frac{\partial p^*_{k,\nu}}{\partial q_{i,\mu}}
    + \frac{\partial\Phi_k}{\partial q_{i,\mu}} \right)-\sum_{a=1}^N\mu_a{\partial \chi_a\over \partial q_{i,\mu}}.
\end{gathered}
}
\]

This procedure guarantees that the individual mass-shell constraints $H_i=0$ are preserved throughout the evolution. Furthermore, when the transverse relative coordinate $q_{T,ij}$ is defined with respect to the global four-velocity $U_\mu$, the momentum derivative of the scalar potential vanishes on the constraint surface owing to the time constraint $q_{T,ij}^\mu U_\mu=0$, as shown in Eq.~\eqref{eq.dq2dp}.

\subsection{Four-nucleon system}
\label{subsec.4nucleon}
In this section, we employ the framework to four-body systems, which can be treated as a simplest symmetric ``heavy-ion'' collisions---deuteron-deuteron collision. This provides a simple test of relativistic dynamics beyond the two-body level and offers a pathway toward extending the framework to the N-body case. It can also serve as an additional probe of the respective roles of scalar and vector potentials in such collisions. 

\begin{figure}[!htb]
\centering
\includegraphics[width=0.45\textwidth]{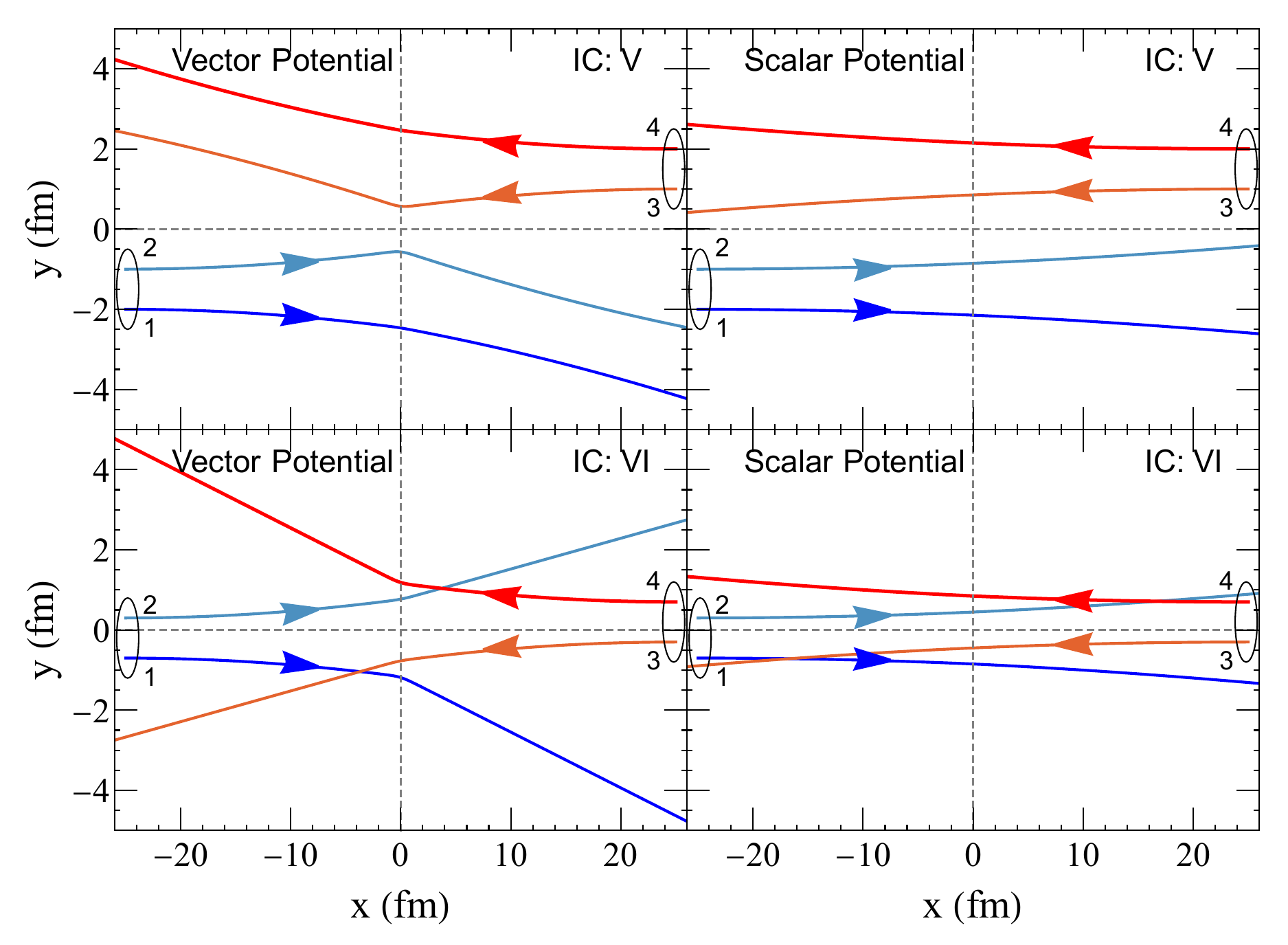}
\caption{Time evolution of the positions and momenta of particles 1 (blue lines), 2 (dark-blue lines), 3 (dark-red line), and 4 (red line) applying the relativistic time evolution equations. The upper panel present the results for the initial condition V (IC: V), while the lower panel show that for the initial condition VI (IC: VI), and in both cases $\kappa=1$. The left panels show the results for a vector potential and right panels that with a scalar potential but the scalar potential is multiplied by -1 to make this potential repulsive.} 
\label{fig.4bodyit12}
\end{figure}

The EoMs for the four-body system can be expressed directly as, 
\begin{eqnarray}
\label{eq.eomrela4}
{dq_i^\mu \over d\tau}&=&\sum_{k=1}^4 \lambda_k \left(2p^{*\nu}_{k}{\partial p^*_{k,\nu} \over \partial p_{i,\mu}}\right)+\sum_{a=1}^4\mu_a{\partial \chi_a\over \partial p_{i,\mu}},\\
{dp_i^\mu \over d\tau}&=&-\sum_{k=1}^4\lambda_k \left(2p_k^{*\nu}{\partial p^*_{k,\nu} \over \partial q_{i,\mu}}+{\partial \Phi_k \over \partial q_{i,\mu}} \right)-\sum_{a=1}^4\mu_a{\partial \chi_a\over \partial q_{i,\mu}}.\nonumber
\end{eqnarray} 
In the following, we focus on the four nucleon system and solve the EoMs~\eqref{eq.eomrela4} numerically with two initial conditions:
\begin{enumerate}[label=\Roman*., start=5]
\item ${\bf x}_1=(-25.0,-2.0,0)~\rm fm$, ${\bf x}_2=(-25.0,-1.0,0)~\rm fm$, \nonumber\\
${\bf x}_3=(25.0,1.0,0)~\rm fm$, ${\bf x}_4=(25.0,2.0,0)~\rm fm$,\nonumber\\
${\bf p}_1=(5,0,0)~\rm GeV$, ${\bf p}_2=(5,0,0)~\rm GeV$, \nonumber\\
${\bf p}_3=(-5,0,0)~\rm GeV$, ${\bf p}_4=(-5,0,0)~\rm GeV$;
\item ${\bf x}_1=(-25.0,-0.7,0)~\rm fm$, ${\bf x}_2=(-25.0,0.3,0)~\rm fm$, \nonumber\\
${\bf x}_3=(25.0,-0.3,0)~\rm fm$, ${\bf x}_4=(25.0,0.7,0)~\rm fm$,\nonumber\\
${\bf p}_1=(5,0,0)~\rm GeV$, ${\bf p}_2=(5,0,0)~\rm GeV$, \nonumber\\
${\bf p}_3=(-5,0,0)~\rm GeV$, ${\bf p}_4=(-5,0,0)~\rm GeV$.
\end{enumerate}
Initial condition V can be considered as vaguely as two deuterons colliding with a large impact parameter, however with the difference that the two body potential between the nucleons is always either attractive of repulsive. To clearly compare the influence of the vector and scalar potentials, we multiply the scalar potential by -1. Then, both the vector and scalar potentials are repulsive. Initial condition VI, which has a smaller impact parameter, can be considered as a collisions between two deuterons for which the impact parameter is smaller than the size of the deuteron. The interactions between nucleons are assumed to be the same as before~\eqref{eq.potentialvandsN}. 
The results are shown in Fig.~\ref{fig.4bodyit12}.
\begin{figure}[!htb]
\centering
\includegraphics[width=0.45\textwidth]{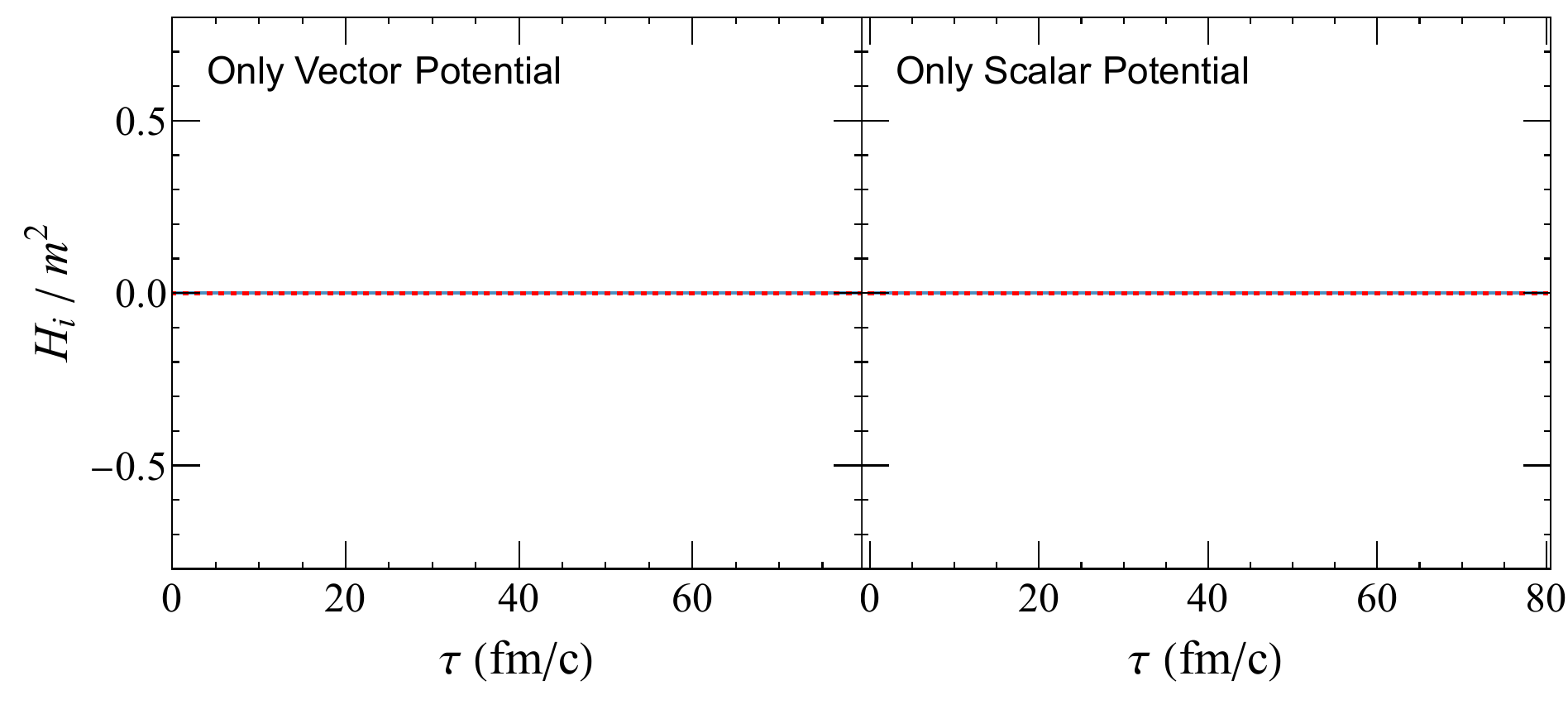}
\caption{Time evolution of the Dirac Hamiltonian of each particle with only vector (left) and scalar potential (right). The initial condition V is used and $\kappa=1$.} 
\label{fig.4bodyHi}
\end{figure}
The top panel, which shows the results of our calculation for a large impact parameter (IC: V), displays again the very different trajectories created by vector and scalar potentials, which have the same strength. Whereas for a scalar potential the repulsive interaction between the nucleons, which move into the same direction, is dominant for the trajectory, for the vector potential, due to the large relative momentum,  the interaction of the particles moving in opposite direction, is equally important and leads to a large angle scattering of both ``deuterons''. 
This is also visible for central collisions (IC: VI) for which the trajectories are displayed in the bottom panel. The vector potential
leads to a deflection of the nucleons of the ``deuteron'' in opposite directions, what is much less the case for a scalar potential. 

The evolution of $H_i$ for each particle is monitored and presented in Fig.~\ref{fig.4bodyHi}. It is evident that the constraint $H_i=0$ is maintained throughout the evolution.

\subsection{Cluster separability}
Besides Lorentz covariance and energy-momentum conservation, another fundamental requirement of relativistic many-body dynamics is the principle of cluster separability. It requires that, when the interaction between two subsystems vanishes, the dynamical evolution of one subsystem should become completely independent of the existence of the other subsystem. Namely, if the total Hamiltonian can be decomposed as
\begin{eqnarray}
H=H_A+H_B,
\end{eqnarray}
where $H_A$ and $H_B$ describe two non-interacting clusters, then the evolution of each cluster should be identical to that obtained by solving the corresponding subsystem independently.

The realization of cluster separability in relativistic constrained dynamics is closely related to the choice of the time constraints. In the original RQMD formulation, Sorge {\it et al.} proposed pairwise equal-time constraints defined in the two-body center-of-mass frame~\cite{Sorge:1989dy},
\begin{eqnarray}
\chi_i&=&\sum_{j\neq i}G_{ij}\,u_{ij,\mu}(q_i^\mu-q_j^\mu)=0, \nonumber\\
\chi_N&=&\sum_{i=1}^N q_i^\mu U_\mu -\tau=0
\end{eqnarray}
where $u_{ij,\mu}=(p_{i,\mu}+p_{j,\mu})/\sqrt{(p_{i,\mu}+p_{j,\mu})^2}$ and $U_\mu$
is the four-velocity of the two-particle and N-particles center-of-mass system, respectively. The weight function is chosen as
\begin{eqnarray}
G_{ij}=
{1\over q_{T,ik}^2/L_c}\exp\!\left(q_{T,ij}^2/L_c\right),
\end{eqnarray}
where
\begin{eqnarray}
q_{T,ij}^\mu=q_{ij}^\mu-(q_{ij}^\nu u_{ij,\nu})u_{ij}^\mu.
\end{eqnarray}
This definition makes each pair interaction depend only on the corresponding two particles instead of total velocity of N-particle system, and therefore improves the locality of the interaction.

The exponential weight plays a crucial role in the original RQMD formulation. Since $\exp(q_{T,ij}^2/L_c)$ decreases rapidly with the invariant separation
between particles. Consequently, only nearby particles contribute
significantly to the synchronization of particle $i$, whereas the
contributions from distant particles are exponentially suppressed. In the
dilute limit, where independent clusters are spatially well separated,
the coupling between different clusters automatically disappears, and
each isolated cluster is synchronized with respect to its own effective
center-of-mass frame. This local synchronization is the key ingredient
that allows the original RQMD formulation to satisfy the cluster
separability condition.

The price for this construction is that the synchronization conditions
must be reconstructed whenever the particle configuration changes. After
each collision or decay, the time constraints are generally violated,
and one must solve a coupled set of nonlinear equations to recover the
constraint surface. As discussed by Nara {\it et al.}~\cite{Nara:2023vrq}, this procedure is
computationally very expensive and the iterative solution does not always
converge. 

For practical applications, the later RQMD/S~\cite{Isse:2005nk} and JAM~\cite{Nara:2023vrq}
transport models adopt a much simpler global synchronization,
\begin{eqnarray}
\chi_i &=& (q_i^\mu-q_N^\mu)U_\mu=0,\nonumber\\
\chi_N &=& q_N^\mu U_\mu-\tau=0,
\end{eqnarray}
where $U_\mu=P_\mu/\sqrt{P^2}$
is the four-velocity associated with the total four-momentum of the
complete system. In practical implementations, the evolution is further
performed in the overall center-of-mass frame, which gives $U^\mu=(1,0,0,0)$.
Consequently, all particles evolve with respect to the same global
coordinate time, and the Lagrange multipliers can be obtained analytically as discussed in Section~\ref{sec.IVb}. When the system separates into several independent clusters,
the center-of-mass frame of each isolated cluster generally differs from
that of the original system. To satisfy the cluster separability condition, the authors manually evolve the cluster with its own synchronization
vector rather than the common synchronization vector of the original system. The authors found that this approximation works remarkably well, yielding results that are nearly identical to those obtained without explicit cluster separation~\cite{Nara:2023vrq}.

The present work adopts the same global synchronization strategy.
Consequently, the present formalism inherits the same limitation
associated with the global synchronization. When the system separates
into several non-interacting clusters, exact cluster separability would
require each isolated cluster to evolve with its own synchronization
vector,
\begin{eqnarray}
U_A^\mu=\frac{P_A^\mu}{\sqrt{P_A^2}},\qquad
U_B^\mu=\frac{P_B^\mu}{\sqrt{P_B^2}},
\end{eqnarray}
instead of the common synchronization vector of the original system.
Equivalently, the global synchronization vector should be reassigned
according to the center-of-mass frame of each isolated cluster before
continuing the evolution independently. However, this reassignment is
not included in the present work and would require dynamically
identifying isolated clusters and reconstructing the corresponding
constraint equations during the evolution.

\section{Summary}
\label{sec.summary}
In this work we derived the relativistic Molecular Dynamics for an N-body system that interacts with both scalar and vector two body interactions. These equations are derived without any approximations, including the momentum-dependent derivative terms, which were not considered in previous studies. Our formulation addressed several fundamental challenges in relativistic many-body dynamics. In particular, we examined the implications of different choices of time constraints, analyzed the non-relativistic limit of the derived equations, tested the frame independence of the system evolution, and disentangled the distinct dynamical roles of scalar and vector potentials in scattering processes. These aspects are investigated in detail for two- and four-body systems. Finally, we present the general form of the equation of motion (EoM) for a N-body system. 

The present formulation provides a solid theoretical foundation for future applications to relativistic heavy-ion collisions at intermediate beam energies, such as those to be explored by CBM at FAIR, HIAF, NICA, and the RHIC Beam Energy Scan program. At these collision energies, the dynamics is dominated by compressed baryonic matter, where mean-field interactions play a decisive role in determining collective flow, stopping, cluster production, and particle correlations. A fully covariant treatment of both scalar and vector interactions is therefore essential for a consistent extraction of the high-density equation of state from experimental observables. The framework developed here can be directly incorporated into relativistic QMD transport models, enabling systematic investigations of the influence of relativistic mean fields on hadronic dynamics and providing a unified description of both equilibrium properties and nonequilibrium evolution. This opens the possibility of studying the microscopic origin of collective phenomena, cluster and hypernuclear production, as well as femtoscopic correlations within a common relativistic transport framework, thereby establishing a closer connection between transport theory and the physics program of upcoming heavy-ion facilities. These topics will be addressed in future studies.

\vspace{0.3cm}
\noindent {\bf Acknowledgment}: 
We acknowledge inspiring discussions with Y. Nara and T. Song.
The computational resources utilized for this work were provided by the Center for Scientific Computing (CSC) at Goethe University Frankfurt.
\vspace{1cm}

\appendix

\section{Time constraints and relativistic kinematics}
\label{app.b}

Time is not universal  in relativistic mechanics . Each particle has its own time (own clock), the forth  component of its position vector. For many particle systems the clock times of the different particles have to be correlated by time constraints. To discuss the influence of the time constraints on the EoMs, we take as an example the time evolution of one particle.
\subsection{Time constraints}
We study here 2 different time constraints and their influence on the equations of motion and on the trajectories of the particle ${\bf x}(t)$, the vector component as a function of the zero component of the position 4-vector.
We start with the  (Lorentz  invariant) time constraint
\begin{equation}
 \chi=q^\mu u_\mu-\tau=0. 
 \label{eq:tcrel}
\end{equation}
For the differential quantity $d\tau$ this constraint yields  
\begin{equation}
d\tau= \gamma(dt-(d{\bf q)v})= \gamma dt(1-{\bf v}^2)=\frac{1}{\gamma}dt.
\label{eq:dtdtau}
\end{equation}
$u^\mu$ is defined as $p_\mu/\sqrt{p_\nu p^\nu}= p_\mu/m=\gamma (1,{\bf v})$. Taking it into Eq.~\eqref{eq.eom1body} yields the EoMs, 
\begin{eqnarray}
{d q^\mu \over d\tau}&=&{p^\mu \over p^\mu u_\mu}={p^\mu \over m},\nonumber\\
{d p^\mu \over d\tau}&=&0.
\label{eq.joerg1}
\end{eqnarray}
Another choice, which is discussed in the literature~\cite{Sudarshan:1981pp} is the non-relativistic time constraint
\begin{eqnarray}  
    \chi=q_0-\tau=t-\tau= 0.
\end{eqnarray}
This time constraint obeys the world line condition for the $N$-body case without interaction but not for an interacting system ~\cite{Sudarshan:1981pp}.
This time constraint leads directly to the EoMs
\begin{eqnarray}
{d {q^\mu} \over d\tau}&=&{d {q^\mu} \over dt}={{p^\mu}\over p_0}, \nonumber \\
{d {p^\mu} \over d\tau}&=&{d {p^\mu} \over dt}=0,
\label{eq:KG}
\end{eqnarray} 
which are not Lorentz invariant, because $p^\mu/p_0$ is no quantity which can be Lorentz transformed.

Even covariant time constraints are not unique. For example, 
\begin{equation}
\chi=2q^\mu U_\mu-\tau=0    
\end{equation}
also fulfills the requirement of a time constraint to connect in a Lorentz invariant way the $dt$ with $d\tau$. This choice gives  ${d q^\mu \over d\tau}={p^\mu \over 2 p^\mu U_\mu}={p^\mu \over 2m}$ and therefore not, as we will see, the right non-relativistic limit. For the time evolution of more than one particles there are, however, several forms of time constraints, which give the right non-relativistic limit. Therefore the time evolution of a relativistic many-body system is not completely determined without fixing the time constraints.

\begin{figure}[!htb]
\centering
\includegraphics[width=0.23\textwidth]{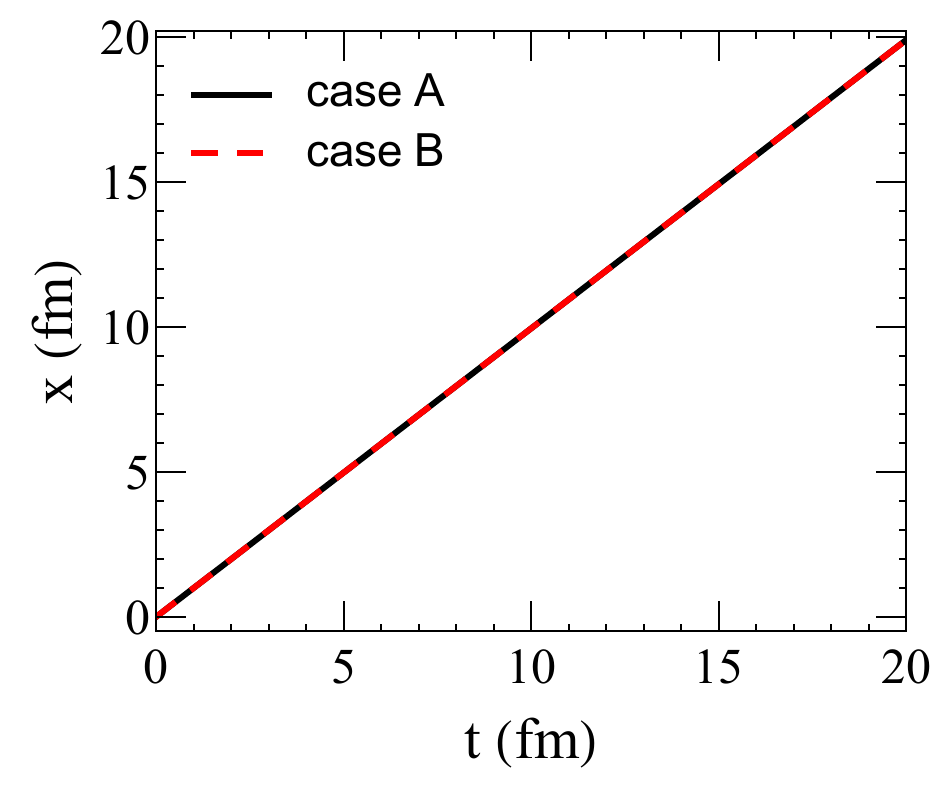}
\includegraphics[width=0.23\textwidth]{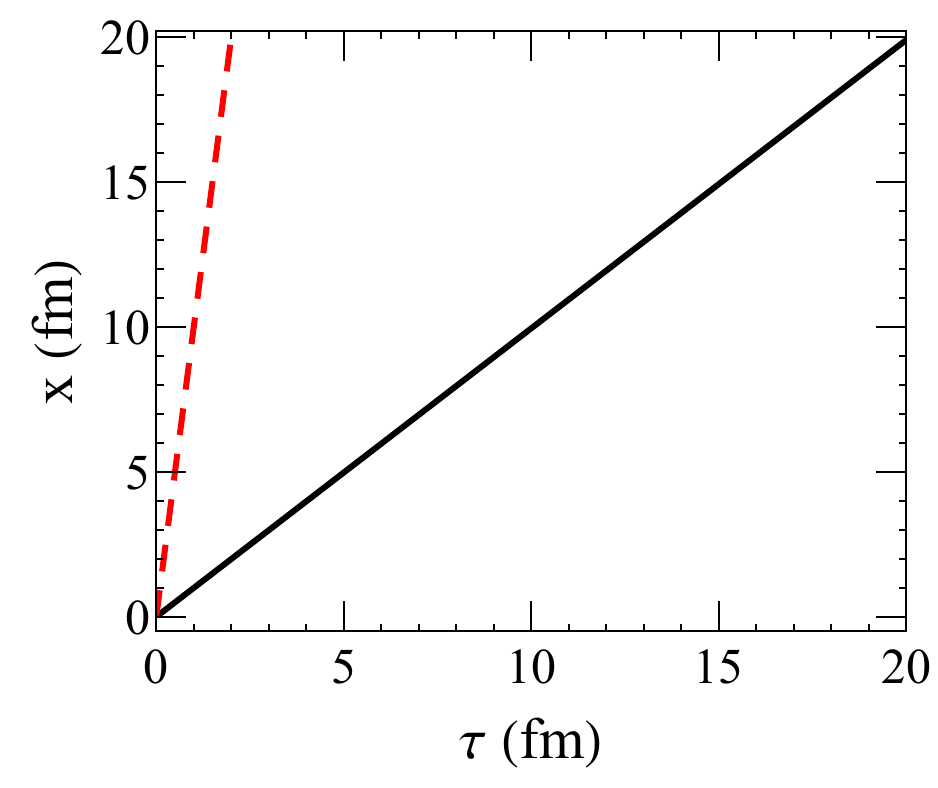}
\caption{Trajectories in two  cases with the time $t$ and the computational time $\tau$.} 
\label{fig.1bodyt}
\end{figure}
Let us now examine the trajectories obtained from the two EoMs Eqs.~\eqref{eq:KG} (called case A)  and \eqref{eq.joerg1} (called case B). For case A we find
\begin{eqnarray}
x^\mu=x_0^\mu+{p^\mu \over p^0}\tau,
\end{eqnarray}
whereas case B yields
\begin{eqnarray}
x^\mu=x_0^\mu+{p^\mu \over m}\tau.
\end{eqnarray}
As a function of $\tau$ the trajectories are rather different but for ${\bf x}(t)$, where $t$ is the zero component of $x^\mu$ we find identical trajectories. This is shown in Fig.~\ref{fig.1bodyt}.
The calculation is performed with the initial position $x=y=z=0$ at $\tau_0=0$, and the initial momentum is chosen as $p_x=10$~GeV, $p_y=p_z=0$, and the particle mass is $m=1$~GeV. This demonstrates that the trajectory itself does not change under different time constraints.

\subsection{Standard derivation of relativisitic EoMs}
It is worthwhile to connect these constraints to the EoMs derived in textbooks for the relativistic kinematics. The non-relativistic equations of motion are given by
\begin{eqnarray}
{d {\bf q} \over dt}&=&{{\bf p}\over m}={ {\bf v} },\nonumber\\
{d {\bf p} \over dt}&=&0.
\label{eq.1bodyceoma}
\end{eqnarray}
To construct 4-vector one extends the three-vectors to the position and momentum 4-vectors. In an arbitrary reference system the particle position and momentum is expressed by the 4-vector  $q^\mu=(q^0,{\bf q})$ and $p^\mu=(p^0,{\bf p})$ with $q^0=t$ and $p^0=\sqrt{m^2+{\bf p}^2}$.

Using the relativistic relations ${\bf p}=\gamma m {\bf v} $ and $p_0 = \gamma m$ the first equation can be rewritten as
\begin{equation}
 {d {\bf q} \over dt}={ {\bf v} }={{\bf p}\over p_0}.
 \label{eq:KG3}   
 \end{equation}
Including the forth components, we obtain the same EoMs as for the non covariant time constraint, Eq.~\eqref{eq:KG}.
It describes the time evolution, seen from a system with respect to which the particle has a velocity ${\bf p}/p_0$.  To make them invariant we introduce the scalar quantity 
\begin{equation}
 (d\tau)^2= dx_\mu dx^\mu =(\frac{1}{\gamma} dt)^2 \underbrace{=}_{\rm particle\ rest\ system} (dt)^2,
\end{equation}
which is the infinitesimal time in the particle rest system. This relation is identical to relativistic time constraint, Eq.~\eqref{eq:tcrel}. Replacing $dt$ by $d\tau$, Eq.~\eqref{eq:KG3} reads as
\begin{equation}
{d {\bf q} \over d\tau}= \gamma {\bf v} ={\bf u} = \frac{\bf p}{m},    
\end{equation}
where ${\bf u}$ are the vector components of the velocity 4-vector $u^\mu= \gamma(1,{\bf v})$ and $p^\mu = m u^\mu$. Including the zero components we have then the time evolution equations, Eqs.~\eqref{eq.joerg1}, which are covariant.  We can therefore conclude that the relativistic EoMs, which we apply in the relativistic constraint dynamics, can be related to the standard results obtained in relativistic kinematics at least for the one-body case.

\section{Numerical method for computing the mechanical momentum and related derivatives}
\label{app.a}

For nuclear systems, the interaction involve both the vector and scalar potential, as shown in Eq.~\eqref{eq.potentialvandsN}. The vector potential is given by 
\begin{eqnarray}
A_i^\mu=\sum_{j\neq i}^Np_j^{*\mu}\rho_{ij}(q_T), \quad i=1,...,N,
\end{eqnarray}
where $p_j^{*\mu}=p_j^\mu-A_j^\mu$.
Aiming at calculating the equations of motion and the parameters $\lambda_i$, one need the explicit form of the mechanical momentum $p_i^*$ for $N$-bodies.  
The above equation can be written as,
\begin{eqnarray}
A_i^\mu+\sum_{j\neq i}^NA_j^{\mu}\rho_{ij}=\sum_{j\neq i}^Np_j^{\mu}\rho_{ij}
\end{eqnarray}
or converted into a matrix form,
\begin{eqnarray}
\bf {D}\cdot {\bf A} = {\bf b},
\end{eqnarray}
where ${\bf A}=(A_1,...,A_N)^T$. The vector ${\bf b}=(b_1,...,b_N)^T$ with components $b_i=\sum_{j\neq i}^Np_j^{\mu}\rho_{ij}$. ${\bf D}$ is a $N\times N$ symmetric matrix with $D_{ii}=1$ and $D_{ij}=D_{ji}=\rho_{ij}$ for $i\neq j$.
Therefore, the vector potential can be obtained as,
\begin{eqnarray}
{\bf A} = {\bf D}^{-1}\cdot {\bf b}.
\end{eqnarray}
The mechanical momentum is $p_i^{*\mu}=p_i^\mu-A_i^\mu$. Now we just need to calculate the inverse of the matrix $\bf D$, which is related to the interaction density.

With $A^\mu$, we can further calculate the other two derivatives. The first one is,
\begin{eqnarray}
{\partial p_{i,\nu}^* \over \partial p_{j,\mu}}&=&\delta_{ij}\delta_{\nu}^{\mu}-{\partial A_{i,\nu} \over \partial p_{j,\mu}}\nonumber\\
&=& \delta_{ij}\delta_{\nu}^{\mu} -\sum_k (D^{-1})_{ik} {\partial b_{k,\nu} \over \partial p_{j,\mu}}.
\end{eqnarray}
From the definition of $b_i$, we have,
\begin{eqnarray}
{\partial b_{k,\nu} \over \partial p_{j,\mu}}=\sum_{l\neq k} \delta_{lj}\delta_{\nu}^{\mu}\rho_{kl}.
\end{eqnarray}
The derivation with respect to the coordinates is given by
\begin{eqnarray}
{\partial p_{i,\nu}^* \over \partial q_{j,\mu}}=-{\partial A_{i,\nu} \over \partial q_{j,\mu}}&=&\sum_{m,n} (D^{-1})_{im} {\partial D_{mn} \over \partial q_{j,\mu}}A_{n,\nu}\nonumber\\
&-&\sum_k (D^{-1})_{ik} {\partial b_{k,\nu} \over \partial q_{j,\mu}},
\end{eqnarray}
where
\begin{eqnarray}
{\partial b_{k,\nu} \over \partial q_{j,\mu}}=\sum_{l\neq k}p_{l,\nu} {\partial \rho_{kl} \over \partial q_{j,\mu}}=\sum_{l\neq k}p_{l,\nu} (\delta_{jk}-\delta_{jl}) {\partial \rho_{kl} \over \partial (q_k-q_l)_\mu }.
\end{eqnarray} 

\bibliographystyle{apsrev4-1.bst}
\bibliography{Ref}

\end{document}